\documentclass[preprint,11pt,a4paper]{elsarticle}

\usepackage{graphicx}

\usepackage{amssymb}
\usepackage{amsmath}

\usepackage{lineno}

\biboptions{compress}

\usepackage[scriptsize,tight,figbotcap]{subfigure}
\usepackage[cm]{fullpage}
\usepackage[title]{appendix}
\usepackage{cases}
\usepackage{siunitx}
\usepackage{doi}
\usepackage{hyperref}
\hypersetup{
    colorlinks,
    linkcolor={red!50!black},
    citecolor={blue!50!black},
    urlcolor={blue!80!black}
}

\journal{Journal of Non-Newtonian Fluid Mechanics}

\begin{document}
\newcommand{\vf}[1]{\underline{#1}}
\newcommand{\tf}[1]{\underline{\underline{#1}}}
\newcommand{\pd}[2]{\frac{\partial #1}{\partial #2}}

\begin{frontmatter}



\title{Viscoplastic flow in an extrusion damper}


\author[up]{Alexandros Syrakos\corref{cor1}}
\ead{syrakos@upatras.gr}

\author[up]{Yannis Dimakopoulos}
\ead{dimako@chemeng.upatras.gr}

\author[dms]{Georgios C. Georgiou}
\ead{georgios@ucy.ac.cy}

\author[up]{John Tsamopoulos}
\ead{tsamo@chemeng.upatras.gr}

\cortext[cor1]{Corresponding author}

\address[up]{Laboratory of Fluid Mechanics and Rheology, Department of Chemical Engineering, University of Patras, 26500 Patras, 
Greece}
\address[dms]{Department of Mathematics and Statistics, University of Cyprus, PO Box 20537, 1678 Nicosia, Cyprus}

\begin{abstract}
Numerical simulations of the flow in an extrusion damper are performed using a finite volume method. The damper is assumed to 
consist of a shaft, with or without a spherical bulge, oscillating axially in a containing cylinder filled with a viscoplastic 
material of Bingham type. The response of the damper to a forced sinusoidal displacement is studied. In the bulgeless case the 
configuration is the annular analogue of the well-known lid-driven cavity problem, but with a sinusoidal rather than constant 
lid velocity. Navier slip is applied to the shaft surface in order to bound the reaction force to finite values. Starting from a 
base case, several problem parameters are varied in turn in order to study the effects of viscoplasticity, slip, damper geometry 
and oscillation frequency to the damper response. The results show that, compared to Newtonian flow, viscoplasticity causes the 
damper force to be less sensitive to the shaft velocity; this is often a desirable damper property. The bulge increases the 
required force on the damper mainly by generating a pressure difference across itself; the latter is larger the smaller the gap 
between the bulge and the casing is. At high yield stresses or slip coefficients the amount of energy dissipation that occurs 
due to sliding friction at the shaft-fluid interface is seen to increase significantly. At low frequencies the flow is in quasi 
steady state, dominated by viscoplastic forces, while at higher frequencies the fluid kinetic energy storage and release also 
come into the energy balance, introducing hysteresis effects.
\end{abstract}


\end{frontmatter}

This article is published in: Journal of Non-Newtonian Fluid Mechanics 232 (2016) 102--124, \\ \doi{10.1016/j.jnnfm.2016.02.011}

\textcopyright 2016. This manuscript version is made available under the CC-BY-NC-ND 4.0 license 
\url{http://creativecommons.org/licenses/by-nc-nd/4.0/}



\section{Introduction}
\label{sec: introduction}

Viscous dampers dissipate mechanical energy into heat through the action of viscous stresses in a fluid. A common design involves 
the motion of a piston in a cylinder filled with a fluid, such that large velocity gradients develop in the narrow gap between 
the piston head and the cylinder, resulting in viscous and pressure forces that resist the piston motion. The potential of such 
dampers can be enhanced by the use of rheologically complex fluids. For example, use of a shear-thinning fluid such as silicon 
oil \cite{Constantinou_1993, Hou_2008} weakens the dependency of the damper reaction force on the piston velocity, which is often 
desirable as it maximises the absorbed energy for a given force capacity. The same effect can be achieved to a higher degree if 
the fluid is viscoplastic.

Viscoplastic materials flow as liquids when subjected to a stress that exceeds a critical value, but respond as rigid solids 
otherwise. More specifically, according to the von Mises yield criterion, flow is assumed to occur when the stress magnitude 
(related to the second invariant of the stress tensor) exceeds a critical value called the yield stress. Such materials are 
usually concentrated suspensions of solid particles or macromolecules. They are classified as generalised Newtonian fluids because 
their viscosity depends on the local shear rate, while they do not exhibit elastic effects. Broad surveys of yield-stress 
materials are given by Bird et al.\ \cite{Bird_1982}, Barnes \cite{Barnes_1999} and Balmforth et al.\ \cite{Balmforth_2014}. The 
simplest viscoplastic materials are Bingham fluids, where the magnitude of the stress increases linearly with the rate of strain 
once the yield stress has been exceeded. Herschel-Bulkley fluids exhibit shear-thinning (or thickening) after yielding.

Damper fluids often exhibit viscoplasticity. For example, electrorheological (ER) and magnetorheological (MR) fluids are 
suspensions of particles that align themselves in the presence of electric or magnetic fields and form structures that provide 
the fluid with a yield stress. They can be modelled as Bingham or Herschel-Bulkley fluids whose rheological parameters depend on 
the strength of the electric or magnetic field \cite{Genc_2002, Susan_2009}. Thus, the operation of ER / MR dampers can be tuned 
by adjusting the field strength. Another example of a damper that works with a yield-stress fluid is the extrusion damper. Here 
the ``fluid'' is actually a ductile solid material which is forced to flow through an annular contraction. Such dampers can 
carry significant loads and have been proposed and used for seismic protection of structures using lead as the 
plastically-deforming material \cite{Robinson_1976, Cousins_1993, Rodgers_2007}. Lead recrystallizes at room temperature and thus 
recovers most of its mechanical properties immediately after extrusion, so that lead extrusion dampers can undergo a large number 
of cycles of operation without performance degradation. The present study was motivated by the participation of the authors in a 
project investigating the design of an extrusion damper employing sand instead of a metallic damping medium. The behaviour of 
sand is nearly temperature independent, whereas the yield strength of lead drops with the temperature rise that is due to the 
energy absorption by the damper \cite{Makris_2013}. Sand is a granular material, and the behaviour of granular materials is 
known to be described well by the Bingham consitutive equation. Nevertheless, the present study is not limited to this 
particular design but aims to provide results of greater generality.

In the literature there exist only a few studies on the flow inside viscous dampers, whether viscoplastic or otherwise. Usually, 
it is assumed that the gap between the piston and the cylinder is narrow enough such that the flow can be approximated by a 
one-dimensional planar Couette - Poiseuille flow. Although simplified, this analysis can offer some insight on how flow 
characteristics such as viscoplasticity \cite{Wereley_1998, Wang_2007}, inertia \cite{Hou_2008, Nguyen_2009, Yu_2013} or 
viscoelasticity \cite{Makris_1996, Hou_2008} affect the damper response. However, it would be desirable to have complete 
simulations of the flow, which can be assumed to be axisymmetric under normal operating conditions. In the damper literature 
there appears to be a lack of such studies, with very limited results given in \cite{Hou_2008, Parlak_2012}. Progress in 
Computational Fluid Dynamics (CFD) has made such simulations feasible at a relatively modest computational cost.

The goal of the present work is to examine in detail the viscoplastic flow inside a damper whose shaft reciprocates sinusoidally. 
The fluid is assumed purely viscoplastic of Bingham type; this helps to isolate the effects of viscoplasticity from other 
phenomena such as shear-thinning, elasticity and thixotropy which may be examined in a future study. The shaft has a protruding 
spherical bulge that acts like a piston, but in order to investigate the effect of this bulge, the ``bulgeless'' configuration is 
also investigated to some extent. The bulged configuration is therefore precisely that of an ``extrusion damper'' 
\cite{Robinson_1976, Cousins_1993, Rodgers_2007}, whereas the bulgeless configuration is simply the flow in an annular cavity 
whose inner cylinder reciprocates sinusoidally. The latter is the annular analogue of the popular lid-driven cavity problem, and 
is of interest on its own. The present simulations span the range of Bingham numbers ($Bn$) from $Bn = 0$ to $Bn = 320$, which 
correspond to relatively soft materials (e.g.\ an ER fluid -- see the next Section for the precise details). This allows for the 
investigation also of inertia effects, which become weaker as the Bingham number is increased. Furthermore, the qualitative 
behaviour of the damper changes very little beyond some value of the Bingham number, and this behaviour is clearly seen in the 
present results for $Bn = 320$, so that increasing the Bingham number beyond that would not offer additional insight while at 
the same time it would significantly increase the computational cost. For simplicity, the effects of temperature increase are 
not examined and the flow is assumed isothermal, although some results on energy dissipation are included because they pertain 
to damper operation.

To the best of our knowledge there do not exist any previous studies for the bulged case, but, rather surprisingly, we have not 
found any studies for the bulgeless case either, with the exception of \cite{Blackburn_2011} which, however, focuses on flow 
instabilities arising at Reynolds numbers significantly higher than those examined here. A related, simpler problem, is the flow 
in a planar lid-driven cavity with sinusoidal lid motion, for which a few Newtonian studies are available. Among them is that of 
Iwatsu et al. \cite{Iwatsu_1992} who performed simulations for a range of Reynolds numbers and oscillation frequencies and found 
that these parameters have a similar effect on the flow as in Stokes' second problem (sinusoidal oscillation of an infinite plate 
in an infinite medium); in particular, at low Reynolds numbers and frequencies the flow is in quasi steady state whereas at high 
Reynolds numbers and frequencies the flow is localised to a thin layer near the lid, while the influence of the lid motion is 
only weakly felt by the fluid that is farther away. Interestingly, in that paper the force that the lid exerts on the fluid is 
calculated (a result that is of great interest in the case of dampers); this force is rarely given in lid-driven cavity studies, 
as the computed value tends to infinity with grid refinement due to the singularities at the lid edges. Therefore, the force 
reported in \cite{Iwatsu_1992} is questionable, but nevertheless the resutls suggest a time lag between the force and the lid 
velocity which increases with the oscillation frequency. A newer study is \cite{Mendu_2013} which reproduces the findings of 
Iwatsu et al. \cite{Iwatsu_1992} and where one can find references to a few other available related published studies.

The oscillating lid-driven cavity problem has not been solved for viscoplastic flow. The closest problem for which we have found 
results is the even simpler oscillating plate problem (Stokes' second problem), which was solved for viscoplastic flow by 
Balmforth et al. \cite{Balmforth_2009} (this problem has also been solved for other types of non-Newtonian flow - see the 
literature review in \cite{McArdle_2012}). We should also mention the study of Khaled and Vafai \cite{Khaled_2004} on 
oscillating plate flow although dealing with Newtonian flow only, because it includes the effects of wall slip; the latter will 
also be employed in the present study in order to overcome the aforementioned infinite force hurdle. The oscillating plate flow 
and the present oscillating annular cavity (or damper) flow are driven by the same sinusoidal boundary motion, yet their 
different geometries lead to significant differences between them. For example, the role of the pressure is trivial in the former 
and very important in the latter.

On the other hand, if one searches for problems that share a similar geometry to our problem rather than a similar driving force, 
then the axial viscoplastic Couette - Poiseuille flow through an annulus naturally comes to mind. This problem differs from our 
own (in the bulgeless case) in that the cylinders extend to infinity rather than form a closed cavity, and the flow is in 
steady-state; but it could be a good approximation to our flow if the length-to-radius ratio of the cavity is relatively large and 
the Reynolds number is small enough such that the flow is in quasi steady state. Early solutions of the annular viscoplastic 
Poiseuille flow (driven by a pressure gradient only) appear in \cite{Laird_1957, Fredrickson_1958}, while more recent 
contributions include \cite{Fordham_1991, Brunn_2007, Kalyon_2012}, the latter including effects of wall slip. The corresponding 
solution for annular Couette flow (driven by the axial motion of the inner cylinder only) for a Bingham fluid can be found in 
\cite{Bird_1982} (see also Appendix \ref{appendix: couette}, where the yield line is given in closed form, something missing from 
the literature). But the combination of these, i.e.\ annular Couette-Poiseuille Bingham flow, has only recently been solved for 
all possible types of flow by Liu and Zhu \cite{Liu_2010} (another notable contribution is \cite{Filip_2003}). Dapr\`{a} and 
Scarpi \cite{Dapra_2011} move a step closer to our problem by providing results for annular Couette-Poiseuille flow where the 
pressure gradient and/or the inner cylinder velocity oscillate sinusoidally; however, their focus is on how the flow rate is 
affected, whereas in order to approximate the flow in a closed annular cavity the instantaneous pressure gradient must be 
adjusted so that the total flow rate is zero.

The rest of the paper is organised as follows. The problem is defined in Section \ref{sec: definition}, where the governing 
equations are also given. Then, in Section \ref{sec: method} an outline of the computational method employed to solve the 
equations is given, together with references where more details can be found. The results then follow in Section \ref{sec: 
results}, where the effects of viscoplasticity, slip, damper geometry, and oscillation frequency on the damper response are 
investigated. Finally, conclusions are drawn in Section \ref{sec: conclusions}.

\section{Problem definition and governing equations}
\label{sec: definition}

\begin{figure}[t]
  \centering
  \subfigure[3-D model]{\label{sfig: layout_3d} \includegraphics[scale=1.00]{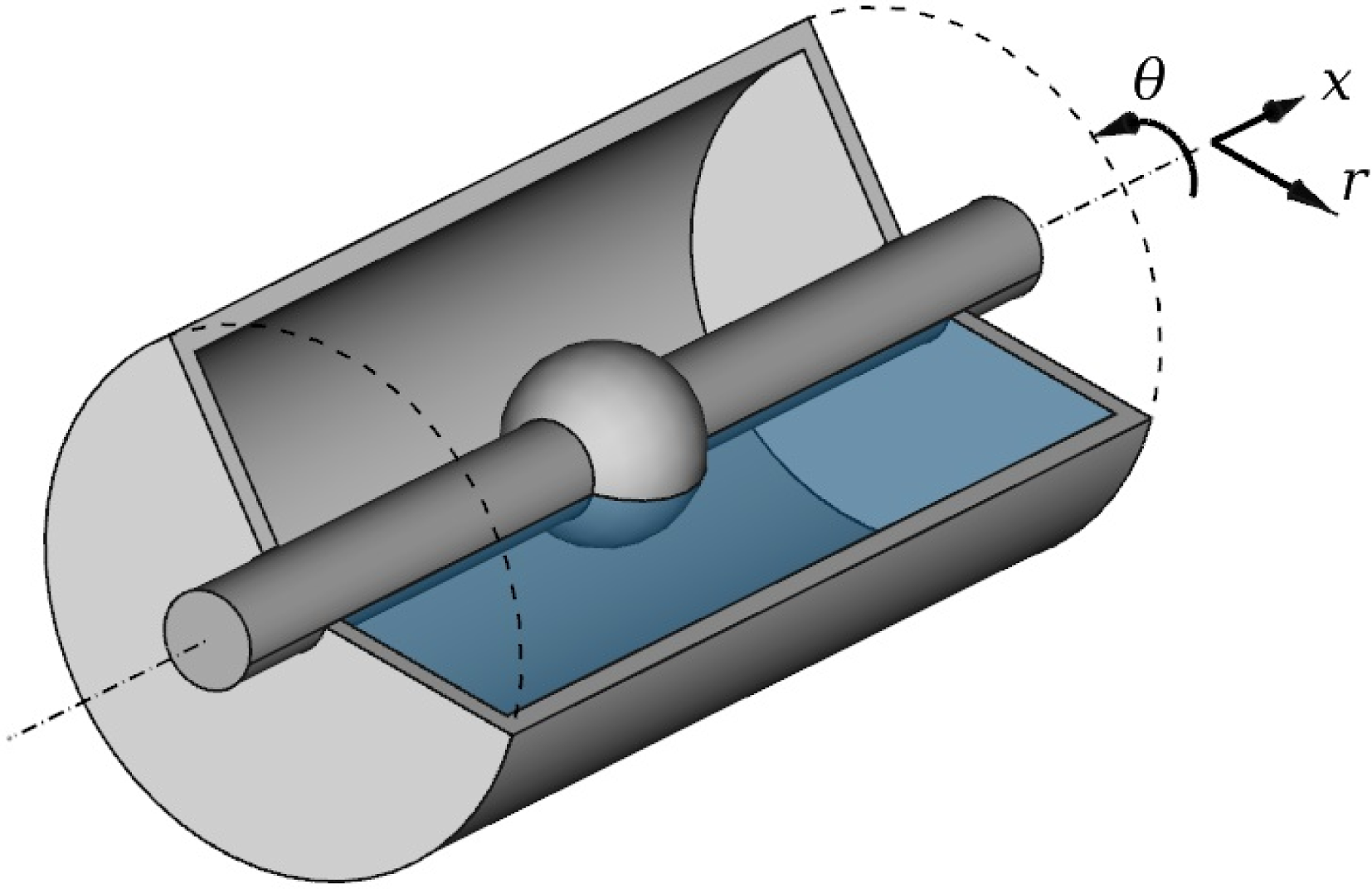}}
  \subfigure[2-D section]{\label{sfig: layout_2d} \includegraphics[scale=1.00]{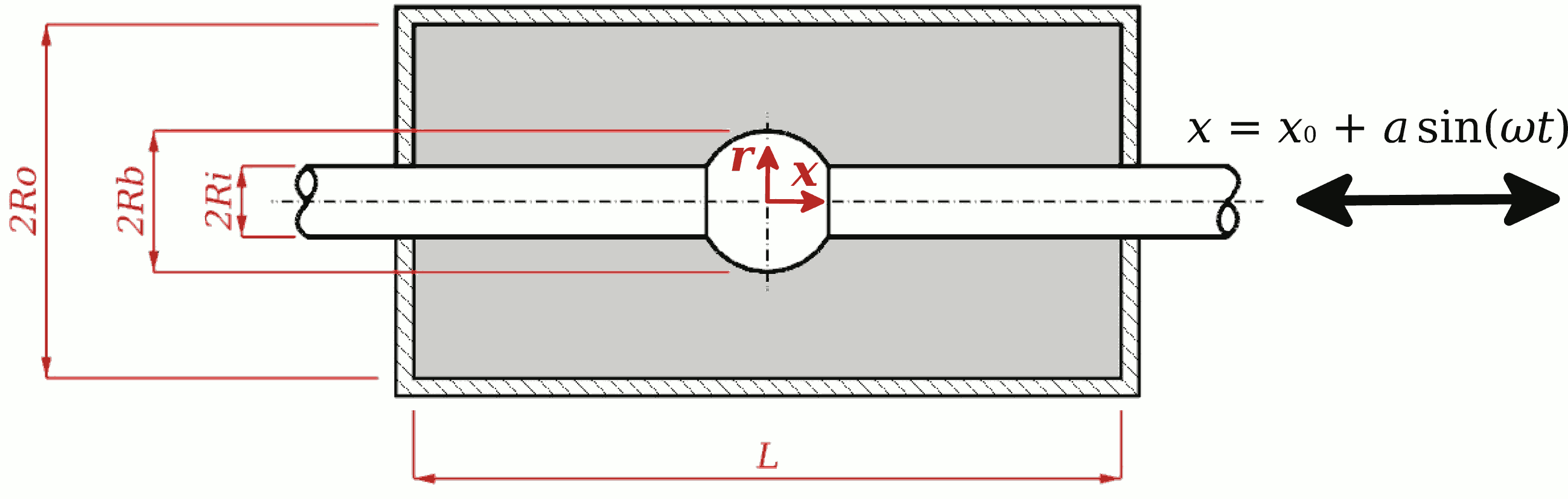}}
  \caption{Layout of the damper. In \subref{sfig: layout_3d} part of the cylinder is removed to reveal the bulged shaft. The 
flow is axisymmetric and can be solved on a single plane $\theta = 0$ (coloured in \subref{sfig: layout_3d}). The viscoplastic 
material is shown shaded in \subref{sfig: layout_2d}.}
  \label{fig: layout}
\end{figure}

The layout of the damper is shown in Fig.\ \ref{fig: layout}. A shaft of radius $R_i$ with a spherical bulge of radius $R_b$ at 
its centre reciprocates sinusoidally inside a cylinder of bore diameter $R_o$ and length $L$, filled with a viscoplastic material 
of Bingham type. A system of cylindrical polar coordinates $(x, r, \theta)$ can be fitted to the problem (Fig.\ \ref{sfig: 
layout_3d}), with $\vf{e}_x$, $\vf{e}_r$, $\vf{e}_{\theta}$ denoting the unit vectors along the coordinate directions. The 
geometry and the flow are assumed to be axisymmetric, so that the solution is independent of $\theta$ and the problem is reduced 
to two dimensions. The fluid velocity is denoted by $\vf{u}$, and its components are denoted by $u = \vf{u} \cdot \vf{e}_x$ and 
$v = \vf{u} \cdot \vf{e}_r$. The azimuthal velocity component, $\vf{u} \cdot \vf{e}_{\theta}$, is zero. Initially, the bulge is 
located midway along the cylinder and the viscoplastic material is at rest. At time $t=0$ the bulged shaft starts to move, 
forcing the confined material to flow. The shaft reciprocates along the axial direction such that the $x-$coordinate of any point 
on the shaft changes in time as

\begin{equation} \label{eq: shaft position}
x(t) = x_0 + \alpha \sin(\omega t)
\end{equation}
where $x_0$ is the position at $t=0$, $\alpha$ is the amplitude of oscillation, and $\omega$ is the angular frequency related to 
the frequency $f$ by $\omega = 2\pi f$. The period of oscillation is $T = 1/f = 2\pi/\omega$. The velocity of the shaft, $dx/dt$, 
is therefore $u_{sh}(t) = U \cos(\omega t)$ where $U = \omega \alpha$ is the maximum shaft velocity. The damper reacts to its 
imposed motion by a reaction force $F_R$ which dissipates the mechanical energy. The force $F_R$ and the associated energy 
dissipation are the quantities of interest.

It is assumed that the properties of the material such as the density $\rho$, the plastic viscosity $\mu$ and the yield stress 
$\tau_y$ are constant. The governing equations are the continuity and momentum balances:

\begin{equation} \label{eq: continuity}
 \pd{\rho}{t} \;+\; \nabla \cdot \left( \rho \vf{u} \right) \;=\; 0
\end{equation}

\begin{equation} \label{eq: momentum}
 \pd{(\rho\vf{u})}{t} \;+\; \nabla \cdot \left( \rho \vf{u} \vf{u} \right) \;=\; -\nabla p \;+\; \nabla\cdot\tf{\tau}
\end{equation}
where $p$ is the pressure and $\tf{\tau}$ is the deviatoric stress tensor. Due to the density $\rho$ being constant, Eqns.\ 
\eqref{eq: continuity} and \eqref{eq: momentum} can be simplified, although these more general forms are shown here. The stress 
tensor is related to the velocity field through the Bingham constitutive equation,

\begin{equation} \label{eq: Bingham_constitutive}
 \left\{ \begin{array}{ll} 
   \tf{\dot{\gamma}} \;=\; \tf{0} \;, \qquad  &  \tau \leq \tau_y
\\
   \tf{\tau} \;=\; \left( \dfrac{\tau_y}{\dot{\gamma}} + \mu \right)\tf{\dot{\gamma}} \;, & \tau > \tau_y
 \end{array} \right.
\end{equation}
where $\tf{\dot{\gamma}}$ is the rate-of-strain tensor, defined as $\tf{\dot{\gamma}} \equiv \nabla\vf{u} + 
(\nabla\vf{u})^{\mathrm{T}}$. The tensor magnitudes, $\tau \equiv ( \frac{1}{2} \tf{\tau}:\tf{\tau}) ^{1/2}$ and $\dot{\gamma} 
\equiv ( \frac{1}{2} \dot{\tf{\gamma}}:\dot{\tf{\gamma}}) ^{1/2}$, also appear in the above equation. Thus, the material flows 
only where the magnitude of the stress tensor exceeds the yield stress.

An aspect of the problem that complicates things is the fact that at the contact points between the shaft and the flat sides of 
the cylinder the velocity jumps discontinuously from non-zero values at the moving shaft to zero at the cylinder. If the no-slip 
boundary condition is used, this results in stress varying as $1/\delta x$ where $\delta x$ is the distance from the 
discontinuity \cite{Batchelor_2000}, and the force exerted on the shaft becomes infinite. This result is spurious, and in fact
molecular dynamics simulations have shown that the no-slip boundary condition is to be blamed, being unrealistic near the 
singularities where an amount of slip is exhibited that bounds the stress and the total force to finite values \cite{Koplik_1995, 
Qian_2005}. In fact, even without the corner singularity, when it comes to viscoplastic flows, wall slip appears to be the rule 
rather than the exception \cite{Sochi_2011}. Navier slip is the simplest alternative to the no-slip boundary condition, but 
nevertheless it is asserted in \cite{Qian_2005} that it is a realistic condition for Newtonian flows with corner singularities. 
It bounds the stress distribution and makes it integrable so that the total force can be calculated, as shown in \cite{he_2009}.

According to the Navier slip condition, the relative velocity between the fluid and the wall, in the tangential direction, is 
proportional to the tangential stress. More formally, for two-dimensional or axisymmetric flows such as the present one, this is 
expressed as follows: Let $\vf{n}$ be the unit vector normal to the wall, and $\vf{s}$ be the unit vector tangential to the wall 
within the plane in which the equations are solved. Let also $\vf{u}$ and $\vf{u}_w$ be the fluid and wall velocities, 
respectively. Then,

\begin{equation} \label{eq: navier slip}
 \left( \vf{u} - \vf{u}_w \right) \cdot \vf{s} \;=\; \beta \left( \vf{n} \cdot \tf{\tau} \right) \cdot \vf{s}
\end{equation}
where the parameter $\beta$ is called the slip coefficient.

For non-Newtonian flows the slip behaviour may be more complex than that described by the Navier slip condition; for example, the 
slip velocity and the wall stress may be related by a power-law relationship \cite{Kalyon_2005}, or there may be a ``slip yield 
stress'', that is, slip may occur only if the wall stress has exceeded a certain value \cite{Hatzikiriakos_2012}. A recent review 
of wall slip possibilities in non-Newtonian flows can be found in \cite{Damianou_2014}. Concerning the present application, 
the interface between the shaft and the extruded material is often lubricated and the shaft surface is polished 
\cite{Robinson_1976, Cousins_1993}. Hence, in the present study, in order not to overly increase the complexity of the problem, 
it was decided to apply the simple Navier slip boundary condition on the polished shaft and the no-slip boundary condition (Eq.\ 
\eqref{eq: navier slip} with $\beta = 0$) on the cylinder bore, whose surface has no special treatment.

It will be useful to express the governing equations in dimensionless form. So, let lengths be normalised by the distance between 
the shaft and the cylinder $H = R_o - R_i$, velocities by the maximum shaft speed $U$, time by the oscillation period $T$, and 
pressure and stresses by a characteristic stress $\tau_{\mathrm{ref}} = \tau_y + \mu U / H$. The latter is composed of a plastic 
component ($\tau_y$) and a viscous component ($\mu U / L$) in order to better represent a typical viscoplastic stress. Then, 
combining Eq.\ \eqref{eq: momentum} with Eq.\ \eqref{eq: Bingham_constitutive}, and using the fact that $\rho$ is constant, one 
obtains for the yielded part of the material 

\begin{equation} \label{eq: momentum nd}
 Re^* \left( \frac{1}{Sr} \pd{\tilde{\vf{u}}}{\tilde{t}} \;+\; \,\tilde{\nabla}\cdot(\tilde{\vf{u}}\tilde{\vf{u}}) \right) \;=\; 
 -\tilde{\nabla} \tilde{p} \;+\;
 \frac{Bn}{Bn\!+\!1} \tilde{\nabla} \cdot \left[ \left( \frac{1}{\tilde{\dot{\gamma}}} \,+\, \frac{1}{Bn} \right) 
\tilde{\dot{\tf{\gamma}}} \right]
\end{equation}
where tildes (\~{}) denote dimensionless variables. Note that the dimensionless rate-of-strain tensor and its magnitude are equal 
to their dimensional counterparts normalised by $U/H$. Equation \eqref{eq: momentum nd} contains three dimensionless numbers. The 
Bingham number $Bn$, defined as

\begin{equation} \label{eq: Bn}
 Bn \;\equiv\; \frac{\tau_y}{\mu U / H}
\end{equation}
is a measure of the viscoplasticity of the flow. The effective Reynolds number $Re^*$ \cite{Nirmalkar_2013} is defined as

\begin{equation} \label{eq: Re*}
 Re^* \;\equiv\; \frac{\rho U^2}{\tau_y + \mu \frac{U}{H}} \;=\; \frac{\rho U^2}{\tau_{\mathrm{ref}}} \;=\; \frac{Re}{Bn+1}
\end{equation}
and is an indicator of the ratio of inertia forces to viscoplastic forces, just like the usual Reynolds number $Re \equiv \rho 
U^2 / (\mu U/H) = \rho U H / \mu$ is an indicator of the ratio of inertia forces to viscous forces. Finally, the Strouhal number 
$Sr$ is defined as

\begin{equation} \label{eq: Sr}
 Sr \;\equiv\; \frac{T}{H/U}
\end{equation}
From $T = 2\pi / \omega$ and $U = \omega \alpha$ it follows that $Sr = 2\pi \alpha / H = 2\pi A$, where $A = \alpha / H$ is the 
dimensionless amplitude of oscillation. Therefore, the fact that the characteristic velocity $U$ is inherently inversely 
proportional to the characteristic time $T$ removes the dependance of $Sr$ on $T$. So, $Sr$ is only a dimensionless expression of 
the amplitude.

The boundary conditions are also expressed in nondimensional form. On the motionless cylinder walls where the no-slip condition 
applies, the boundary condition is just $\tilde{\vf{u}} = 0$. The dimensionless shaft velocity is $\tilde{u}_{sh} = \cos(2\pi 
\tilde{t})$, and can be seen not to depend on any of the dimensionless numbers. However, another dimensionless number enters 
through the dedimensionalisation of the Navier slip condition \eqref{eq: navier slip}, which is applied on the shaft:

\begin{equation} \label{eq: navier slip nd}
 \left( \tilde{\vf{u}} - \tilde{\vf{u}}_{sh} \right) \cdot \vf{s} \;=\; \tilde{\beta} \left( \vf{n} \cdot \tilde{\tf{\tau}} 
\right) 
\cdot \vf{s}
\end{equation}
The dimensionless Navier slip coefficient is given by:

\begin{equation} \label{eq: slip coefficient nd}
 \tilde{\beta} \;\equiv\; \frac{\beta \mu}{H} \left( Bn + 1 \right) \;=\; \tilde{l} \left( Bn + 1 \right)
\end{equation}
where $l = \beta \mu$ is the slip (or extrapolation) length and $\tilde{l} = l/H$ is its dimensionless counterpart. In Newtonian 
flows the use of the slip length is preferred to the use of the slip coefficient, and therefore in the simulations we will 
occasionally mention the values of the slip length as well.

Finally, three additional dimensionless numbers are needed to determine the boundary geometry. Different choices are possible. One 
such choice leads to the following set of seven dimensionless variables that define the problem: $Re^*$, $Sr$, $Bn$, 
$\tilde{\beta}$, $R_i / R_o$, $R_b / R_o$ and $L / H$.

The code used for the simulations solves the dimensional equations. However, the results will be presented mostly in dimensionless 
form because this form offers greater insight into the phenomena and greater generality. An important result for the present 
application is the force acting on the shaft, which in the present study is dedimensionalised by a reference force 
$F_{\mathrm{ref}}$:

\begin{equation} \label{eq: Fref tau_ref}
 F_{\mathrm{ref}} \;=\; (2\pi R_i L) \tau_{\mathrm{ref}}
 \qquad \mathrm{where} \qquad
 \tau_{\mathrm{ref}} = \tau_y + \mu U/H
\end{equation}
Thus the reference force is that which results from the reference stress $\tau_{\mathrm{ref}}$ acting on the whole shaft 
surface, of area $2\pi R_i L$, in the absence of a bulge.

\begin{table}[b]
\caption{Values of the dimensional and dimensionless parameters defining the base case.}
\label{table: base case}
\begin{center}
\renewcommand\arraystretch{1.25}   
\begin{tabular}{ r | l }
\hline
 Fluid properties & $\rho = 1000$ \si{kg/m^3}, \hspace{0.2cm} $\mu = 1.0$ \si{Pa.s}, \hspace{0.2cm} $\tau_y = 31.416$ \si{Pa}
\\
 Geometry         & $R_i = 10$ mm, \hspace{0.2cm} $R_o = 50$ mm, \hspace{0.2cm} $R_b = 20$ mm, \hspace{0.2cm} $L = 200$ mm \\
 Oscillation      & $f = 0.5$ Hz, \hspace{0.2cm} $\alpha = 20$ mm \\
 Slip             & $\beta$ = \num{4.7619e-5} \si{m/Pa.s} \hspace{0.2cm} ($l$ = \num{4.7619e-5} m) 
\\
\hline
 \textbf{Dimensionless parameters} & $Re^* = 0.12$ \hspace{0.2cm} ($Re = 2.51$), \hspace{0.2cm} $Sr = 3.14$, \hspace{0.2cm} 
                                     $Bn = 20$, \\
                                   & $\tilde{\beta} = 0.025$ \hspace{0.2cm} ($\tilde{l}$ = \num{1.19e-3}), \\
                                   & $R_i/R_o = 0.2$, \hspace{0.2cm} $R_b/R_o = 0.4$, \hspace{0.2cm} $L/H = 5$\\
\hline
\end{tabular}
\end{center}
\end{table}

Due to the large number of parameters, it was decided to set a base case, which is defined in Table \ref{table: base case}, and 
then vary several of the problem parameters, each in turn, in order to investigate their effect on the damper response. The base 
case was defined using typical values for the parameters, choosing values that lie in the parameter range of the damper 
literature cited in Section \ref{sec: introduction} and result in ``nice'' (rounded) values of the dimensionless parameters. In 
particular: The geometry is of the ``extrusion'' type and is closer to the compact design of \cite{Rodgers_2007} rather than the 
older, bulky designs of \cite{Robinson_1976, Cousins_1993}. The oscillation amplitude is near the average of that found in the 
referenced studies (e.g.\ \cite{Constantinou_1993, Symans_1999, Wang_2007, Yu_2013}), while the base frequency is near the low 
end of the spectrum of frequencies in the referenced studies. For example, for seismic applications frequencies in the range 0.1 
-- 2.5 Hz are reported in \cite{Symans_1999}, but they can be as high as 10 Hz for short buldings \cite{Makris_1996}. Here the 
chosen base frequency is 0.5 Hz but numerical experiments with frequencies of up to 8 Hz are performed in Section \ref{sec: 
results}. The rheological properties resemble those of an ER or MR fluid, modelled as a Bingham fluid, \cite{Makris_1996, 
Wereley_1998, Parlak_2012}, with yield stresses of up to 500 Pa employed in Section \ref{sec: results}.

Finally, we note that the present results do not apply to lead extrusion dampers since metal extrusion is governed by other 
constitutive equations, where plasticity is dominant. However, such equations are not much different than viscoplastic 
constitutive equations such as the Bingham equation in the limit of high plasticity; for example, in \cite{Basic_2005} metal 
extrusion is modelled using a regularised Levy-Mises flow rule to which the Bingham equation reduces when $\mu = 0$. In fact, the 
methodology and finite volume solution method of that study are very similar to those employed in the present study. The Bingham 
constitutive equation, which is adopted here, allows the investigation of inertial, viscous, and plastic effects and therefore 
gives more generality to the results. As will be seen in Section \ref{sec: results}, in the present numerical experiments, at the 
higher Bingham numbers tested here plasticity is also dominant over other flow mechanisms (inertia, viscosity). Another study 
where the finite volume methodology is applied to solve metal extrusion problems is \cite{Williams_2010}, where further 
references can be found.

\section{Numerical Method}
\label{sec: method}

The problem defined in Section \ref{sec: definition} was solved using a finite volume method, which will be described in the 
present section only briefly, providing pertinent references. A detailed description of the method will be presented in a 
separate publication. It is an extension of that presented in \cite{Syrakos_06a, Syrakos_13}, with extensions for transiency, 
axisymmetry, grid motion and the slip boundary condition.

The method employs second-order accurate central differences for both the convective and viscous terms, with correction terms 
included to account for grid non-orthogonality, skewness and stretching \cite{Syrakos_06a}. All variables are stored at the 
volume centres and spurious pressure oscillations are avoided by the use of momentum interpolation \cite{Rhie_1983, Syrakos_06a}. 
The gradients of the flow variables are calculated using a least-squares procedure \cite{Muzaferija_1997}. These gradients 
are needed for the calculation of the aforementioned correction terms and also for the calculation of the magnitude of the 
rate-of-strain tensor and hence of the viscosity (eq.\ \eqref{eq: Papanastasiou constitutive} below). Time derivatives are 
approximated by a fully implicit, second-order accurate three-time-level backward differencing scheme \cite{Ferziger_02}. To 
account for axisymmetry, the original planar finite volume code was adjusted as described in \cite{Ferziger_02}, and, in 
addition, the calculation of the magnitude of the rate-of-strain tensor must take into account that the component 
$\dot{\gamma}_{\theta \theta} = 2v/r$ is, in general, non-zero.

\begin{figure}[b]
  \centering
  \subfigure[]{\label{sfig: grid t0} \includegraphics[scale=1.00]{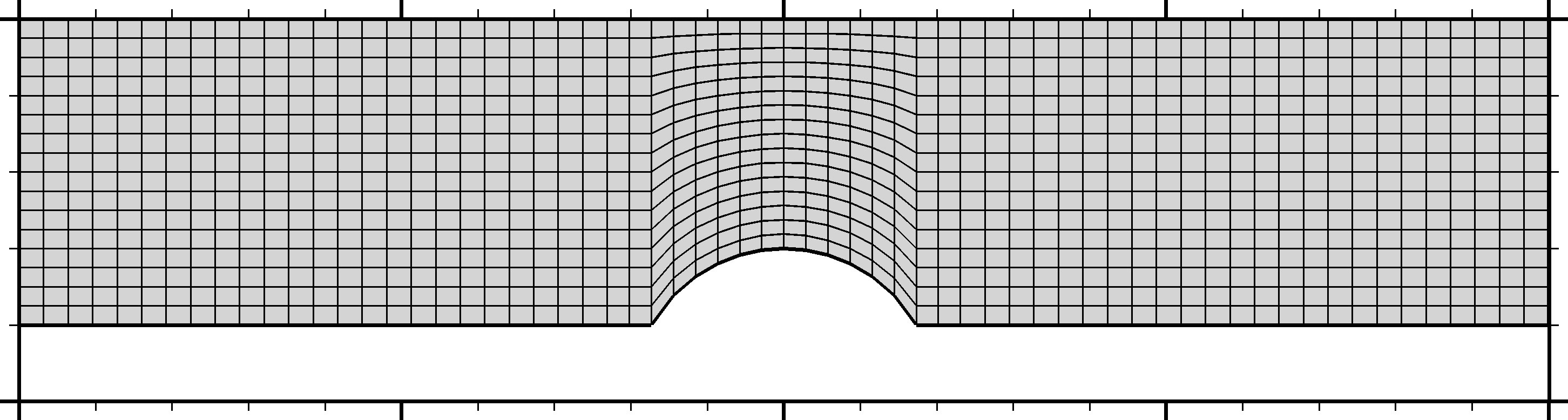}}\\
  \subfigure[]{\label{sfig: grid t1} \includegraphics[scale=1.00]{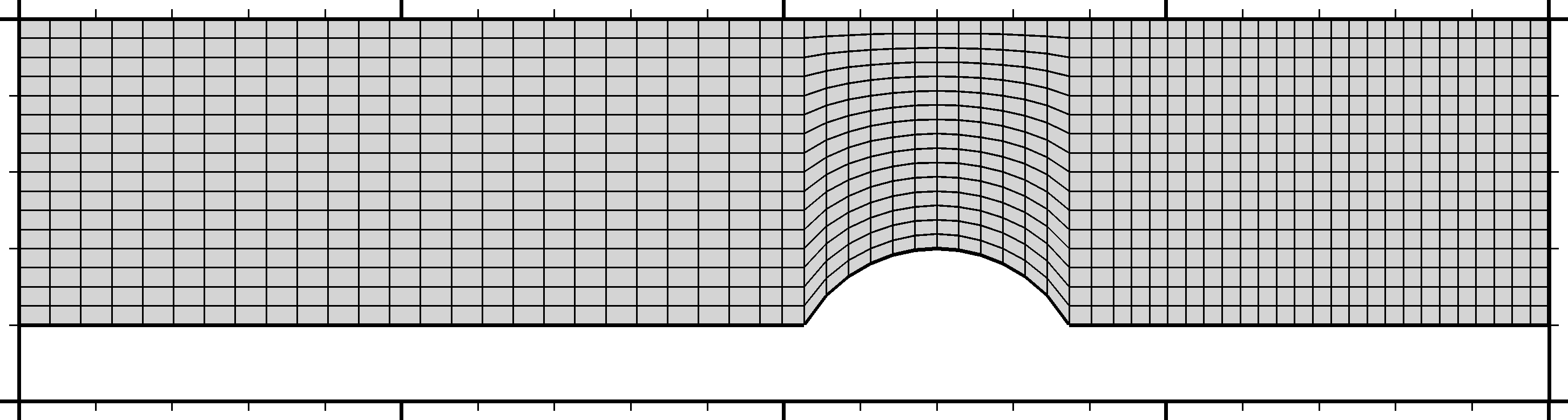}}
  \caption{Sample grids, \subref{sfig: grid t0} at a time instant when the bulge is located midway along the cylinder, and
\subref{sfig: grid t1} at a time when the bulge is offset towards the right. For clarity, coarse $64 \times 16$ volume grids are 
shown instead of the $1024 \times 256$ grids actually used.}
  \label{fig: grid}
\end{figure}

Since it is axisymmetric, the problem can be solved in any single $x-r$ plane. Such a plane is partitioned into a number of 
finite volumes, using grids of sizes of $1024 \times 256$ volumes in the $x$- and $r$-directions, respectively. When the shaft is 
bulgeless, the grid is stationary; but when a bulged shaft is used the grid changes in time to follow the deformation of the 
domain due to the bulge motion. Figure \ref{fig: grid} shows a couple of coarser grids at different time instances. The grid 
above the bulge and small margins on either side of it along the $x$-direction remains fixed, while the rest of the grid is 
compressed / expanded accordingly as the bulge moves.

Grid motion requires that the convective terms of the equations use the relative velocity of the fluid relative to the volume 
faces, rather than the absolute fluid velocity. In the present method this is implemented by a scheme that ensures that the 
so-called \textit{space conservation law} \cite{Demirdzic_1988} is obeyed, i.e.\ that the fluid volume ``swallowed'' by the faces 
during a time step is equal to the volume increase of the cell that owns the faces. The details will be presented in a future 
publication, but we note that in the present case such a specialised scheme is not really necessary as the fact that only one 
set of grid lines move ensures space conservation anyway \cite{Demirdzic_1988}.

A difficulty with simulations involving yield stress fluids is that the domain of application of each branch of their constitutive 
equation, such as \eqref{eq: Bingham_constitutive}, is not known in advance. A popular approach to overcoming this difficulty is 
to approximate Eq.\ \eqref{eq: Bingham_constitutive} by a regularised equation which is applicable throughout the material 
without branches. Several such regularised equations have been proposed; some of them are compared in \cite{Frigaard_2005}. In 
the present work we adopt the one proposed by Papanastasiou \cite{Papanastasiou_87}, which is perhaps the most popular and has 
been used successfully for simulating many flows of practical interest (see, e.g., \cite{Karapetsas_2006, Dimakopoulos_2007, 
Tsamopoulos_2008, Papaioannou_2009}, among many others). It is formulated as follows: 

\begin{equation}
 \label{eq: Papanastasiou constitutive}
 \tf{\tau} \;=\; \left[ \frac{\tau_y}{\dot{\gamma}} \left( 1-\textrm{e}^{-m\dot{\gamma}} \right) \,+\, \mu \right] 
\tf{\dot{\gamma}} \;=\; \eta(\dot{\gamma}) \tf{\dot{\gamma}}
\end{equation}
or, in non-dimensional form:

\begin{equation}
 \label{eq: Papanastasiou constitutive nd}
 \tilde{\tf{\tau}} \;=\; \frac{Bn}{Bn + 1} \left[ \frac{1-e^{-M\tilde{\dot{\gamma}}}}{\tilde{\dot{\gamma}}} \,+\,
                                                  \frac{1}{Bn} \right] \tilde{\tf{\dot{\gamma}}}
 \;=\; \tilde{\eta}(\dot{\gamma}) \tilde{\tf{\dot{\gamma}}}
\end{equation}
where the term in square brackets in Eq.\ \eqref{eq: Papanastasiou constitutive}, $\eta$, is the effective viscosity and $m$ is 
a stress growth parameter which controls the quality of the approximation: the larger this parameter the better Eq.\ \eqref{eq: 
Papanastasiou constitutive} approximates \eqref{eq: Bingham_constitutive}. This parameter is nondimensionalised as $M = m U / H$. 
Increasing the value of $M$ also makes the equations stiffer and harder to solve, so a compromise must be made. In our previous 
study for the lid-driven cavity test case \cite{Syrakos_13} it was found that increasing $M$ beyond 400 caused numerical 
problems. However, in the present case it was possible to use a value of $M$ = 1000. Thus, Eq.\ \eqref{eq: Papanastasiou 
constitutive} assumes all of the material to be a generalised Newtonian fluid whose effective viscosity is given by the term in 
square brackets, and the unyielded material is approximated by assigning very high values to the viscosity. To identify the 
unyielded material we employ the usual criterion $\tau < \tau_y$, or, in terms of dimensionless stress, $\tilde{\tau} < 
\tilde{\tau}_y = \tau_y / \tau_{\mathrm{ref}} = Bn/(Bn+1)$ (the ratio $Bn/(Bn+1)$ is sometimes called the \textit{effective 
Bingham number} $Bn^*$ \cite{Nirmalkar_2013}); see \cite{Burgos_99, Syrakos_2014} for discussions on the use of this criterion.

The use of a regularised constitutive equation is also justified by the fact that experiments have not shown definitively that the 
transition from solid-like to fluid behaviour is completely sharp \cite{Barnes_1999}. In this respect, Eq.\ \eqref{eq: 
Papanastasiou constitutive} could be regarded as a more realistic constitutive equation. Nevertheless here it will be considered 
an approximation to Eq.\ \eqref{eq: Bingham_constitutive}. The accuracy of the regularisation approach to solving viscoplastic 
flows is discussed in \cite{Frigaard_2005, Dimakopoulos_2013}; their main disadvantage is the difficulty sometimes exhibited in 
accurately capturing the yield surfaces, but for the present application this is not of main concern. For alternative approaches, 
see \cite{Dimakopoulos_2013, Glowinski_2011}.

To ensure that the value $M = 1000$ is sufficient to obtain an accurate solution, a series of steady-state simulations was 
performed with varying values of $M$, where a shaft without a bulge moves at a constant velocity equal to the maximum velocity of 
the base case (Table \ref{table: base case}; $U = 2\pi f \alpha$). The dimensionless numbers for this steady-state problem have 
the same values as in Table \ref{table: base case}, except for the Strouhal number which is infinite, due to the problem being 
steady-state. So, solving this steady state problem we obtain values of 2.01988, 2.02156, 2.02241 and 2.02286 for the 
nondimensional force $\tilde{F}$ exerted on the shaft, for $M$ = 125, 250, 500 and 1000, respectively. The dependency of the 
force on $M$ appears to be weak, with the force values $F(M)$ converging towards a value, say $F^*$. If it is assumed that 
convergence of $F(M)$ to $F^*$ follows the formula $F(M) = F^* + cM^{-q}$, where $q$ is the order of convergence and $c$ a 
constant, then $q$ can be estimated using the results from using three different values of $M$ related through a fixed ratio, say 
$M$, $2M$ and $4M$, to give:

\begin{equation} \label{eq: convergence order}
 q \;=\; \frac{\log\left(\frac{F(M)-F(2M)}{F(2M)-F(4M)}\right)}{\log(2)}
\end{equation}
Applying this formula to the above values gives $q \approx 1$, which means that doubling the value of $M$ reduces the error 
$cM^{-q}$ to half. This result can be used to estimate the error $F^* - F(M) \approx [F(M)-F(M/2)]/(2^q-1) = F(M)-F(M/2)$ (for 
$q=1$). Therefore, the error due to regularisation at $M = 1000$ is about 0.02\%, which is very small. In fact, for most 
engineering applications, even lower values of $M$ would provide acceptable accuracy.

The system of non-linear algebraic equations that arises from the discretisation is solved using the SIMPLE algorithm 
\cite{Patankar_72} with multigrid acceleration. The Navier slip condition can be easily accounted for in SIMPLE using the deferred 
correction approach of Khosla and Rubin \cite{Khosla_1974}. The details will be provided in a forthcoming publication focusing 
on the numerical method employed here. Alternative treatments, including treatments for more complicated slip conditions, can be 
found in \cite{Ferras_2013}.

An important issue concerning numerical solutions is grid convergence: the grid must be fine enough so that the solution is 
sufficiently accurate. The existence of singularities at the grid corners does not pose problems concerning the bulk of the flow; 
thus, the accuracy of the present method for grids of resolution comparable to the present case is demonstrated in 
\cite{Syrakos_13, Syrakos_2014}. But the present study examines a new result, the force exerted on the shaft, and it turns out 
that the accuracy of this result is heavily affected by the singularities at the shaft endpoints. The reason is the following. As 
discussed in Section \ref{sec: definition}, the Navier slip boundary condition results in finite stress and pressure at the shaft 
endpoints for any $\beta \neq 0$. However, the smaller the value of $\beta$ the larger the stress and pressure there, and the 
larger the overall force on the shaft, tending to infinity as $\beta \rightarrow 0$. So, by varying the slip parameter the shaft 
force can obtain values in the whole range from zero to infinity. Using smaller values of $\beta$ results in steeper rise of the 
stress near the shaft ends, which requires finer grids to maintain an accurate calculation of the force. The present section 
therefore ends with a grid convergence study.

\begin{figure}[!b]
  \centering
  \subfigure[]{\label{sfig: steady forces Bn=20 per grid} \includegraphics[scale=1.10]{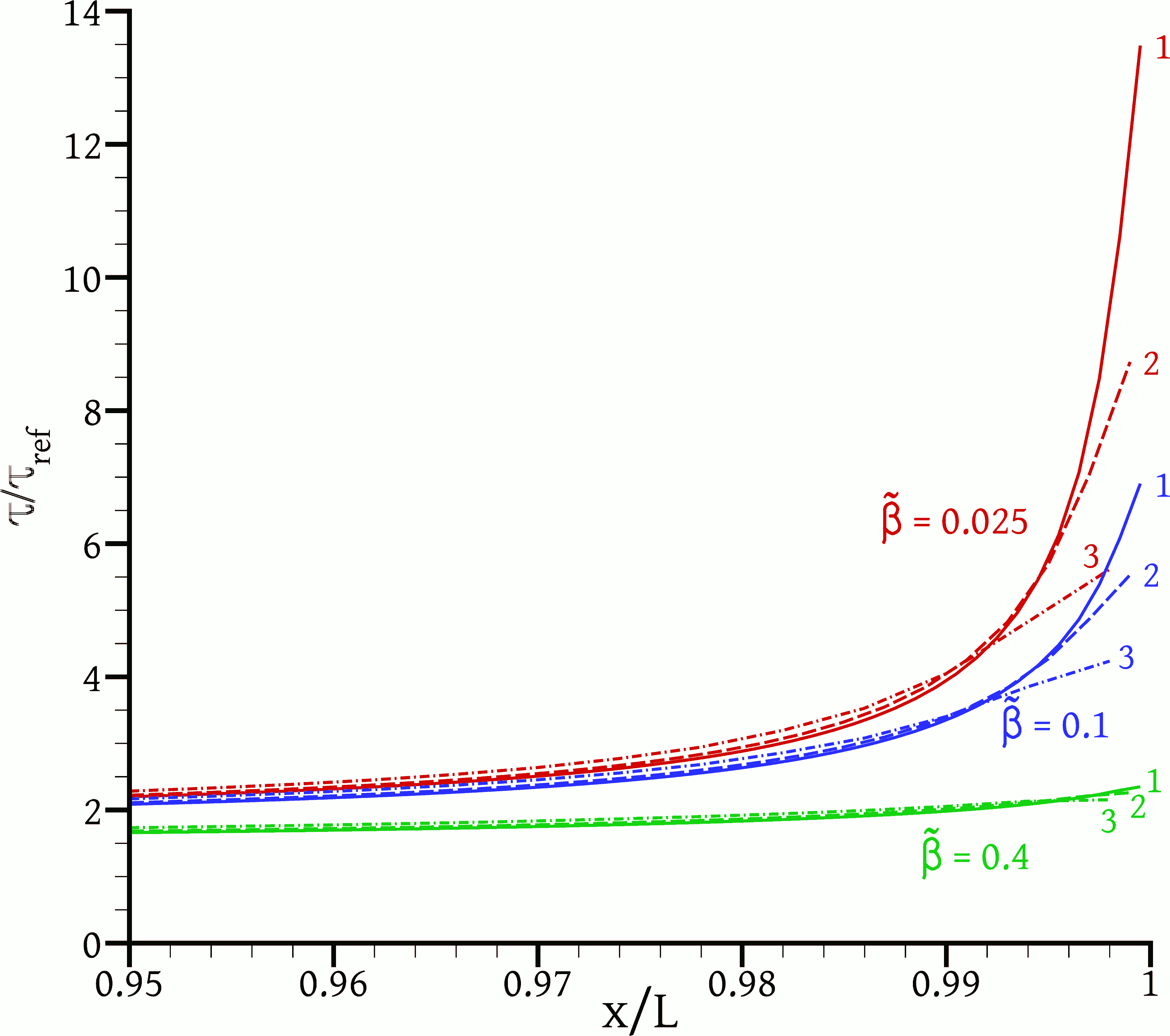}}
  \subfigure[]{\label{sfig: steady forces  Bn=20 per slip} \includegraphics[scale=1.10]{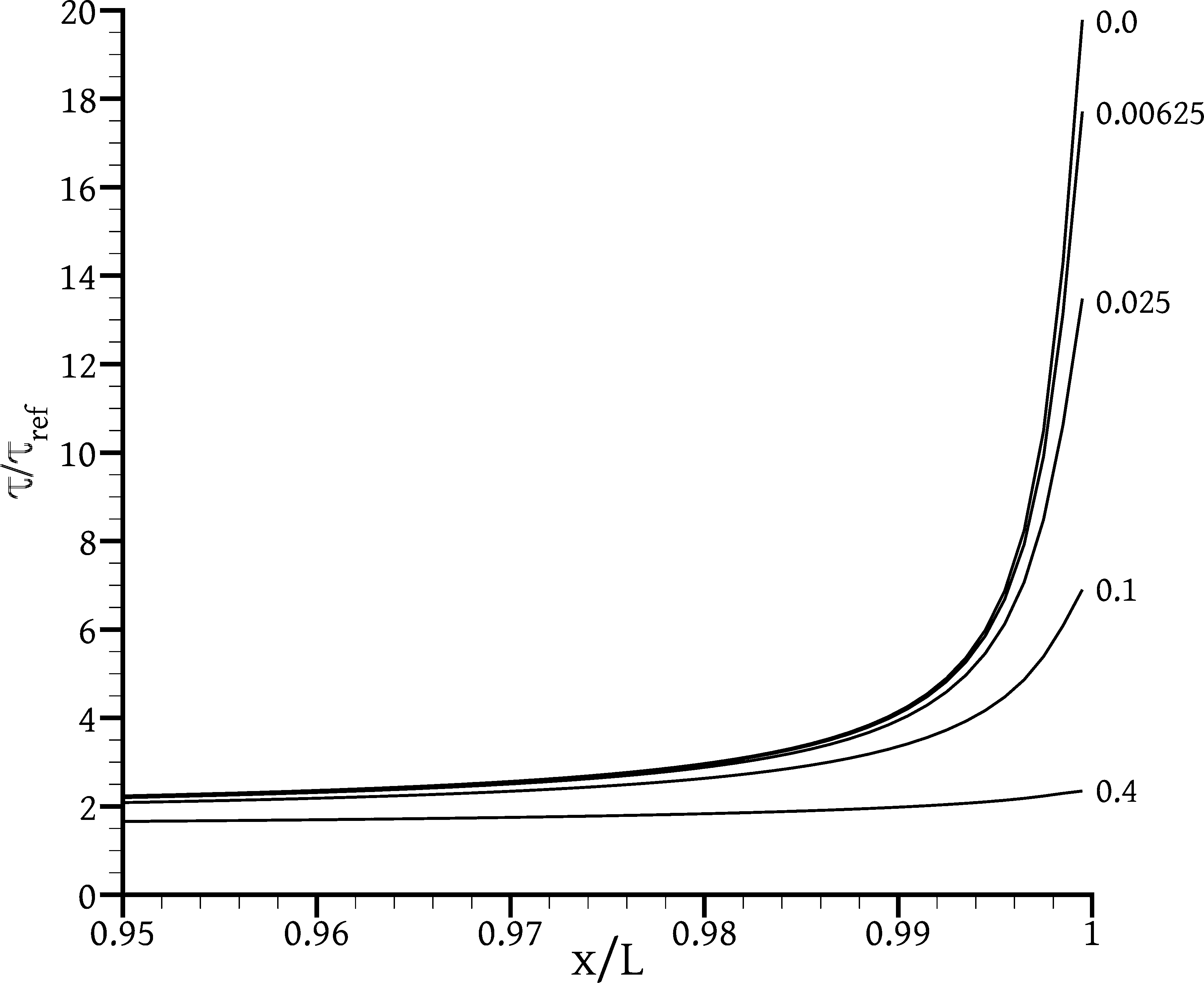}}
  \caption{Nondimensional shear stress distributions $\tilde{\tau}_{rx}$ along the shaft surface near the end of the shaft, for 
a 
steady state problem (shaft velocity = $U = 2\pi f \alpha$) without a bulge. The problem parameters are as displayed in Table 
\ref{table: base case}, except for the slip coefficient, which is varied to obtain the non-dimensional values indicated on each 
figure. In \subref{sfig: steady forces Bn=20 per grid} the stress is plotted for $\tilde{\beta}$ = 0.025 (red), 0.1 (blue) and 
0.4 (green), as calculated on grids 1 ($1024 \times 256$ volumes), 2 ($512 \times 128$ volumes) and 3 ($256 \times 64$ volumes). 
In \subref{sfig: steady forces  Bn=20 per slip} the stress is plotted for different values of $\tilde{\beta}$, indicated on each 
curve, calculated on a $1024 \times 256$ grid.}
  \label{fig: steady forces Bn=20}
\end{figure}

Figure \ref{sfig: steady forces Bn=20 per grid} demonstrates that grid convergence of the stress distribution near the corners is 
much faster when $\tilde{\beta}$ is large than when it is small. Column ``$Bn = 20$'' of Table \ref{table: grid convergence} 
lists the computed values of force on grids of varying density for a steady-state variant of the base case of Table \ref{table: 
base case} without a bulge, along with the order of grid convergence. Up to the $512 \times 128$ grid the force decreases with 
grid refinement and appears to converge with order $q = 2$, but on the $1024 \times 256$ grid the force increases slightly. This 
behaviour can be explained, with reference to Fig.\ \ref{sfig: steady forces Bn=20 per grid}, by the fact that grid refinement 
causes the computed stress to decrease over most of the length of the shaft, except near the ends where it increases due to the 
singularities. At high grid densities the stress increase near the corners dominates over the stress decrease over the rest of 
the shaft because the latter has already converged, whereas the former has not. The need therefore arises to estimate the force 
error on the $1204 \times 256$ grid, and assess whether it is acceptable.

\begin{table}[!t]
 \caption{Computed values of (dimensional) force on the shaft as a function of the grid density, for a couple of steady state 
problems. The problem parameters are listed in Table \ref{table: base case}, with a constant shaft velocity of $U = 2\pi f 
\alpha$, and there is no bulge. In addition, the second problem involves Newtonian flow ($\tau_y = Bn = 0$). The grid 
convergence index $q$ is calculated from formula \eqref{eq: convergence order} if $M$ is replaced by the number of volumes in 
either the $x$- or the $r$-direction.}
 \begin{center}
 \renewcommand\arraystretch{1.25}   
 {%
\newcommand{\mc}[3]{\multicolumn{#1}{#2}{#3}}
\begin{center}
\begin{tabular}{c | c c | c c}
\hline
\textbf{grid size} & $Bn = 20$ & $q$    & $Bn = 0$ & \textbf{$q$} \\
\hline
$128 \times 32$    & 0.8643    &        & 0.1372   &              \\
$256 \times 64$    & 0.8436    &        & 0.1475   &              \\
$512 \times 128$   & 0.8379    & 1.85   & 0.1565   & 0.21         \\
$1024 \times 256$  & 0.8385    & -      & 0.1635   & 0.34         \\
$2048 \times 512$  &           &        & 0.1683   & 0.56         \\
$4096 \times 1024$ &           &        & 0.1710   & 0.84         \\
$8192 \times 2048$ &           &        & 0.1722   & 1.12         \\
\hline
 \end{tabular}
 \end{center}
}%

 \end{center}
 \label{table: grid convergence}
\end{table}

\begin{figure}[!b]
  \centering
  \includegraphics[scale=1.10]{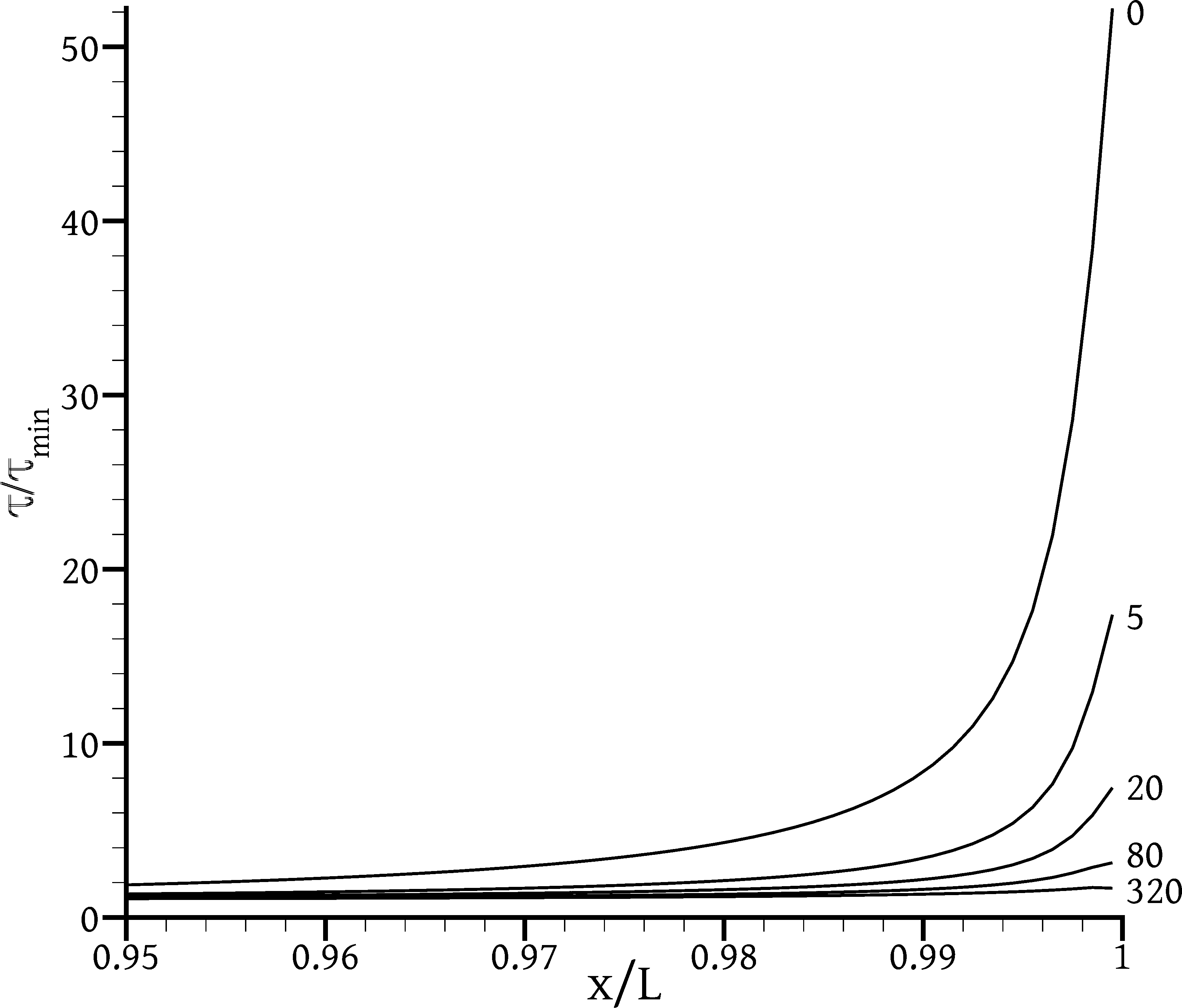}
  \caption{Shear stress $\tilde{\tau}_{rx}$ along the shaft surface near the end of the shaft, for a steady state problem 
without 
a bulge, for various Bingham numbers, as computed on the $1024 \times 256$ grid. For each $Bn$, the stress is normalised by its 
minimum value along the shaft. The problem parameters are as displayed in Table \ref{table: base case}, except that the shaft 
velocity is constant at $U = 2\pi f \alpha$, there is no bulge, and the value of the yield stress is chosen so as to obtain the 
Bingham numbers shown.}
  \label{fig: steady forces per Bn}
\end{figure}

The error can be estimated by comparing against the solution on a finer grid, but due to the deterioration of the 
SIMPLE/multigrid algorithm on Bingham problems which is discussed in \cite{Syrakos_13} this is not practical. A more appropriate 
treatment would be to use adaptively refined grids with large densities near the corners, using techniques such as those described 
in \cite{Syrakos_2014}. However, the adaptive mesh refinement algorithm has not yet been extended to time-dependent problems and 
moving grids in the available code. So, it was decided instead to solve a Newtonian steady-state problem (which is easier to 
solve) on a series of very fine grids with up to $8192 \times 2048$ volumes, and estimate the error of that problem. The results 
are also listed in Table \ref{table: grid convergence}, and they show that the force value does converge, although the full 
second-order rate of convergence has not yet been attained even on the finest grid. Assuming that the rate of convergence on grid 
$8192 \times 2048$ is approximately first order, the error on grid $1204 \times 256$ is estimated at about 0.01 \si{N}, or 0.01 / 
0.17 = 6\% which is rather large. But for higher Bingham numbers this percentage drops, as Fig.\ \ref{fig: steady forces per Bn} 
suggests. The large stresses near the shaft ends are due to steep velocity gradients, which induce stress components related to 
fluid deformation, $\mu \tf{\dot{\gamma}}$. When the yield stress $\tau_y$ is increased, the proportion of the deformation-induced 
component $\mu \tf{\dot{\gamma}}$ within the total stress $\tau_y (\tf{\dot{\gamma}} /\dot{\gamma}) + \mu \tf{\dot{\gamma}}$ 
falls. So, assuming that the component $\mu \tf{\dot{\gamma}}$ does not change much between the $Bn = 0$ and $Bn = 20$ cases, 
the force error on grid $1024 \times 256$ for the $Bn = 20$ case of Table \ref{table: grid convergence} would also be about 0.01 
\si{N}, or 0.01 / 0.84 = 1.2\%, which is acceptable. So, the value $\tilde{\beta} = 0.025$ selected for the base case offers 
accurate computation of the shaft force on the $1024 \times 256$ grid, while at the same time Fig.\ \ref{sfig: steady forces  
Bn=20 per slip} suggests that it results in a flow field that is negligibly different from that of the no-slip condition, except 
very close to the ends of the shaft.

For the temporal discretisation a time step of $\Delta t = T/400$ was used; it will be shown in Section \ref{sec: results} that 
the time step size has a small effect on the accuracy, because for most of the test cases studied the temporal term in the 
momentum equation is small compared to the other terms.

\section{Results}
\label{sec: results}

\begin{figure}[!p]
  \centering
  \subfigure[$t = T/4$]            {\label{sfig: base flow 100} \includegraphics[scale=1.0]{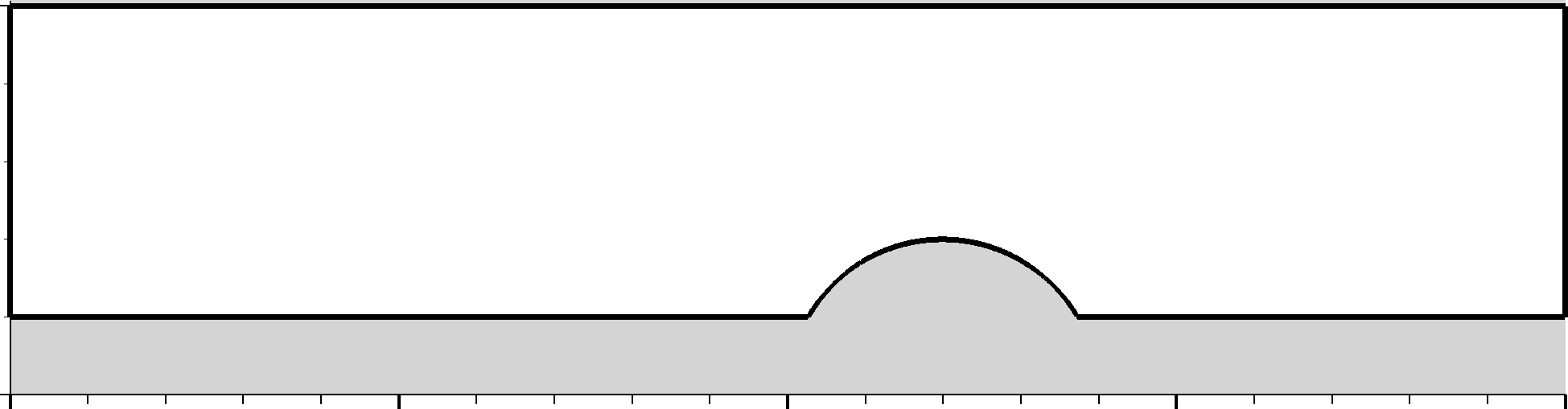}}
  \subfigure[$t = 3T/4: -/+$]      {\label{sfig: base flow 300} \includegraphics[scale=1.0]{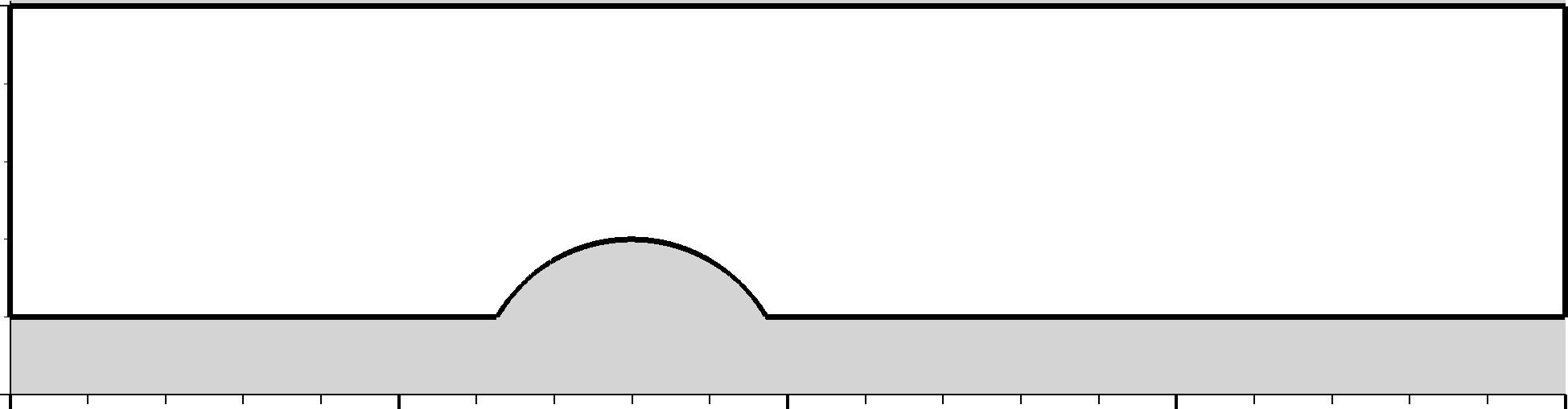}}
  \subfigure[$t = 1.25(T/4)$]      {\label{sfig: base flow 125} \includegraphics[scale=1.0]{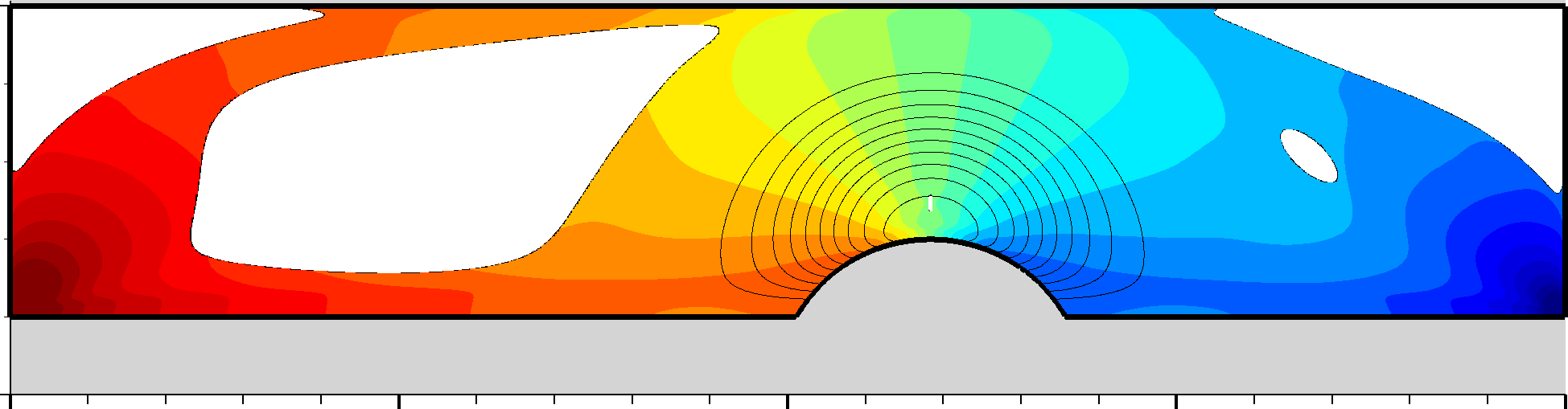}}
  \subfigure[$t = 0.75(T/4): +/-$] {\label{sfig: base flow 075} \includegraphics[scale=1.0]{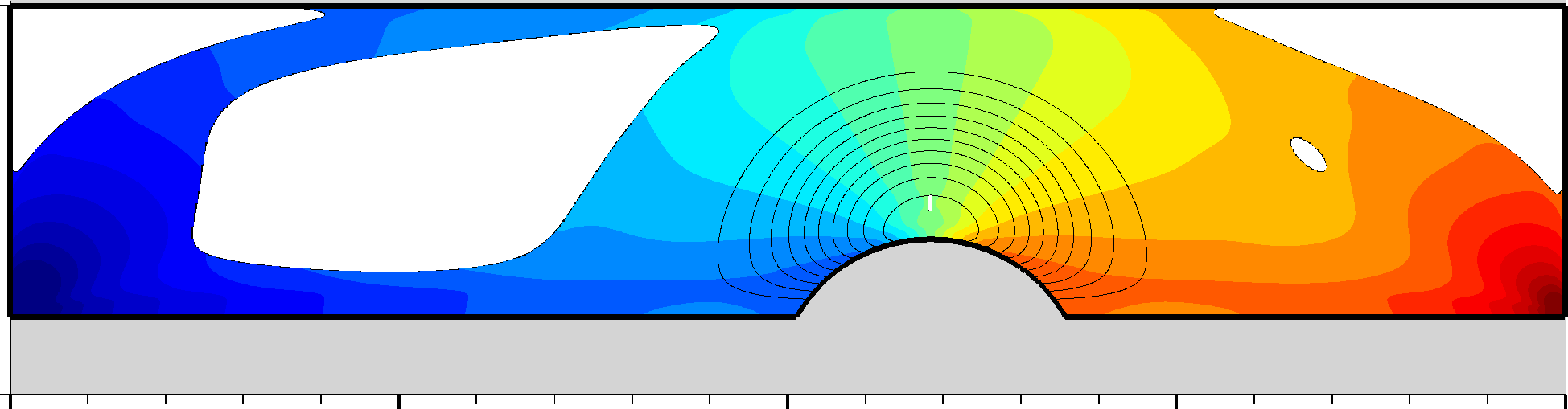}}
  \subfigure[$t = 1.5(T/4)$]       {\label{sfig: base flow 150} \includegraphics[scale=1.0]{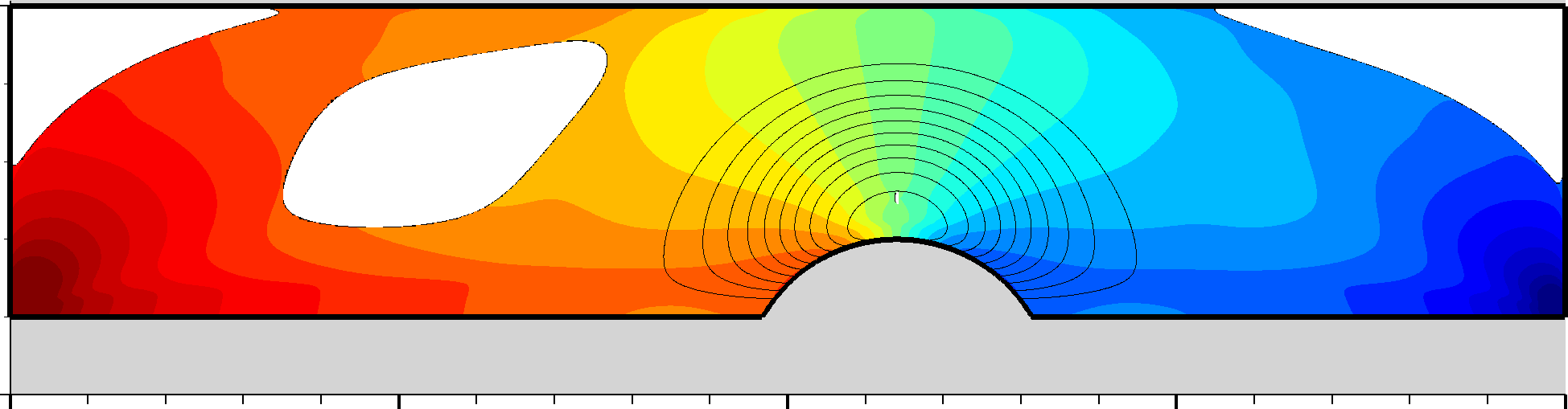}}
  \subfigure[$t = 2.5(T/4): -/+$]  {\label{sfig: base flow 250} \includegraphics[scale=1.0]{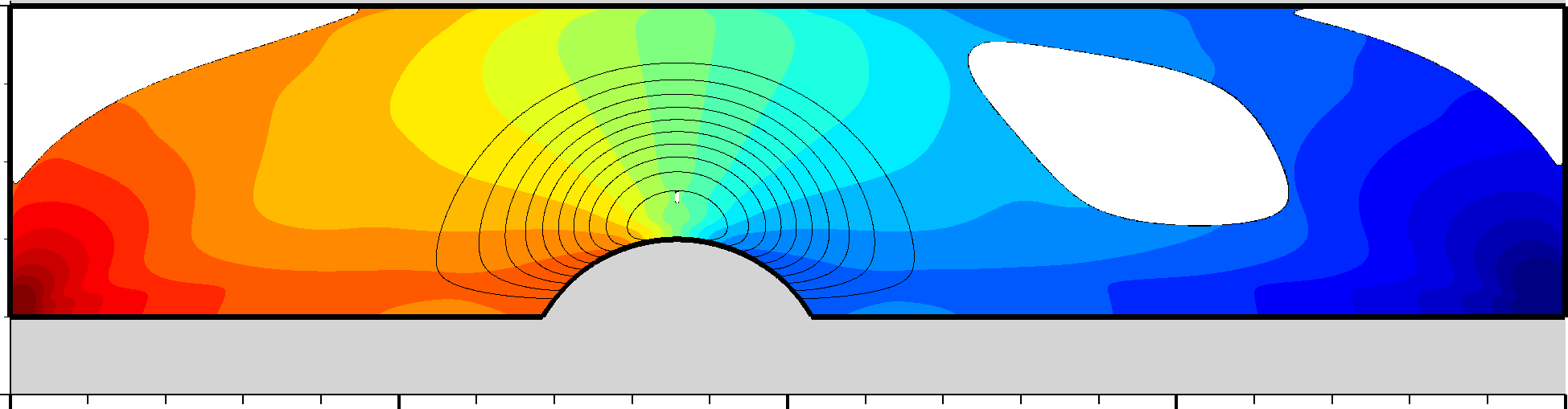}}
  \subfigure[$t = 1.75(T/4)$]      {\label{sfig: base flow 175} \includegraphics[scale=1.0]{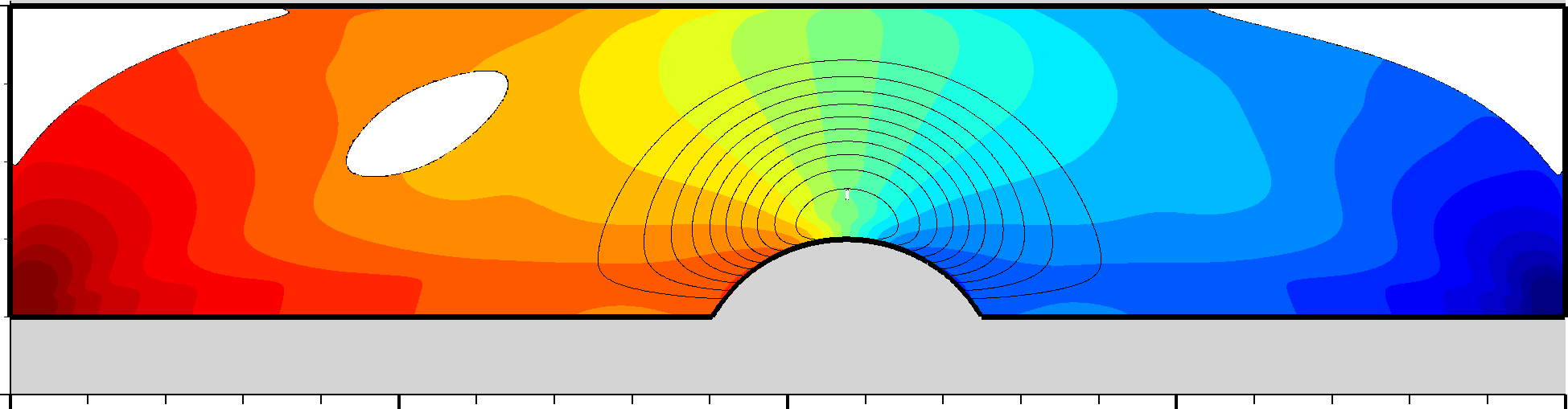}}
  \subfigure[$t = 3.75(T/4): -/-$] {\label{sfig: base flow 375} \includegraphics[scale=1.0]{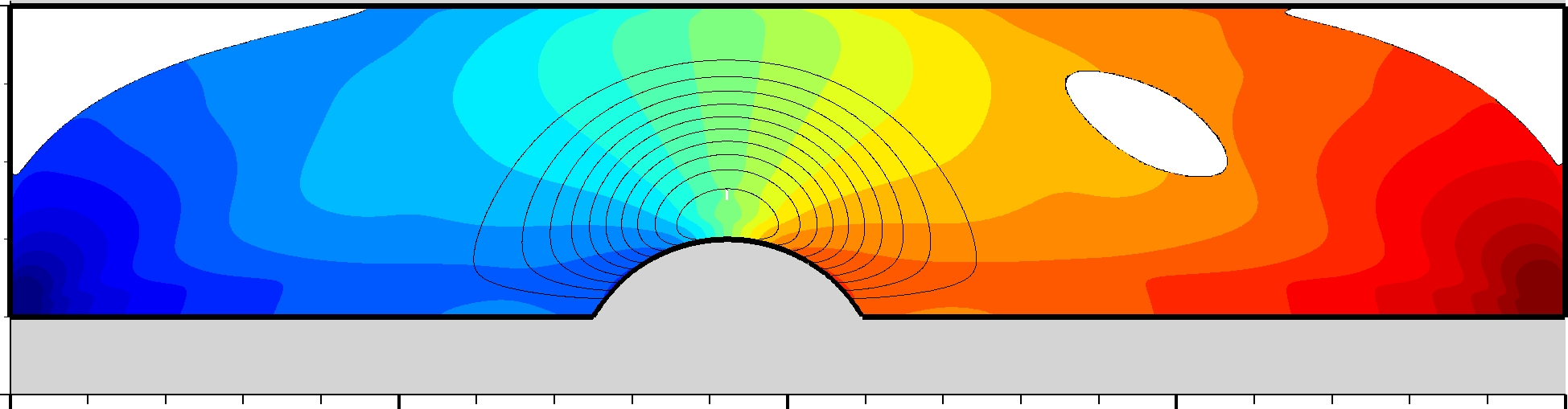}}
  \subfigure[$t = T/2$]            {\label{sfig: base flow 200} \includegraphics[scale=1.0]{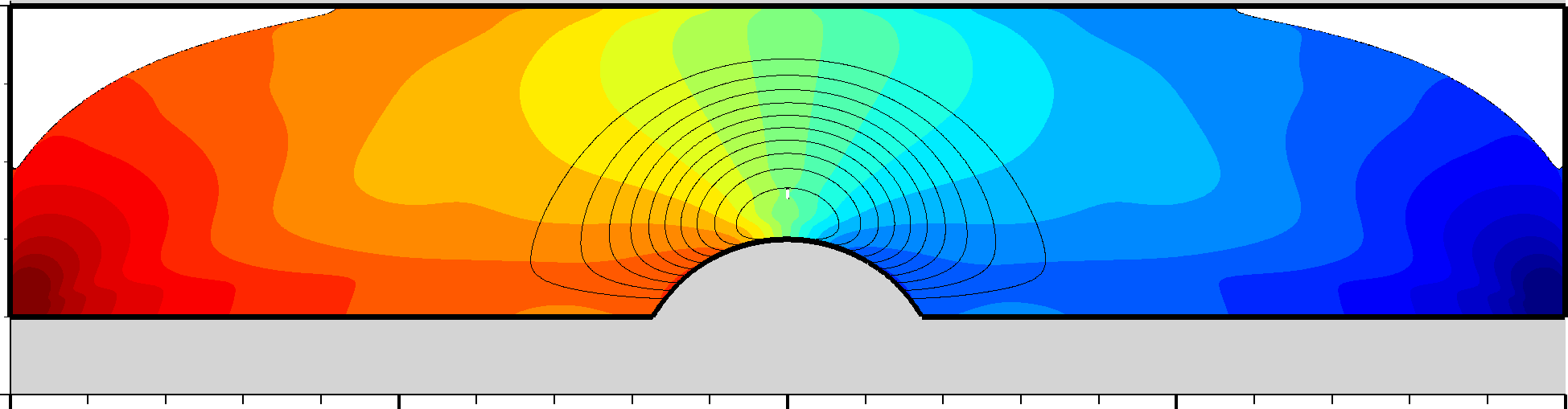}}
  \subfigure[$t = T: +/-$]         {\label{sfig: base flow 400} \includegraphics[scale=1.0]{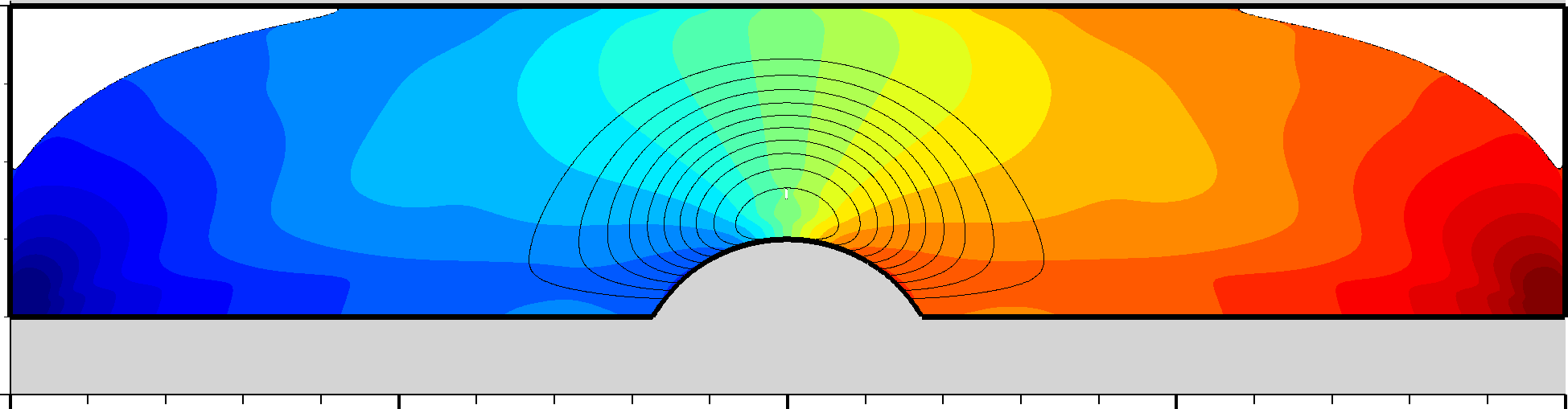}}
  \caption{Snapshots of the flow field for the base case (Table \ref{table: base case}). The shaft is shown in grey. The black 
lines are streamlines; within each figure they correspond to equispaced values of the streamfunction, from zero (at motionless 
walls) to the instantaneous maximum value. The colour contours represent dimensionless pressure $\tilde{p} = p / 
\tau_{\mathrm{ref}}$, in the range from $-10$ (blue) to $+10$ (red) with a step of $0.8$. Regions of unyielded material ($\tau < 
\tau_y$) are shown in white. The left column of figures are in chronological order from top to bottom, starting with the shaft 
motionless in its extreme right position at time $t = T/4$ \subref{sfig: base flow 100} and ending with the shaft in the middle 
position, moving with maximum velocity, at time $t = T/2$ \subref{sfig: base flow 200}. The right column of figures is not in 
chronological order, but each figure exhibits some sort of symmetry compared to the figure immediately to its left. This is 
indicated with a pair of signs in the caption of each figure in the form D/U, where D is a ``$+$'' if the displacement of the 
shaft is the same as in the figure immediately on the left, and a ``$-$'' if it is opposite to that, and U gives the same 
information for the shaft velocity.}
  \label{fig: base flow}
\end{figure}

We start with a general description of the flow inside the damper for the base case, which is visualised in Fig.\ \ref{fig: base 
flow}. A first observation is that at the extreme points of the shaft motion, Figs.\ \ref{sfig: base flow 100} and \ref{sfig: base 
flow 300}, when the shaft velocity is zero, the fluid velocity is also zero and the material is in fact completely unyielded. The 
same observation has been made for almost all test cases studied in the present work, except for Newtonian flow and flow at high 
frequency. Therefore, there is no point in extending the duration of each simulation beyond a single period $T$, as the flow has 
already reached a periodic state from $t = T/4$. For a few exceptional cases we extended the simulation duration to two periods, 
although the results which will be presented in this paragraph show that the periodic state is reached sooner.

As the shaft retracts from its extreme right position and accelerates (left column of snapshots in Fig.\ \ref{fig: base flow}), 
increasingly more of the material yields. The amount of yielded material becomes maximum when the shaft is at its central 
position and its velocity is maximum (Fig.\ \ref{sfig: base flow 200}); at this point the unyielded material is restricted only to 
the outer corners of the cylinder. Yet, most of the flow occurs in the vicinity of the bulge, as shown by the density of the 
streamlines. The bulge motion causes the fluid immediately downstream of it to be pushed out of the way, and following a 
circular-like path it is transported behind the bulge. Away from the bulge, the fluid, although mostly yielded, moves extremely 
slowly. 

The right column of snapshots in Fig.\ \ref{fig: base flow} corresponds to time instances when the shaft displacement and 
velocity are either equal or opposite to that of the snapshot immediately to the left. It is evident from comparing the left and 
right columns of the figures that symmetry or equality in the instantaneous boundary conditions implies also symmetry or equality 
of the flow field. The flow history does not play a significant role; it is mostly the instantaneous boundary conditions that 
determine the flow field. This can be attributed to the low Reynolds number of the base case, $Re^* = 0.12$ (Table \ref{table: 
base case}) which makes the left hand-side of the momentum Eq.\ \eqref{eq: momentum nd}, i.e. the inertia forces, very small 
compared to the right hand side (pressure and viscoplastic forces). The time derivative term in the left-hand side of Eq.\ 
\eqref{eq: momentum nd} thus plays an insignificant role and the flow is in a quasi steady state where at each time instance the 
flow field is determined by the instantaneous boundary conditions and not by the history of the flow. As a side-effect, the 
accuracy of the simulation depends only weakly on the time step $\Delta t$.

\begin{figure}[t]
  \centering
  \subfigure[$t = 1.25(T/4)$]      {\label{sfig: base flow 125 nb} \includegraphics[scale=1.0]{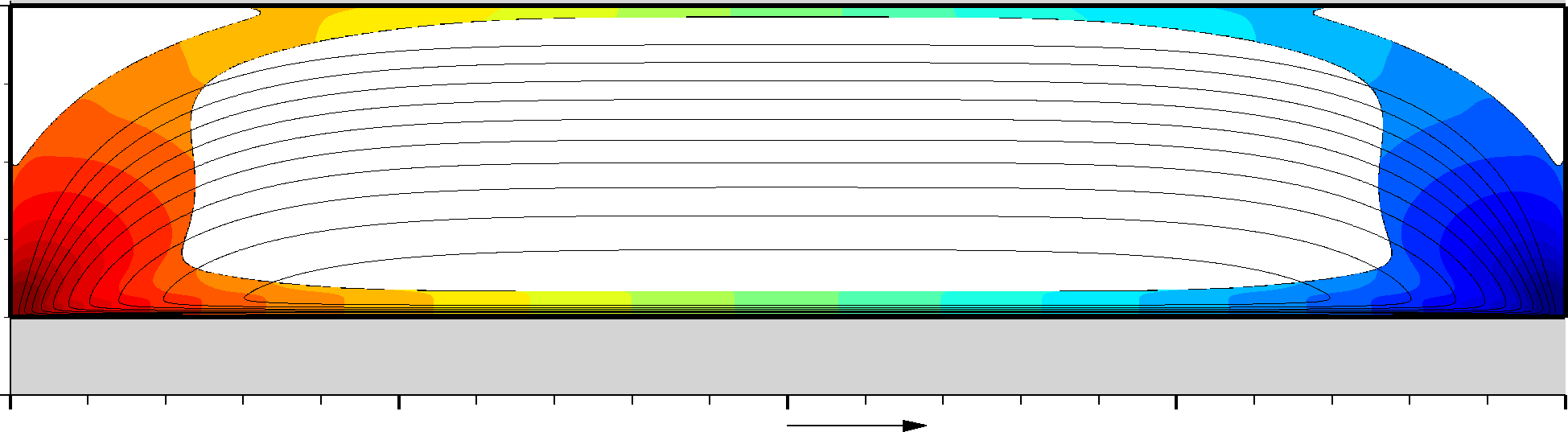}}
  \subfigure[$t = 1.5(T/4)$]       {\label{sfig: base flow 150 nb} \includegraphics[scale=1.0]{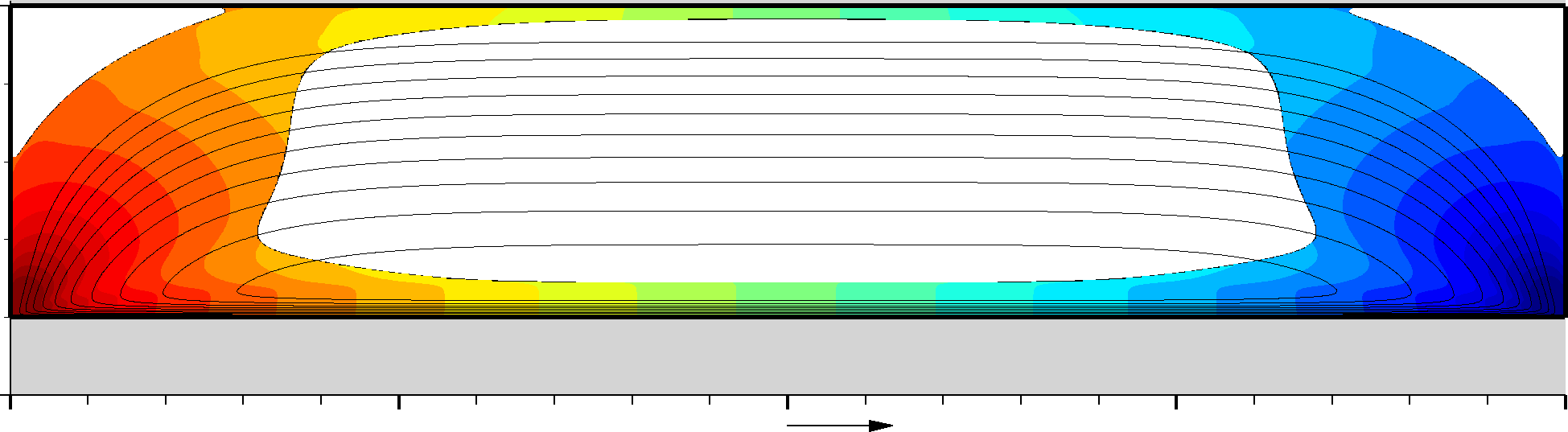}}
  \subfigure[$t = 1.75(T/4)$]      {\label{sfig: base flow 175 nb} \includegraphics[scale=1.0]{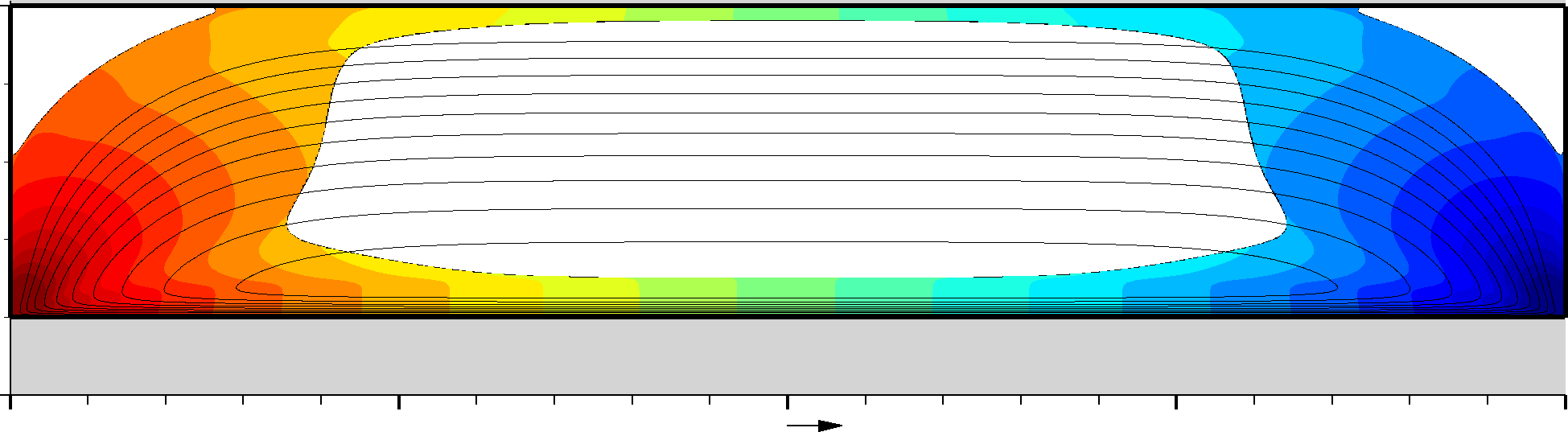}}
  \subfigure[$t = T/2$]            {\label{sfig: base flow 200 nb} \includegraphics[scale=1.0]{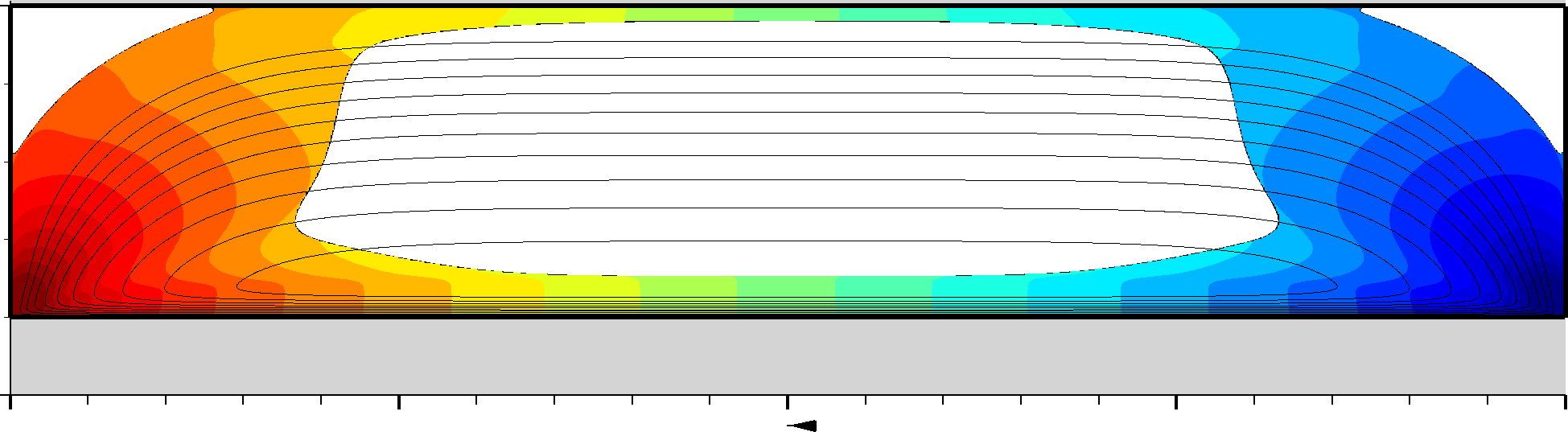}}
  \caption{Snapshots of the flow field for the base case (Table \ref{table: base case}) in the absence of a bulge. See caption 
of Fig.\ \ref{fig: base flow} for details. The arrow below each figure indicates the shaft displacement. The snapshots 
correspond to the times of the left column of figures in Fig.\ \ref{fig: base flow}. At $t = T/4$ the material is completely 
unyielded.}
  \label{fig: base flow no bulge}
\end{figure}

A feel of the effect of the bulge can be obtained by comparing the snapshots of Fig.\ \ref{fig: base flow no bulge}, obtained 
also for the parameters of the base case but in the absence of a bulge, against those in the left column of Fig.\ \ref{fig: base 
flow}. The bulge causes larger stresses in its vicinity, causing more of the material to yield. The streamline pattern shows that 
it also causes significant flow around it as it moves. On the contrary, in the absence of a bulge the streamline pattern shows 
that the motion of the material is concentrated in a very thin layer close to the shaft, whereas in the rest of the domain the 
material moves very slowly. Also, the variation of the size and shape of the unyielded regions during the oscillation is weaker 
than in the bulged shaft case; in fact in the bulgeless case, throughout the oscillation, the central unyielded plug zone 
extends over most of the domain leaving only thin yielded layers over the shaft and outer cylinder, while its axial extent 
varies weakly with time.

In the paragraphs that follow, the effect of various parameters on the flow is examined.

\subsection{Effect of viscoplasticity}
\label{ssec: results viscoplasticity}

\begin{figure}[!p]
  \centering
  \subfigure[] {\label{sfig: Ftot vs dx per Bn} \includegraphics{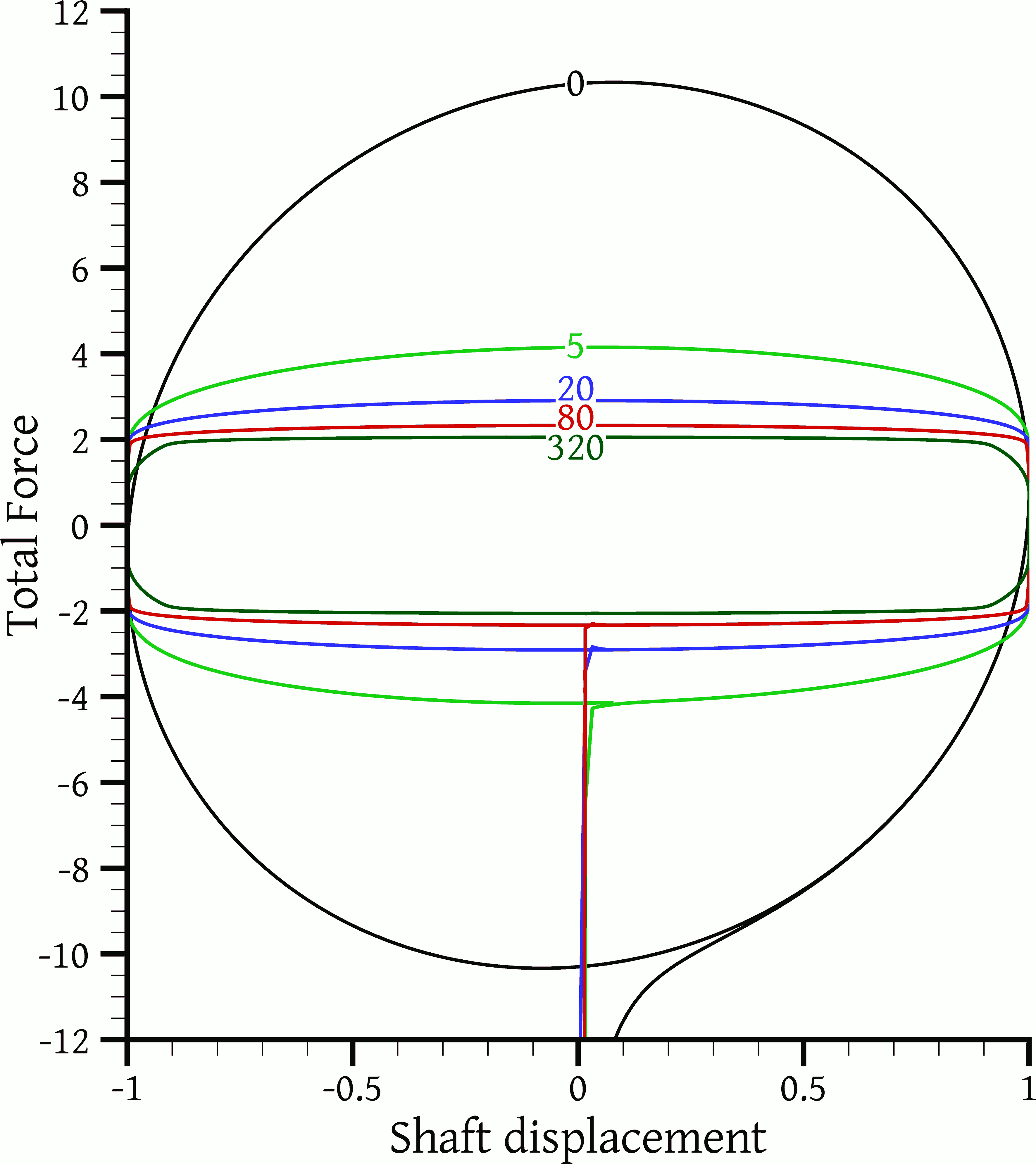}}
  \subfigure[] {\label{sfig: Fvisc vs dx per Bn} \includegraphics{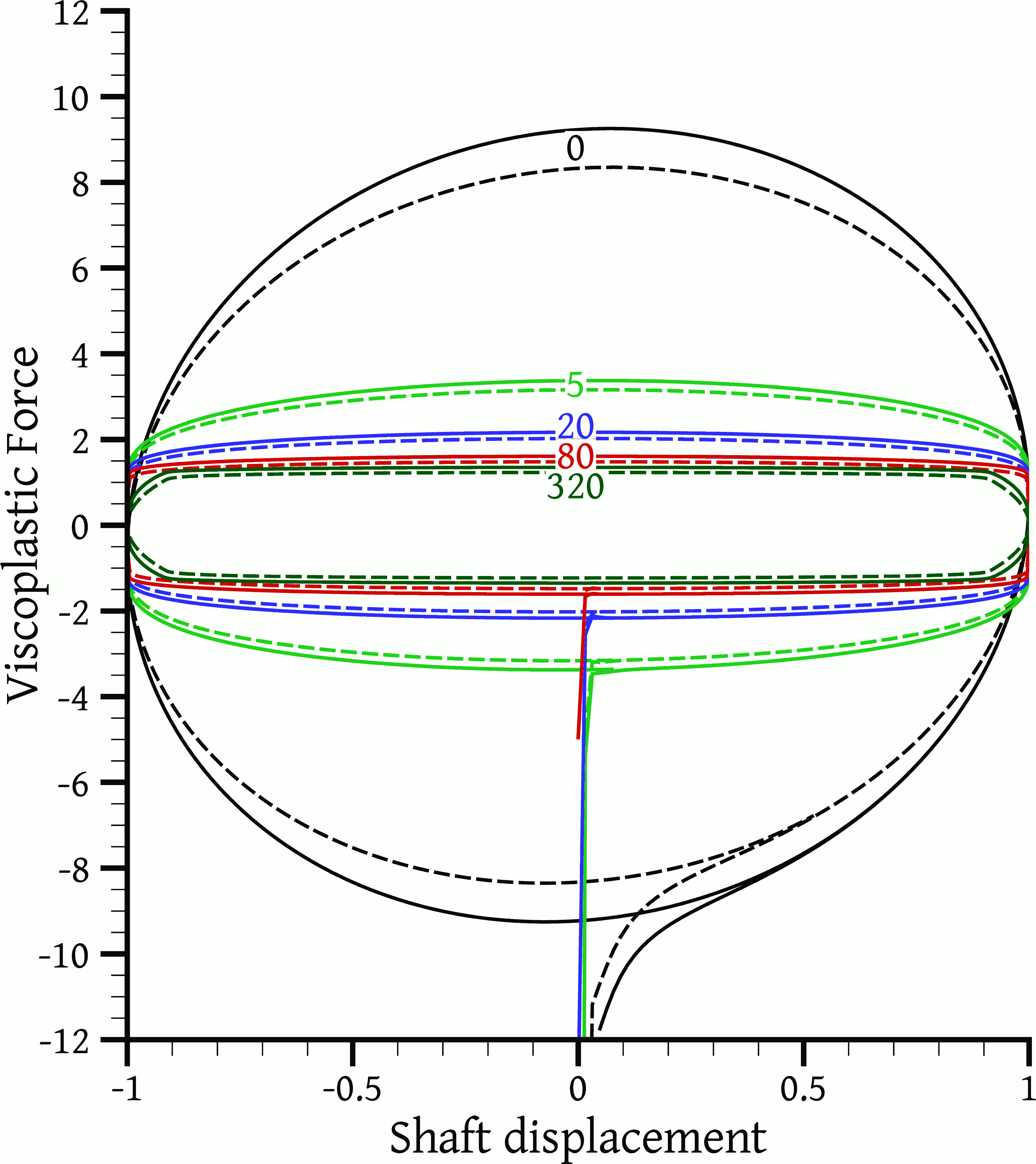}}
  \subfigure[] {\label{sfig: Fpres vs dx per Bn} \includegraphics{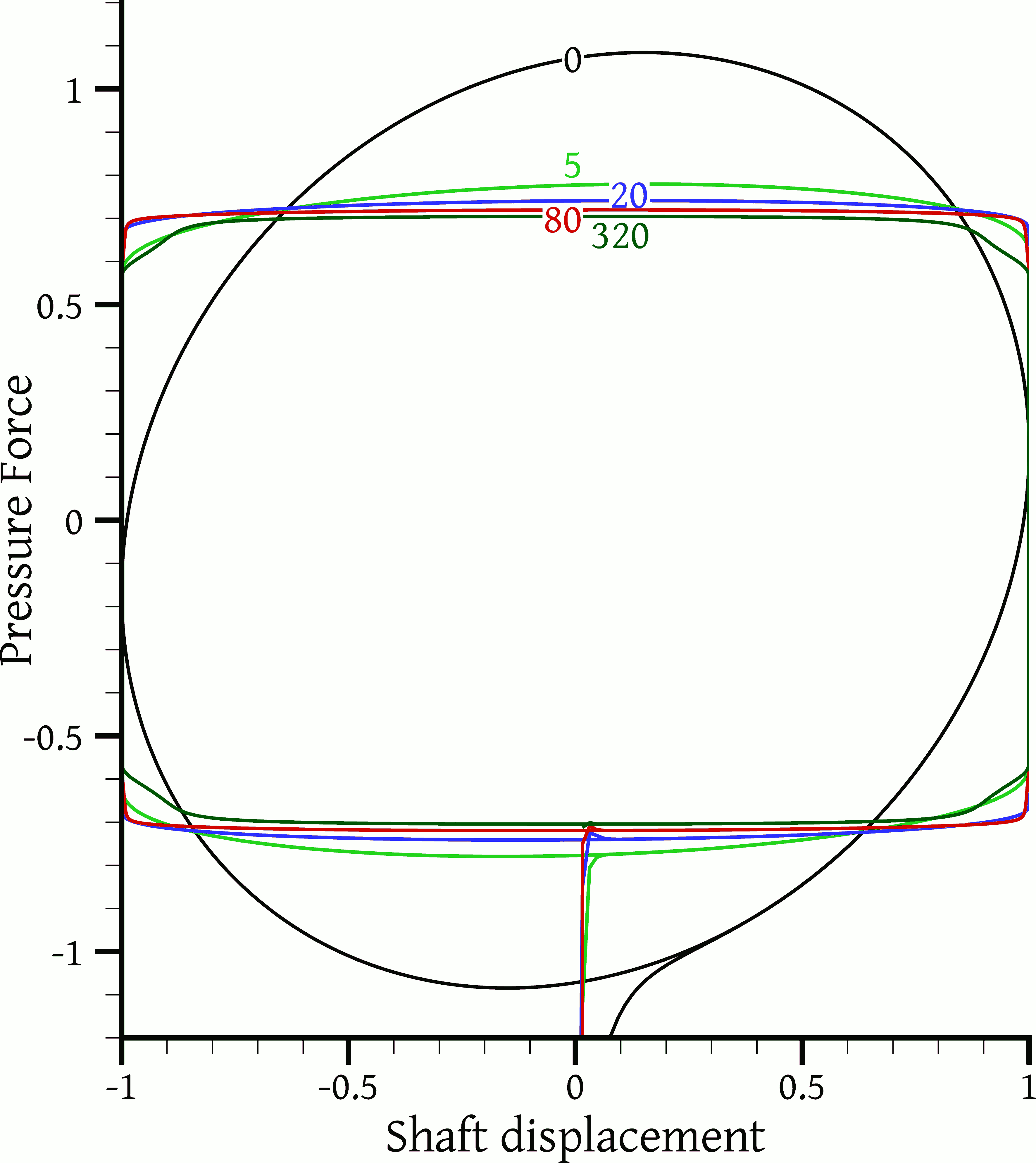}}
  \subfigure[] {\label{sfig: Ftot vs V per Bn} \includegraphics{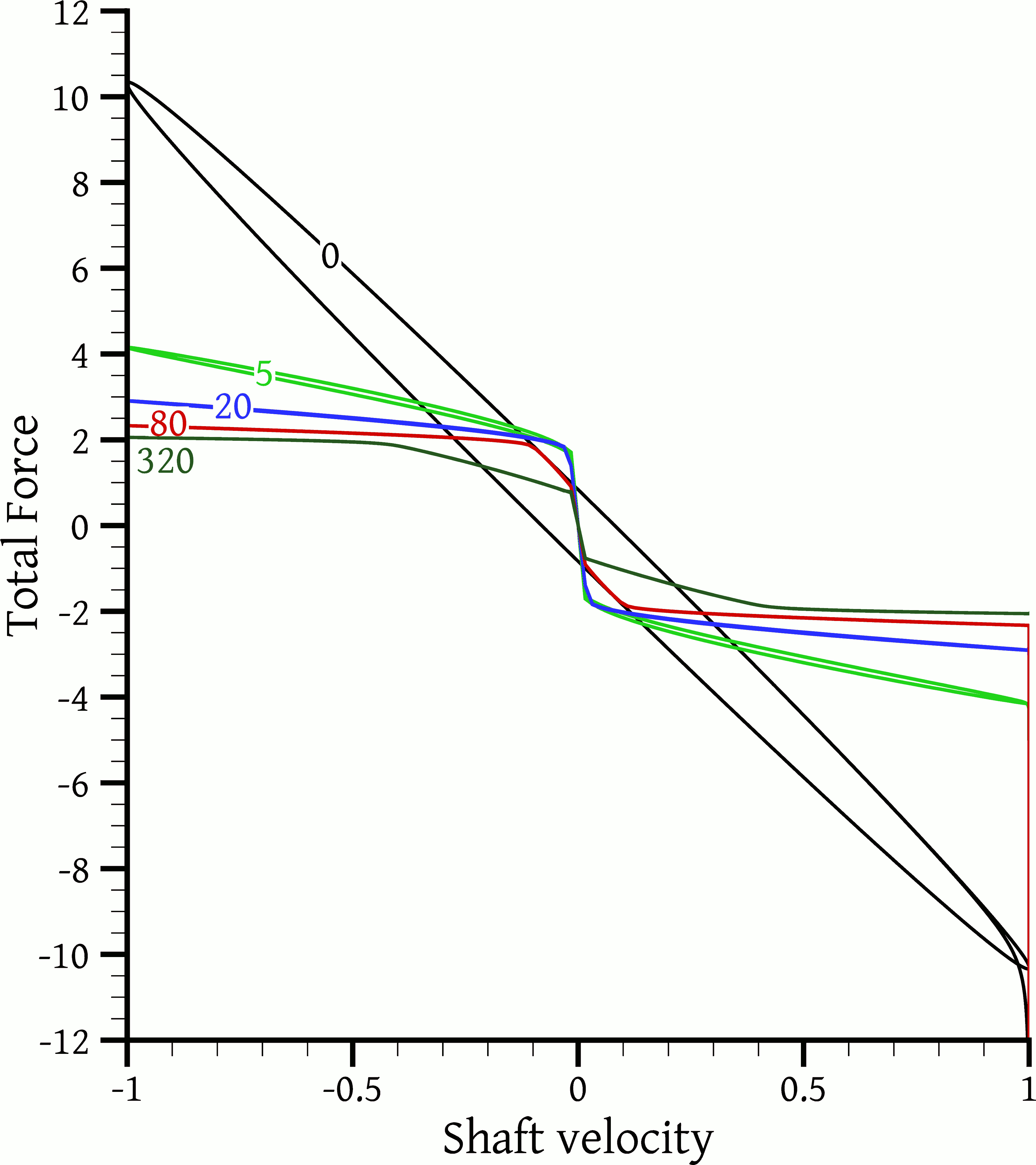}}
  \caption{Various components of force as a function of either shaft displacement or shaft velocity, for several different 
Bingham 
numbers which are indicated on each curve. Forces are dedimensionalised by $F_{\mathrm{ref}}$, displacement by the oscillation 
amplitude $\alpha$, and velocity by $U$. The dimensional parameters of each experiment are as shown in Table \ref{table: base 
case}, except that the yield stress $\tau_y$ has been adjusted to obtain the Bingham numbers shown. All simulations have a 
duration of one period $T$, except the Newtonian case which has a duration of $2T$. Each curve is traversed in a 
counterclockwise 
sense with respect to time. In \subref{sfig: Fvisc vs dx per Bn}, the dashed lines indicate the force in the absence of a bulge.}
  \label{fig: force per Bn}
\end{figure}

The most important result of the simulations is the damper reaction force $F_R$ as a function of the shaft displacement or 
velocity. This force can be analysed into two components, a viscoplastic component $\int_{sh} \vf{n} \cdot \tf{\tau} \, 
\mathrm{d}\!A$ and a pressure component $\int_{sh} -p \vf{n} \, \mathrm{d}\!A$, where integration is over all the shaft surface 
and $\vf{n}$ is the unit vector normal to this surface. Due to symmetry, the net force is in the axial direction $\vf{e}_x$ 
only. Figure \ref{fig: force per Bn} shows how the total force and its separate viscoplastic and pressure components are affected 
by the viscoplasticity of the material. The different curves correspond to materials with different yield stress, while the rest 
of the material properties are the same, as listed in Table \ref{table: base case}. The Bingham number, being representative of 
the viscoplasticity of the material, is used to differentiate between the curves, but by changing the yield stress other 
dimensionless numbers change as well: the effective Reynolds number $Re^*$ (but not the usual $Re$) decreases as $\tau_y$ 
increases (Eq.\ \eqref{eq: Re*}), reflecting the fact that by increasing the yield stress the viscoplastic forces become more 
dominant over inertia; and the slip coefficient $\tilde{\beta}$ (Eq.\ \eqref{eq: slip coefficient nd}) increases as $\tau_y$ 
increases, reflecting the fact that, for a given shaft velocity $U$, increasing $\tau_y$ generally increases the overall levels 
of stress in the domain leading to more slip at the walls.

The effect of viscoplasticity is summarised in Fig.\ \ref{sfig: Ftot vs dx per Bn}. At time $t=0$ the material is initially at 
rest, but the shaft suddenly starts to move towards the right at a finite velocity of $U = \alpha \omega$. This creates a very 
large initial inertial reaction force (towards the left, i.e.\ with negative sign) whose magnitude drops very rapidly due to the 
small value of $Re^*$; this drop is illustrated by the nearly vertical part of the curves at zero displacement. The test case 
where the relative importance of inertia is greatest is the Newtonian case ($Bn = 0$) where indeed the initial force drop can be 
seen to be more gradual, but nevertheless the periodic state is quickly attained when the displacement is about $0.4\alpha$ and 
from that point on the force at time $t$ is indistinguishable from that at time $t+T$ (the Newtonian case was solved for a 
duration of $2T$). The force-displacement curves for $Bn = 0$, and to a lesser extent for $Bn = 5$, are skewed; that is, the 
reaction force is smaller when the shaft is approaching an extreme position $x = \pm \alpha$ and decelerating, than when it is 
retracting from it and accelerating. This is due to inertia: when the shaft is accelerating then it also has to accelerate the 
surrounding fluid, whereas when it is decelerating it does not have to do so because the fluid has already acquired momentum in 
the direction of motion. This effect is insignificant when the viscous forces greatly surpass the inertia forces, i.e. at low 
$Re^*$ numbers. This is shown more clearly in the force-velocity diagram, Fig.\ \ref{sfig: Ftot vs V per Bn}, where the curves 
for $Bn$ = 0 and 5 exhibit some hysteresis, i.e. the force does not depend only on the current velocity but also on the flow 
history. For each shaft velocity there are two values of force: a higher one, when the shaft is accelerating, and a lower one, 
when it is decelerating. On the contrary, for $Bn \geq 20$ no such hysteresis is observable, and the curves are symmetric with 
respect to the zero displacement line in Fig.\ \ref{sfig: Ftot vs dx per Bn}. Thus for these cases $Re^*$ is so small that the 
inertia terms in Eq.\ \eqref{eq: momentum nd} are negligible. The weakening of the hysteresis effect with increasing the Bingham 
number can be seen in the experimental results reported in \cite{Nguyen_2009}.

In the diagrams of Fig.\ \ref{fig: force per Bn} the force is dedimensionalised by $F_{\mathrm{ref}}$, Eq.\ \eqref{eq: Fref 
tau_ref}, and so increasing $\tau_y$ makes the force appear smaller whereas in fact it becomes larger. The obvious effect of 
increasing $\tau_y$ is to make the force curves flatter, i.e.\ the larger $\tau_y$ the less the force varies during the motion of 
the shaft. For some applications this is considered an advantage of the damper, since it maximises the energy absorbed for a 
given force capacity. The explanation is simple: the total viscoplastic stresses consist of two components, one of constant 
magnitude (plastic) and one of variable magnitude (viscous), $\tau = \tau_y + \mu \dot{\gamma}$. The Bingham number is an 
indicator of the ratio of the constant to the variable component. Thus, the larger $Bn$ the smaller the variation of $\tau$ and 
of the resulting force during the shaft motion. Hence, the circular shape of the force vs.\ displacement cycle in the Newtonian 
case tends to a rectangular one as the Bingham number is increased. Of course, the force also has a pressure component, but the 
momentum equation suggests that pressure forces behave similarly to viscoplastic ones when $Re^*$ is small. These theoretical 
findings are confirmed experimentally, see e.g.\ \cite{Symans_1999, Nguyen_2009}. In fact, force-displacement diagrams for 
varying Bingham numbers can be found in most ER and MR damper studies, as they are obtained for different strengths of the 
electric or magnetic fields. However, usually it is the dimensional forces that are plotted, under the same scale, and since even 
at low field strengths the Bingham number is rather high, it is difficult to discern the differences in the curvature of the 
plots (for example, in Fig.\ \ref{sfig: Ftot vs dx per Bn} the differences between the curves for $Bn$ = 20, 80 and 320 would not 
be easily discernable had the dimensional forces been plotted instead).

Figures \ref{sfig: Fvisc vs dx per Bn} and \ref{sfig: Fpres vs dx per Bn} show the viscoplastic and pressure contributions to 
the total force. In Fig.\ \ref{sfig: Fvisc vs dx per Bn} the force that would result had there been no bulge is also plotted with 
dashed lines. It can be seen that the presence of the bulge increases the viscoplastic force only slightly. On the other hand, 
the pressure force is due solely to the bulge; in its absence there is no pressure force in the $x$-direction, since the 
projection of the shaft's surface in that direction is zero. Figure \ref{sfig: Fpres vs dx per Bn} shows that the pressure 
force, normalised by $F_{\mathrm{ref}}$, is almost independent of the Bingham number, having a value of about 0.7. This will be 
shown later to depend on the damper geometry. The viscoplastic force on the other hand does depend on $Bn$ for lower values of 
$Bn$, but tends to unity as $Bn$ is increased. The pressure forces appear rather small compared to the viscoplastic forces, but 
become more important as the Bingham number is increased. This implies also that the role of the bulge becomes more important as 
$Bn$ is increased; however, as noted, asymptotically as $Bn$ is increased, the ratio of viscoplastic to pressure forces tends to 
a certain limit.

It is interesting to examine what happens when the shaft reaches an extreme position and momentarily stops ($x = \pm \alpha$, 
shaft velocity = 0). It is most clearly seen in Fig.\ \ref{sfig: Ftot vs V per Bn} that in the Newtonian case the fluid continues 
to flow, resulting in a non-zero reaction force, but in all the viscoplastic cases shown the fluid stops (actually it becomes 
completely unyielded) and the reaction force $F_R$ becomes zero. Nevertheless, even the slightest shaft motion causes non-zero 
rates of deformation and therefore yielding of the fluid, with the stress magnitude jumping from zero to the yield stress. Thus 
the force also immediately jumps from zero to some non-zero value, and then gradually increases further as the rate of deformation 
increases due to shaft acceleration. A departure from this behaviour can be noticed for the $Bn = 320$ case both in Fig.\ 
\ref{sfig: Ftot vs V per Bn} and in Fig.\ \ref{sfig: Ftot vs dx per Bn}, where a relatively smaller jump in $F_R$ occurs relative 
to the smaller $Bn$ cases, followed by a more gradual increase of $F_R$ until it reaches a nearly constant value. This is due to 
the Navier slip boundary condition and will be explained in the following subsection.

Another interesting quantity that would help shed more light on the damper operation is the rate of dissipation of mechanical 
energy to thermal energy inside the fluid. For generalised Newtonian fluids, this rate, per unit volume, is given by the 
dissipation function

\begin{equation} \label{eq: dissipation function}
 \phi \;\equiv\; \tf{\tau} : \nabla \vf{u} \;=\; \eta \dot{\gamma}^2
\end{equation}
The second equality is valid for generalised Newtonian fluids, for which $\tf{\tau} = \eta \tf{\dot{\gamma}}$. The term 
$\tf{\tau} : \nabla \vf{u}$ gives the rate of work done in deforming the fluid, per unit volume. For viscoelastic fluids, some of 
this work is stored as elastic energy in the material, but for generalised Newtonian fluids, including the Bingham material 
considered here, this work concerns only conversion of mechanical energy into heat \cite{Winter_1987}. The dissipation function 
is dedimensionalised here by

\begin{equation} \label{eq: phi ref}
 \phi_{\mathrm{ref}} = \tau_{\mathrm{ref}} U/H \;=\; \tau_y \frac{U}{H} \;+\; \mu \frac{U^2}{H^2}
\end{equation}

\begin{figure}[t]
  \centering
  \includegraphics[scale=1.00]{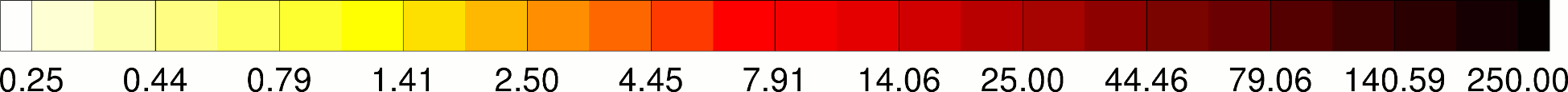}\\[0.15cm]
  \subfigure[$Bn = 0$] {\label{sfig: dissipation 200 Bn0} 
      \includegraphics[scale=1.0]{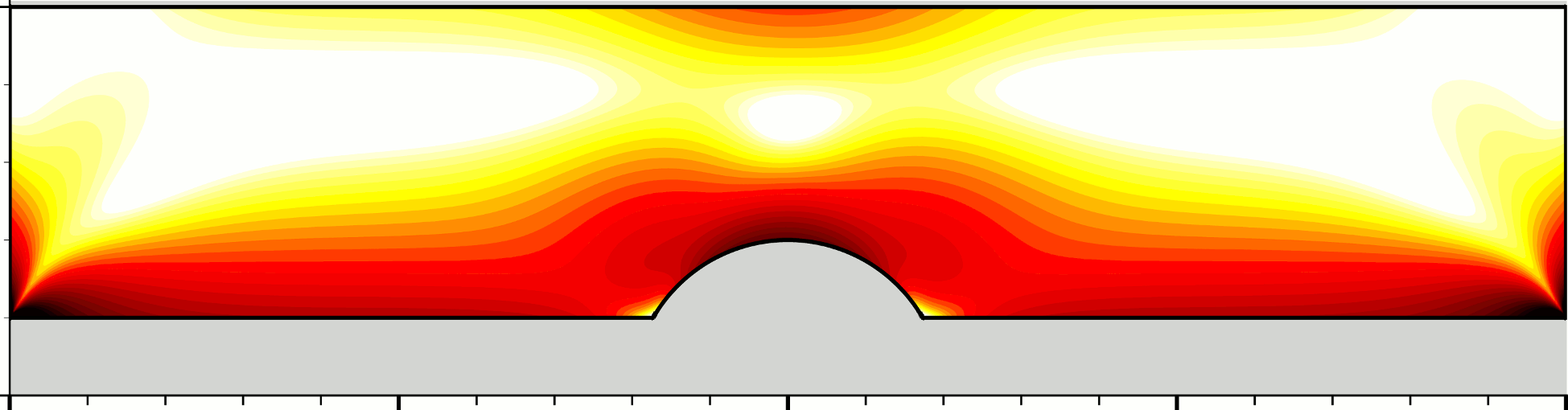}}\\
  \subfigure[$Bn = 5$] {\label{sfig: dissipation 200 Bn5} 
      \includegraphics[scale=1.0]{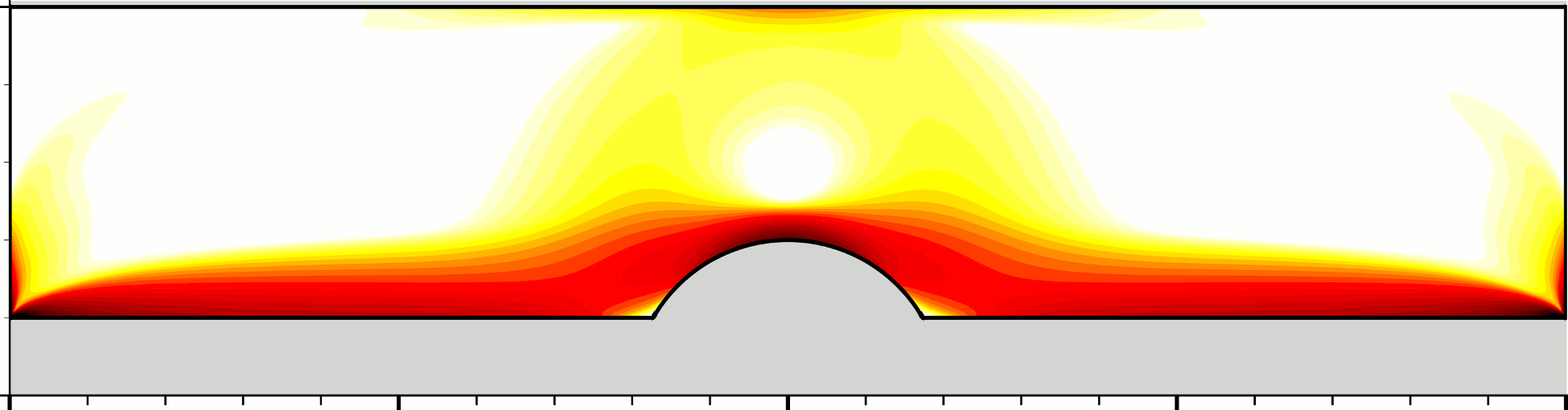}}\\
  \subfigure[$Bn = 20$] {\label{sfig: dissipation 200 Bn20} 
      \includegraphics[scale=1.0]{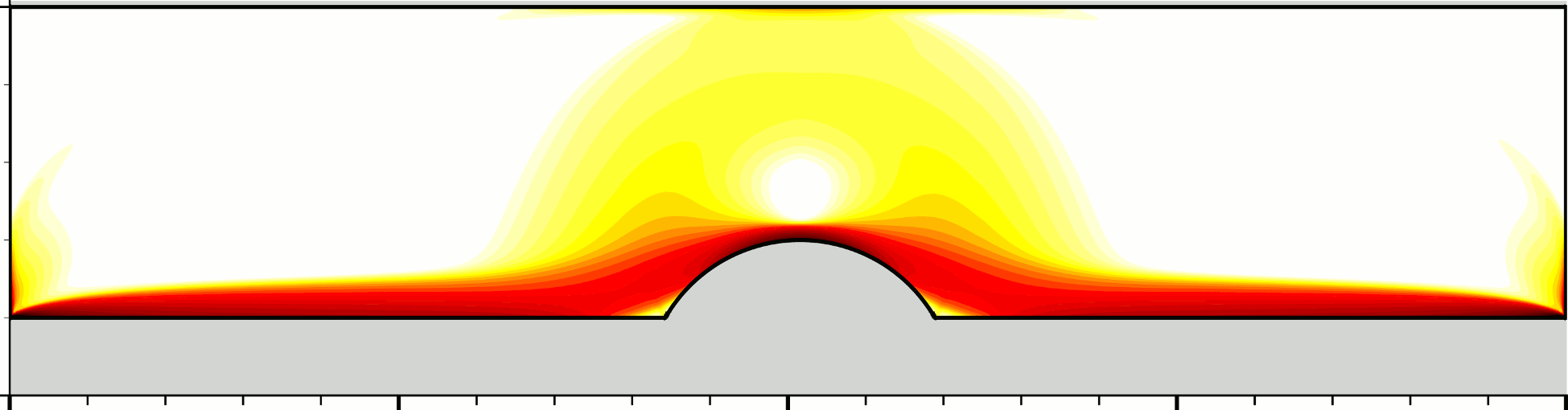}}\\
  \subfigure[$Bn = 80$] {\label{sfig: dissipation 200 Bn80} 
      \includegraphics[scale=1.0]{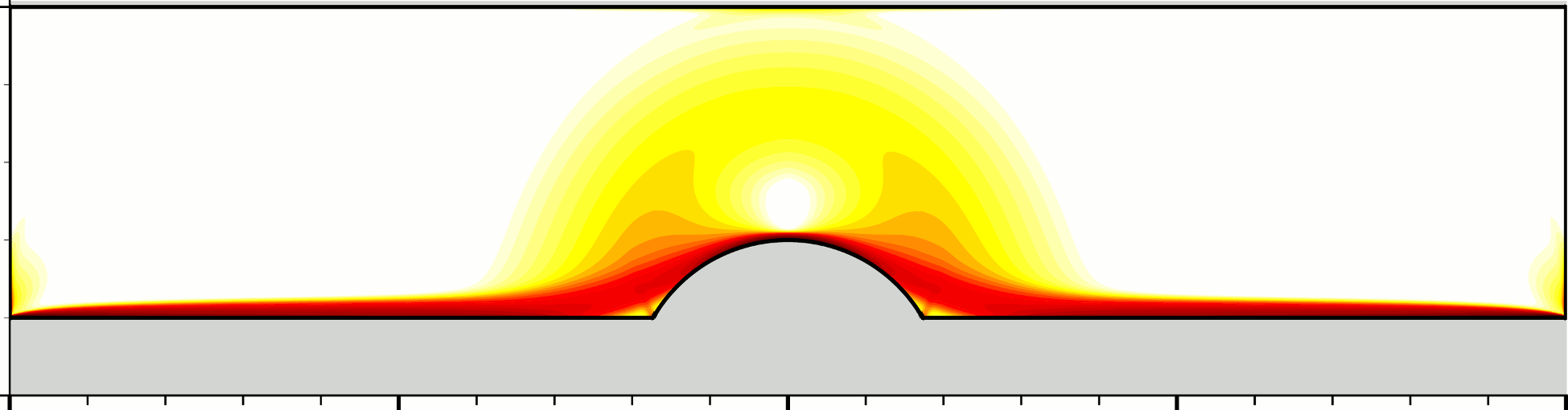}}\\
  \subfigure[$Bn = 320$] {\label{sfig: dissipation 200 Bn320} 
      \includegraphics[scale=1.0]{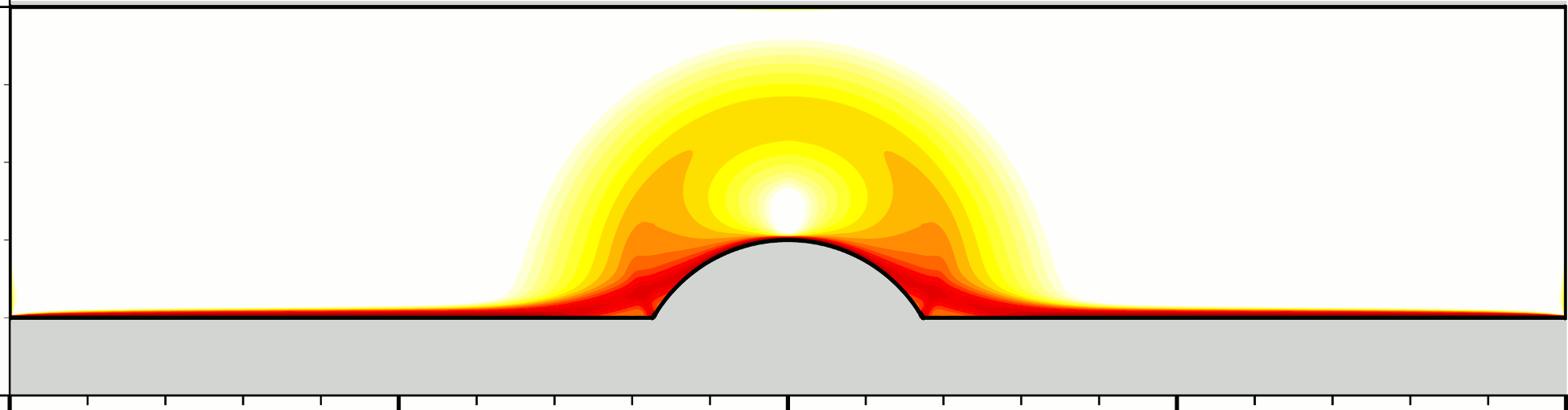}}
  \caption{Plots of the dimensionless dissipation function $\tilde{\phi} = \phi/\phi_{\mathrm{ref}}$ (Eq.\ \eqref{eq: 
dissipation 
function}) for the test cases of Fig.\ \ref{fig: force per Bn}, at $t = T/2$.}
  \label{fig: dissipation per Bn}
\end{figure}

Figure \ref{fig: dissipation per Bn} shows plots of the dissipation function for the various cases, at a time instance when the 
shaft velocity is maximum. It is evident that as the viscoplasticity of the material increases, energy dissipation becomes more 
localised, confined to a thin layer of fluid surrounding the shaft and to a ring of rotating fluid between the bulge and the 
outer cylinder. The maximum energy dissipation appears to occur at the endpoints of the shaft, where it meets the outer damper 
casing, and at the top of the bulge. For low Bingham numbers, the energy dissipation at the shaft endpoints is very significant, 
and it is due to the very large velocity gradients there, despite the Navier slip boundary condition. At higher Bingham numbers 
the contribution of these areas to the overall energy dissipation diminishes -- see also Fig.\ \ref{fig: steady forces per Bn}. 
The ring of material that rotates between the bulge and the outer cylinder also decreases in size as the yield stress is 
increased. For $Bn = 80$ (Fig.\ \ref{sfig: dissipation 200 Bn80}) and $320$ (Fig.\ \ref{sfig: dissipation 200 Bn320}) the ring 
does not extend all the way up to the outer cylinder. This suggests that for these and higher Bingham numbers the chosen radius 
$R_o$ of the enclosing cylinder has a negligible effect on the produced force, and that using a larger radius would not change 
the magnitude of the reaction force.

\subsection{Effect of slip}
\label{ssec: results slip}

\begin{figure}[!t]
  \centering
  \subfigure[with bulge]    {\label{sfig: Ftot vs V per slip}  \includegraphics{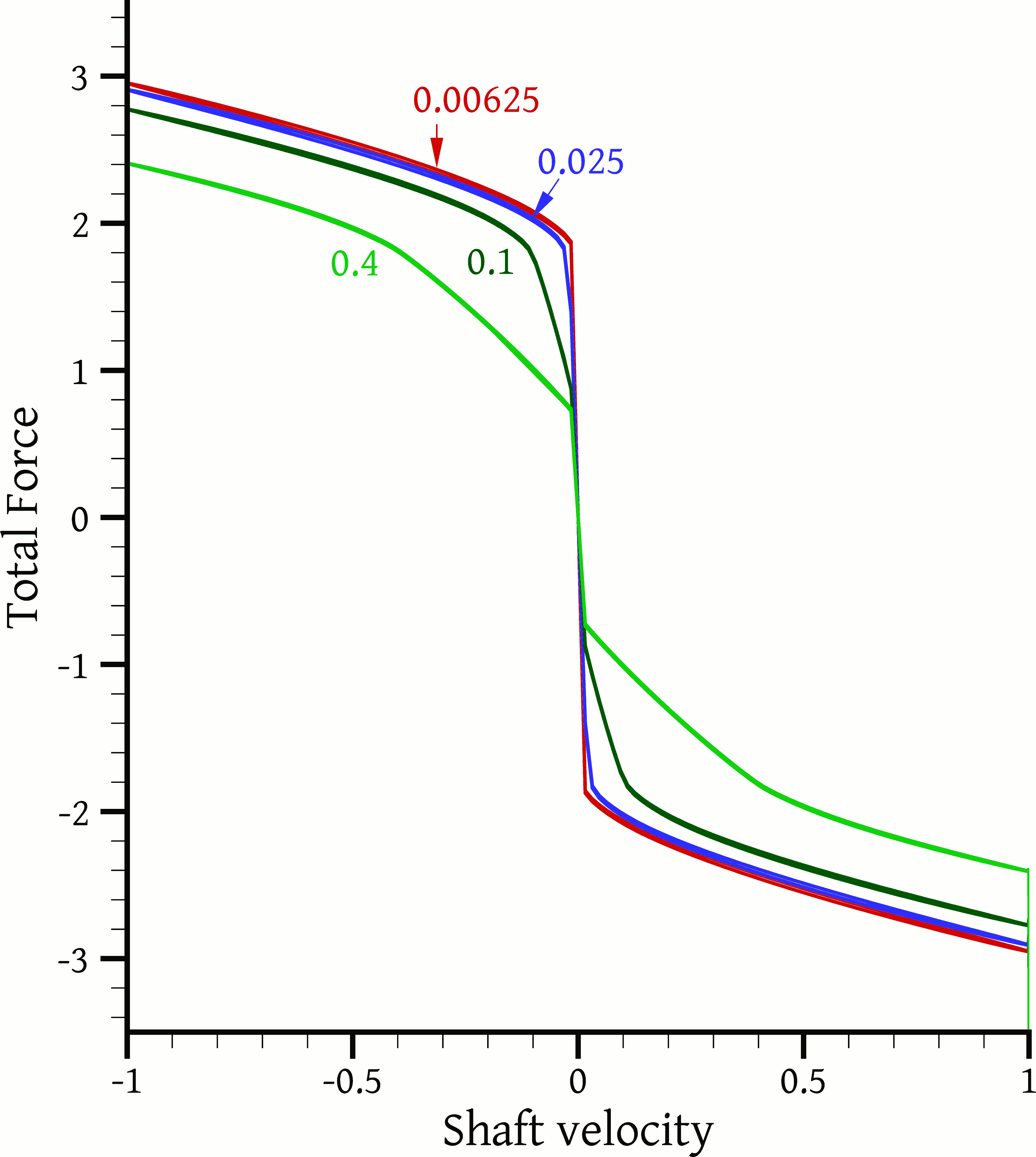}}
  \subfigure[with bulge]    {\label{sfig: Fvisc vs V per slip} \includegraphics{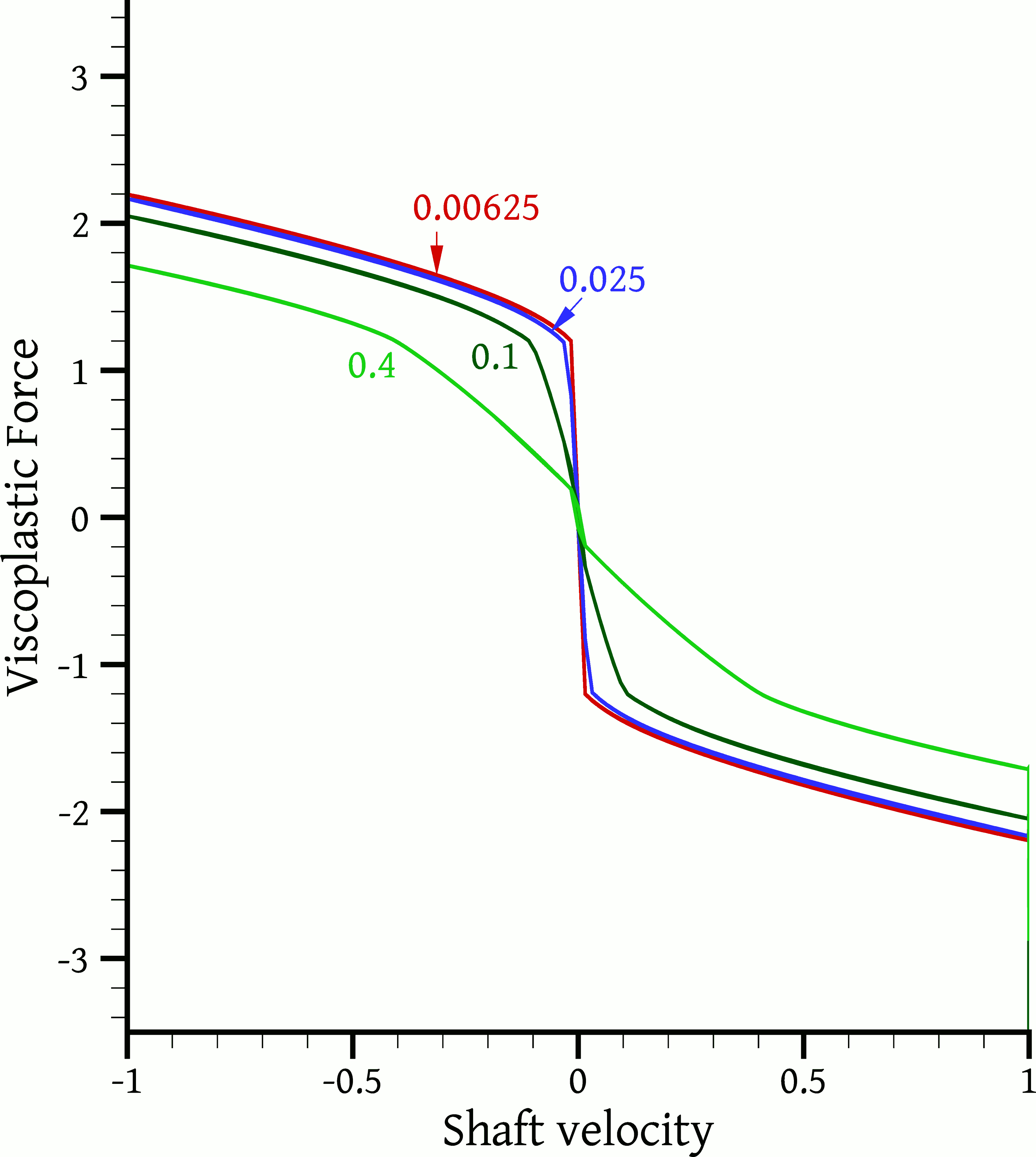}}
  \subfigure[with bulge]    {\label{sfig: Fpres vs V per slip} \includegraphics{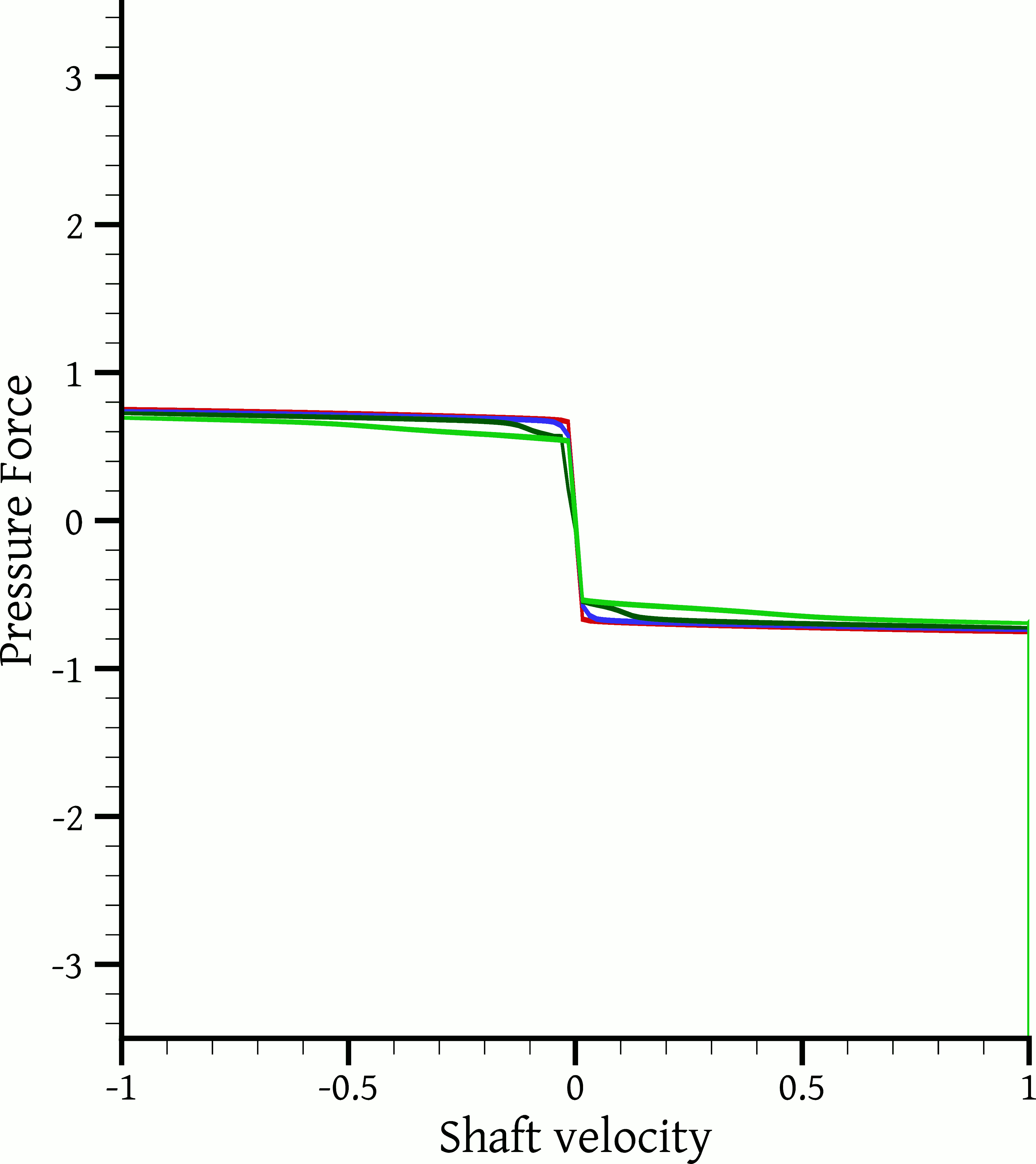}}
  \subfigure[without bulge] {\label{sfig: F vs V nb per slip}  \includegraphics{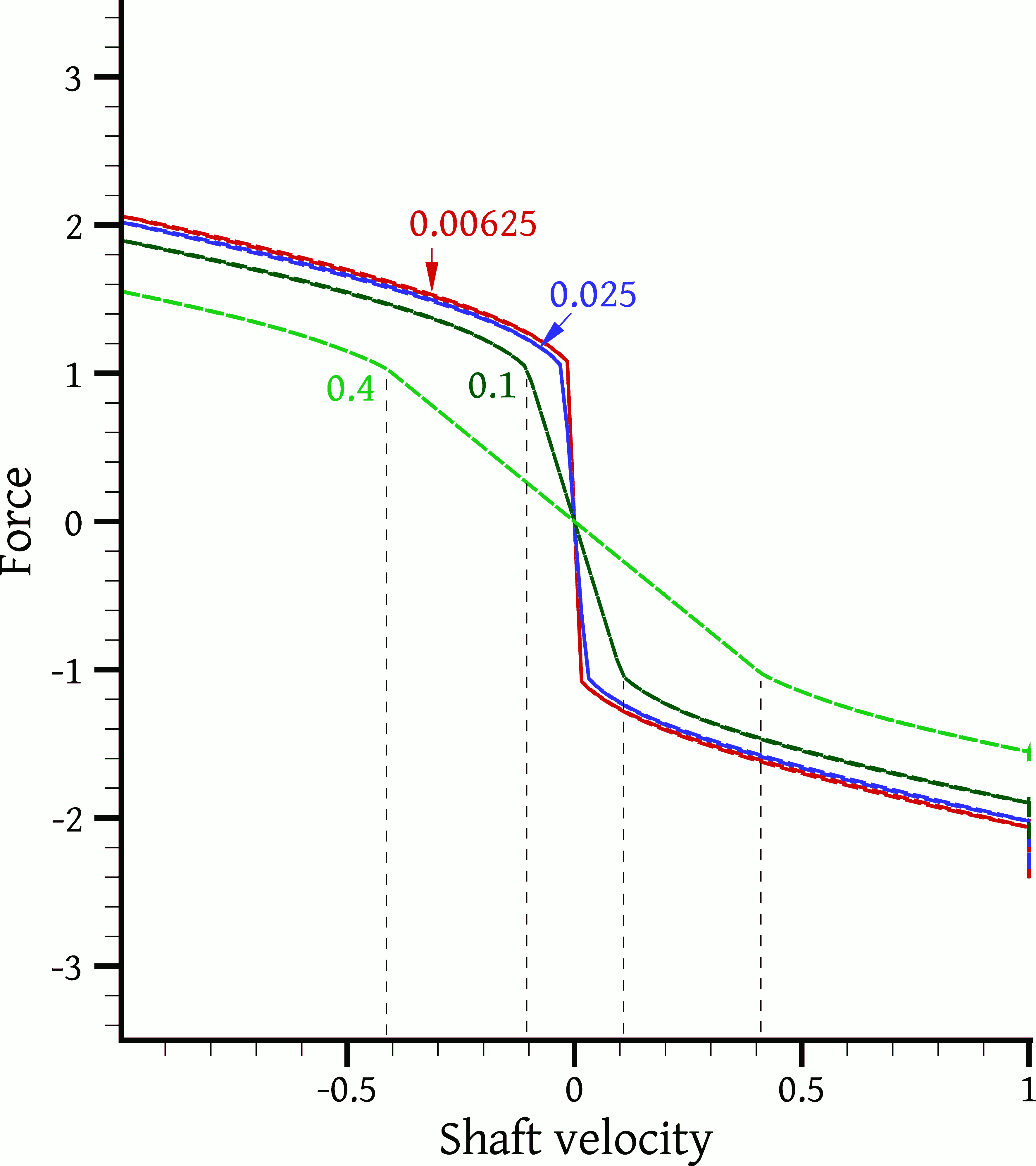}}
  \caption{Various force components as a function of shaft velocity, for different values of the dimensionless slip parameter 
$\tilde{\beta}$, indicated on each curve. Forces are dedimensionalised by $F_{\mathrm{ref}}$ and velocity by $U$. The 
dimensional 
parameters of each experiment are as shown in Table \ref{table: base case}, except that the slip parameter  $\beta$ has been 
adjusted to obtain the indicated values of $\tilde{\beta}$. Each curve is traversed in a counterclockwise sense with respect to 
time. Fig.\ \subref{sfig: F vs V nb per slip} refers to the viscous force in an experiment without a bulge (this is also the 
total 
force, because the pressure force is zero in this case).}
  \label{fig: force per slip}
\end{figure}

In the next series of simulations we start from the base case (Table \ref{table: base case}) and change the slip coefficient 
$\beta$. The only dimensionless number affected is $\tilde{\beta}$. Figure \ref{fig: force per slip} shows how the force 
$\tilde{F}_R$ and its components vary with the shaft displacement or velocity, for various values of this coefficient. As 
expected, increasing the slip coefficient decreases the reaction force and its components. Note that since the reference force 
$F_{\mathrm{ref}}$ used for the dedimensionalisation does not depend on the slip coefficient, the forces in the diagrams of Fig.\ 
\ref{fig: force per slip} are directly comparable, unlike those of Fig.\ \ref{fig: force per Bn}. All graphs in Fig. \ref{fig: 
force per slip} show that the forces resulting from $\tilde{\beta} = 0.025$ (the base case) and $\tilde{\beta} = 0.00625$ are 
nearly identical. This suggests that for the base case, the slip coefficient is too small for the slip to have a significant 
impact on the flow. But for larger slip coefficients the flow is affected significantly.

We first turn our attention to Fig.\ \ref{sfig: F vs V nb per slip} which refers to a bulgeless configuration and shows how the 
force varies with respect to shaft velocity (in the absence of a bulge the total force is equal to the viscoplastic force). For 
low $\tilde{\beta}$ the force behaves as expected: When the shaft velocity is zero, all of the material is unyielded and the 
force is also zero; then, the slightest movement of the shaft causes fluid deformation and therefore yielding and the stress 
jumps to $\tau_y$, resulting in a sharp rise of the reaction force to $\tilde{F}_R \approx 1$. From that point on, $\tilde{F}_R$ 
continues to rise more slowly as the shaft velocity increases and so does the component of stress that is proportional to fluid 
deformation. Actually, the increase of force when the shaft starts to move is very sharp but not completely vertical. Of course, 
this could be attributed to regularisation \eqref{eq: Papanastasiou constitutive}, but one cannot help but notice that this force 
increase becomes much more gradual as $\tilde{\beta}$ is increased. The explanation lies in the Navier slip boundary condition 
\eqref{eq: navier slip nd}. Taking a closer look at what happens when the shaft starts to move from a still position, we note 
that initially the material is completely unyielded. Suppose that after a small time the shaft has acquired a small velocity 
$\tilde{u}_{sh}$. According to the Navier slip condition \eqref{eq: navier slip nd} this causes the shaft to impose a stress 
$\tilde{\tau} = \tilde{\beta}^{-1}(\tilde{u}_{sh}-\tilde{u}) \leq \tilde{\beta}^{-1}\tilde{u}_{sh}$ on the viscoplastic 
material. If this is smaller than the yield stress then the material will remain unyielded, and thus motionless. The shaft then 
simply slides over the motionless material without moving it, and the stress that develops in the shaft / material interface is 
due to the friction between them. Since the material is motionless, $\tilde{u} = 0$ and the boundary condition is $\tilde{\tau} 
= \tilde{\beta}^{-1}\tilde{u}_{sh}$. As the shaft accelerates and $\tilde{u}_{sh}$ increases, the stress $\tilde{\tau}$ also 
increases proportionally and eventually it reaches the yield stress $\tilde{\tau}_y = Bn / (Bn+1)$. This is the onset of 
yielding, and occurs at a critical shaft velocity of

\begin{equation} \label{eq: yield velocity}
 \tilde{u}_{sh}^y \;=\; \tilde{\beta} \frac{Bn}{Bn + 1}
\end{equation}
These theoretical results are confirmed by Fig.\ \ref{sfig: F vs V nb per slip}. Indeed, since $Bn/(Bn+1) \approx 1$ for $Bn = 
20$, the material should yield when the shaft velocity has reached $\tilde{u}_{sh}^y \approx \tilde{\beta}$, i.e. 
$\tilde{u}_{sh}^y \approx 0.1$ for $\tilde{\beta} = 0.1$ and $\tilde{u}_{sh}^y \approx 0.4$ for $\tilde{\beta} = 0.4$. This is 
confirmed by Fig.\ \ref{sfig: F vs V nb per slip}. Furthermore, up to the yield point the dimensionless force should be 
proportional to $\tilde{\tau} = \tilde{\beta}^{-1}\tilde{u}_{sh}$, which is again confirmed by the linear variation of force in 
Fig.\ \ref{sfig: F vs V nb per slip}, for velocities of magnitude $|\tilde{u}_{sh}| \leq \tilde{u}_{sh}^y$. For shaft velocities 
larger than $\tilde{u}_{sh}^y$ the material yields so that the slip velocity $\tilde{u}_{sh}-\tilde{u}$ increases more slowly 
than before (as now $\tilde{u} \neq 0$), causing the slope of the force curves to decrease.

\begin{figure}[t]
  \centering
  \subfigure[$\tau_y = 31.4 \mathrm{Pa}$ ($Bn = 20$),
             $\beta = 7.619\!\cdot\!10^{-4}\; \mathrm{m}/\mathrm{Pa}\!\cdot\!\mathrm{s}$ ($\tilde{\beta} = 0.4$),
             $t = 0.7(T/4)$]
             {\label{sfig: flow 70 s0.4} \includegraphics{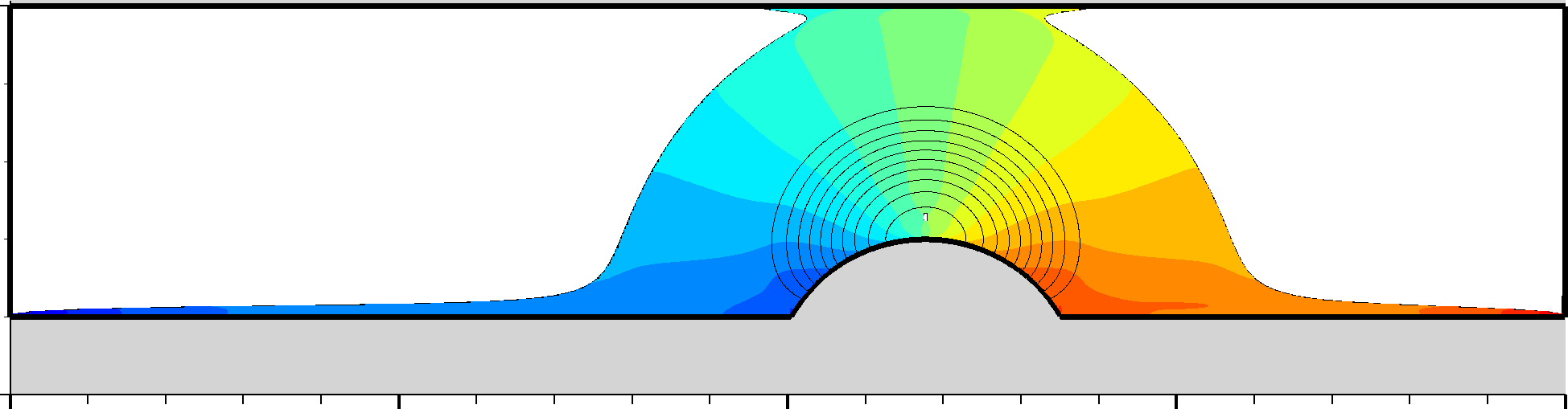}} \\
  \subfigure[Same as \subref{sfig: flow 70 s0.4}, but at $t = 0.8(T/4)$]
             {\label{sfig: flow 80 s0.4} \includegraphics{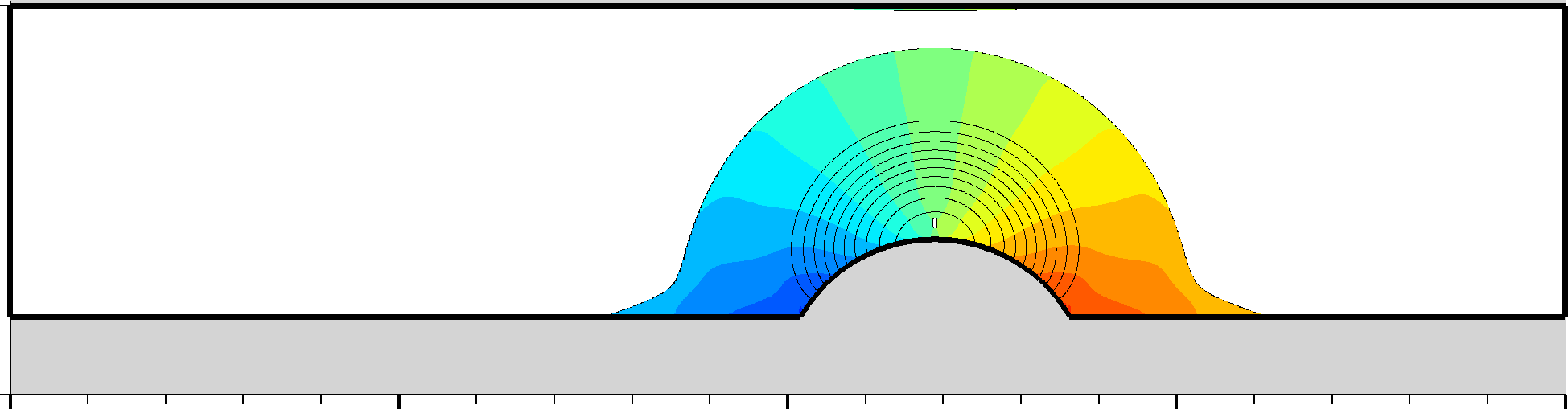}} \\
  \subfigure[$\tau_y = 502.7 \mathrm{Pa}$ ($Bn = 320$),
             $\beta = 4.762\!\cdot\!10^{-5}\; \mathrm{m}/\mathrm{Pa}\!\cdot\!\mathrm{s}$ ($\tilde{\beta} = 0.38$),
             $t = 0.8(T/4)$]
             {\label{sfig: flow 80 Bn320} \includegraphics{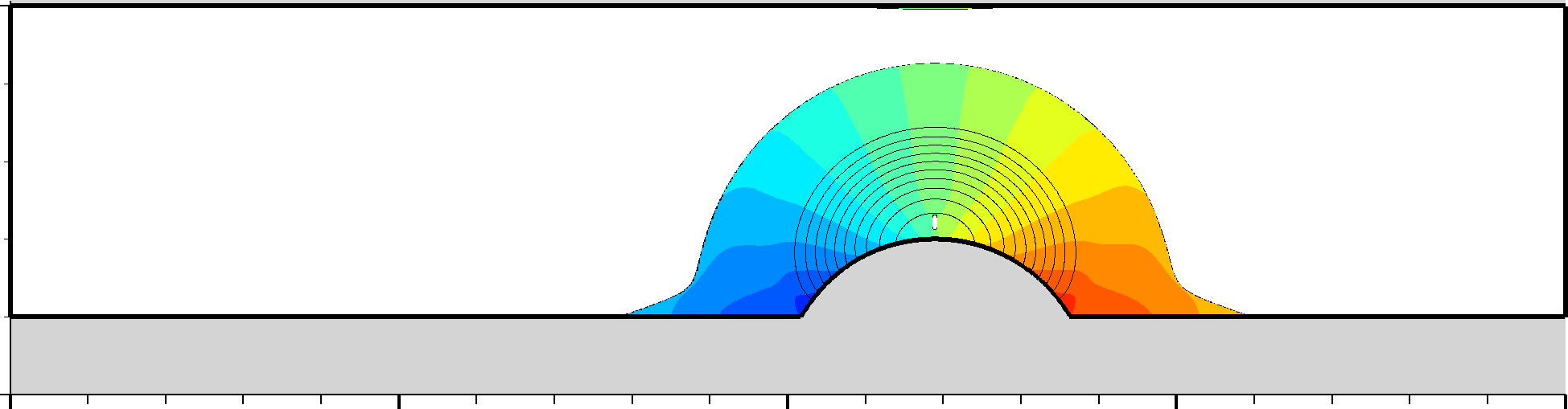}}\\
  \subfigure[Same as \subref{sfig: flow 80 Bn320}, but the dissipation rate is plotted]
             {\label{sfig: dissi 80 Bn320} \includegraphics{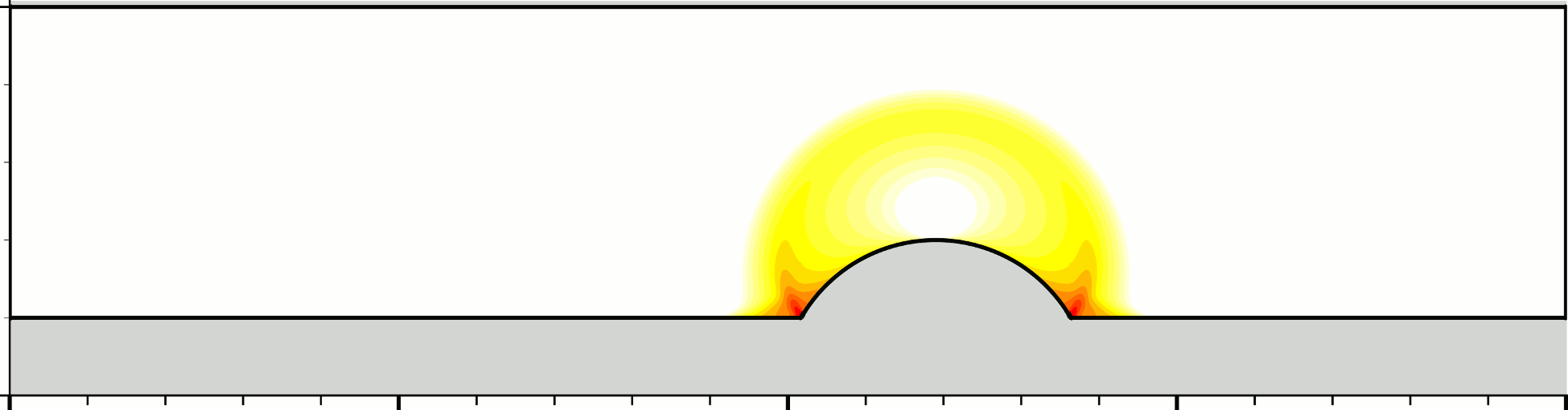}}
  \caption{Flow field snapshots at times when the wall stress is close to the yield stress of the material. The parameters of 
the 
problems are as in Table \ref{table: base case} unless otherwise indicated. Figures \subref{sfig: flow 70 s0.4} and 
\subref{sfig: flow 80 s0.4} correspond to the ``$\tilde{\beta} = 0.4$'' case of Fig.\ \ref{fig: force per slip}, while Figs.\ 
\subref{sfig: flow 80 Bn320} and \subref{sfig: dissi 80 Bn320} correspond to the ``$Bn = 320$'' case of Fig.\ \ref{fig: force 
per Bn}. See captions of Figs.\ \ref{fig: base flow} and \ref{fig: dissipation per Bn} for visualisation details.}
  \label{fig: flow with slip below yield}
\end{figure}

The existence of the bulge has the consequence that even the slightest shaft motion changes the domain shape and thus causes 
deformation and yielding of the material. Therefore, the only instance when the material may be completely unyielded is when the 
shaft velocity is zero. To see what happens, Figs.\ \ref{sfig: flow 70 s0.4} and \ref{sfig: flow 80 s0.4} show snapshots of the 
flow field as the shaft is decelerating, at two time instances, when its velocity is just above $\tilde{u}_{sh}^y$ (Fig.\ 
\ref{sfig: flow 70 s0.4}) and when it is just below $\tilde{u}_{sh}^y$ (Fig.\ \ref{sfig: flow 80 s0.4}). In Fig.\ \ref{sfig: 
flow 70 s0.4} the stress at the shaft / material interface is everywhere above the yield stress and the shaft is everywhere 
surrounded by a layer of yielded material. In Fig.\ \ref{sfig: flow 80 s0.4} the stress at the shaft / material interface is 
mostly below the yield stress so that the shaft is in direct contact with, and sliding on, unyielded material over most of its 
length; however, the bulge is surrounded by a bubble of yielded material. The consequences of this on the force can be seen by 
comparing the $\tilde{\beta} = 0.4$ curves of Figs.\ \ref{sfig: F vs V nb per slip} and \ref{sfig: Fvisc vs V per slip}; when 
there is a bulge, at the onset of shaft motion there is an immediate albeit relatively small increase in the force, contrary to 
the no-bulge case, due to yielding of the material surrounding the bulge. Thus, slip may obscure the viscoplastic nature of the 
material by causing apparent flow that hides the existence of a yield stress, but it cannot do so completely if the shaft has a 
bulge. Note that this phenomenon will occur for any finite value of the slip coefficient; it occurs also for $\tilde{\beta}$ = 
0.00625 and 0.025 in Fig.\ \ref{fig: force per slip} only that it is difficult to discern because the corresponding values of 
$\tilde{u}_{sh}^y$ are very small. Slip is known to introduce such increased complexity to viscoplastic flows, see e.g. 
\cite{Damianou_2014, Damianou_2014b} for other examples. As far as the pressure force is concerned, Figure \ref{sfig: Fpres vs V 
per slip} shows that it is relatively independent of the slip coefficient.

These phenomena become more pronounced not only when the slip coefficient is increased, but also when the yield stress is 
increased; in the latter case a higher shaft velocity is required for yielding to occur. This is reflected on the dimensionless 
slip coefficient, Eq.\ \eqref{eq: slip coefficient nd}, which depends not only on $\beta$ but also on $Bn$. Thus the slip 
phenomena are more pronounced for $Bn = 320$ in Fig.\ \ref{fig: force per Bn} than for lower $Bn$ numbers. In fact the $Bn = 320$ 
case of Fig.\ \ref{fig: force per Bn} has $\tilde{\beta} = 0.38$ which is very close to the $\tilde{\beta} = 0.4$ case of Fig.\ 
\ref{fig: force per slip}, and so they have very similar yield shaft velocities $\tilde{u}_{sh}^y$. Figure \ref{sfig: flow 80 
Bn320} shows a snapshot of the $Bn = 320$ case at the same time instance as for Fig.\ \ref{sfig: flow 80 s0.4}; the two flow 
fields can be seen to be very similar. Also, the dissipation function is plotted in Fig.\ \ref{sfig: dissi 80 Bn320} and, as 
expected, it can be seen to be non-zero only within the yielded ``bubble'' surrounding the bulge.

\begin{figure}[!p]
  \centering
  \includegraphics[scale=1.00]{figures/dissi_legend.png}\\[0.15cm]
  \subfigure[base case, $t = T$] {\label{sfig: dissipation 400 base} 
      \includegraphics[scale=1.0]{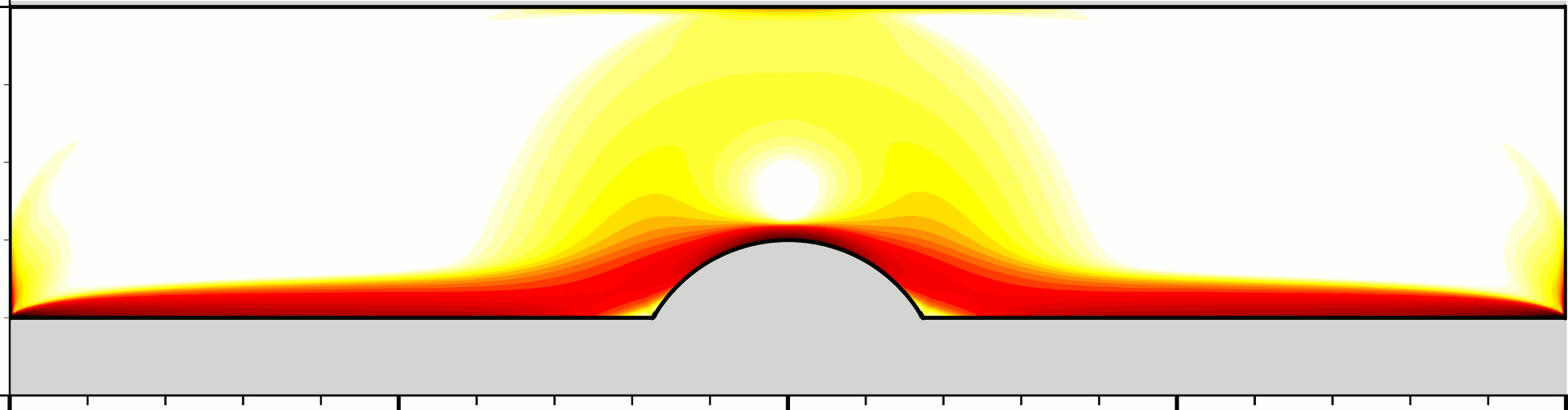}}\\
  \subfigure[$\tilde{\beta}$ = 0.4, $t = T$] {\label{sfig: dissipation 400 s0.4} 
      \includegraphics[scale=1.0]{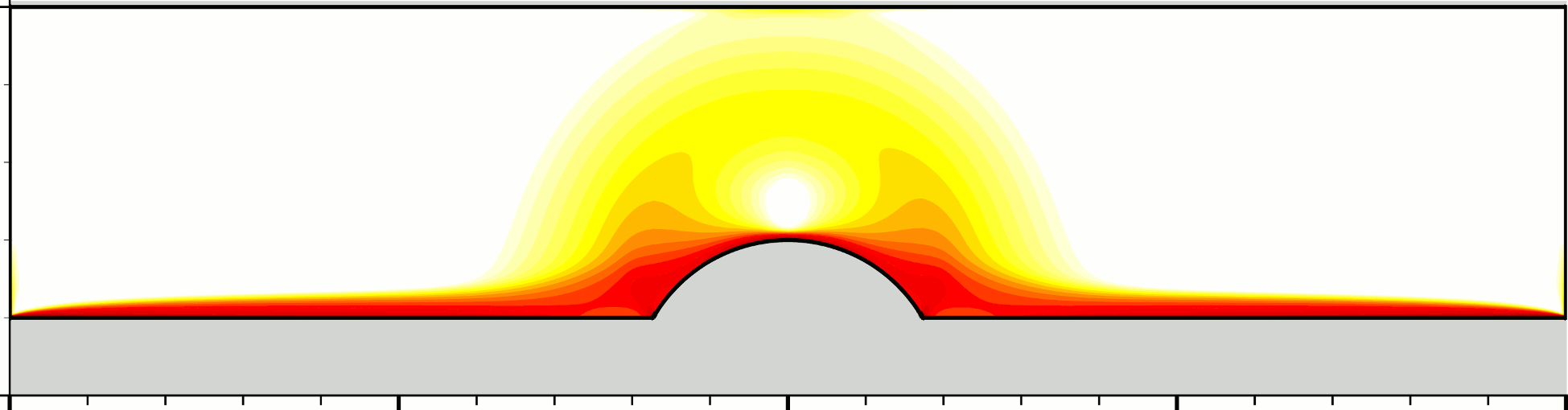}}\\
  \subfigure[$R_o$ = 25.6 mm, $t = T$] {\label{sfig: dissipation 400 Ro256} 
      \includegraphics[scale=1.0]{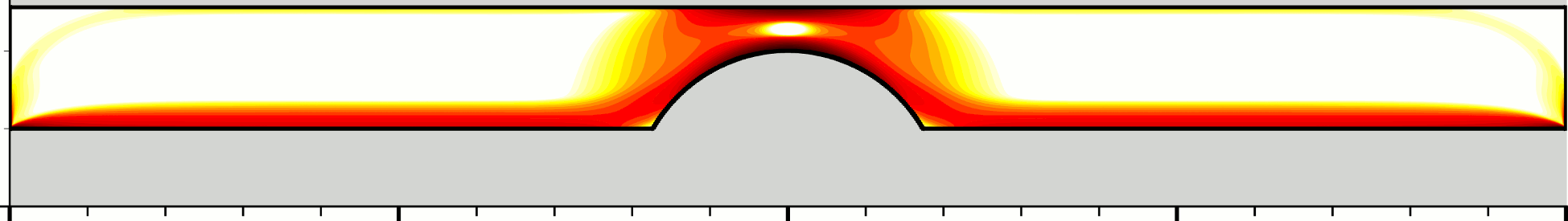}}\\
  \subfigure[$R_b$ = 30 mm, $t = T$] {\label{sfig: dissipation 400 Rb30} 
      \includegraphics[scale=1.0]{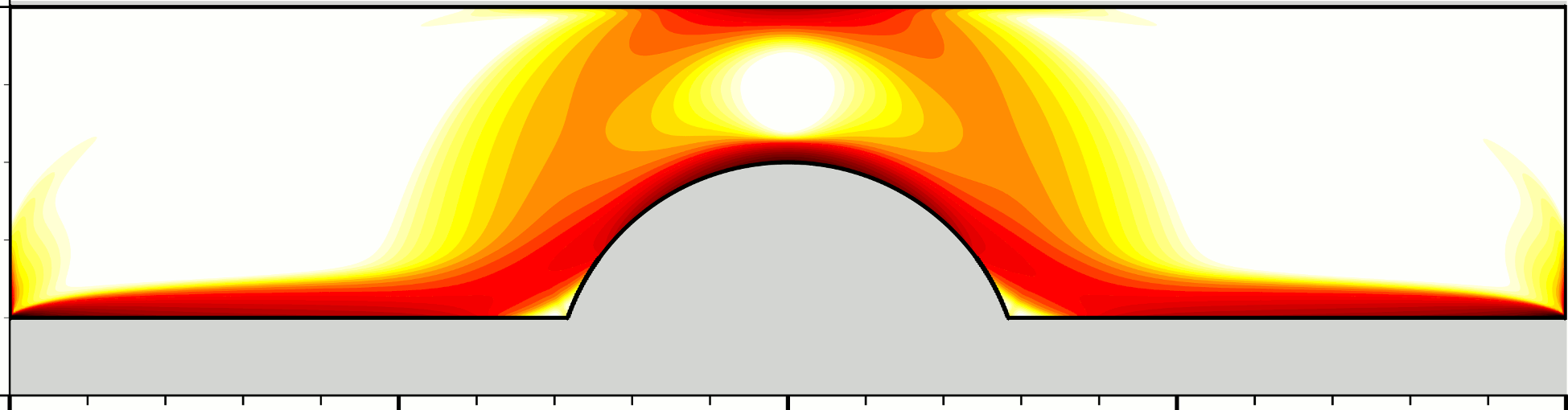}}\\
  \subfigure[$f$ = 8 Hz, $t = 2T$ ] {\label{sfig: dissipation 1600 f8} 
      \includegraphics[scale=1.0]{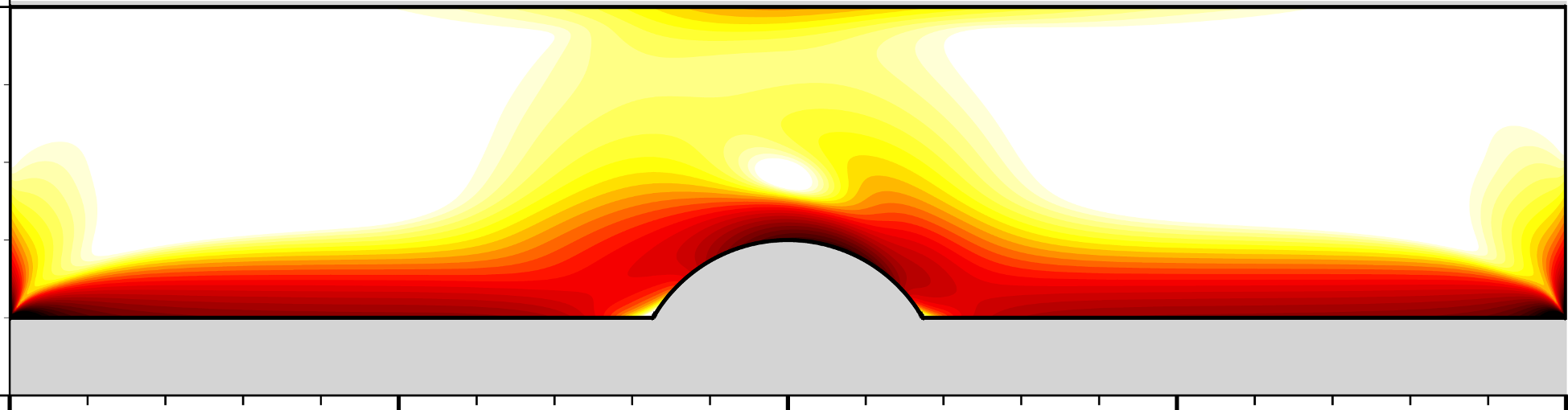}}\\
  \subfigure[$f$ = 8 Hz, $t = 1.75T$ ] {\label{sfig: dissipation 1400 f8} 
      \includegraphics[scale=1.0]{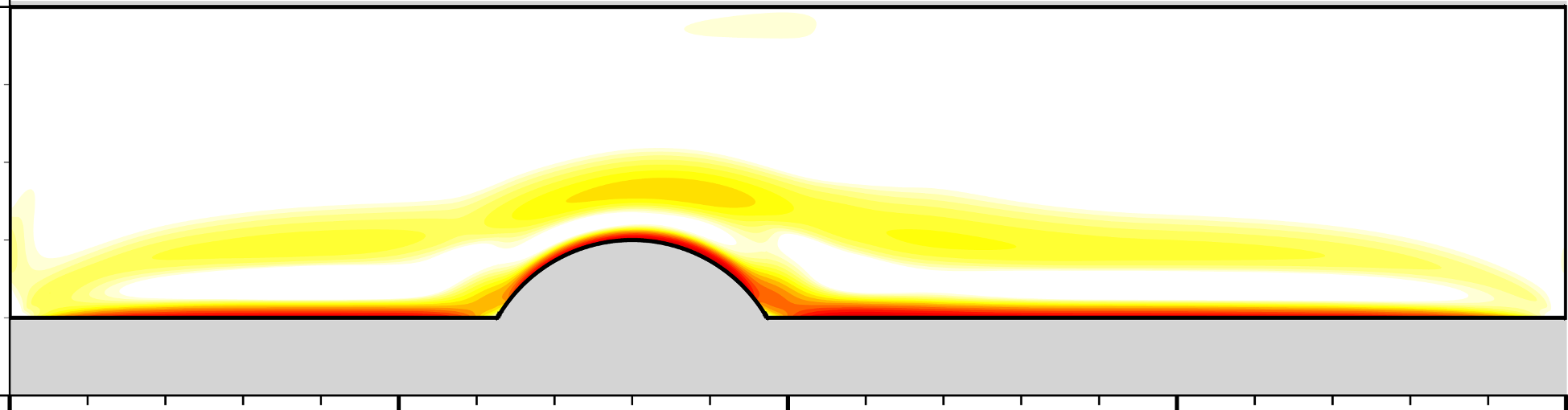}}
  \caption{Plots of the dimensionless dissipation function $\tilde{\phi} = \phi/\phi_{\mathrm{ref}}$ (Eq.\ \eqref{eq: phi ref}) 
for various cases. The dimensional parameters have the values listed in Table \ref{table: base case}, unless otherwise indicated 
in each figure caption.}
  \label{fig: dissipation various}
\end{figure}

Other differences in the dissipation function distribution that are due to slip can be seen by comparing Figs.\ \ref{sfig: 
dissipation 400 base} and \ref{sfig: dissipation 400 s0.4}. Increasing slip can be seen to reduce energy dissipation in the bulk 
of the material, especially at the shaft ends and at the tip of the bulge, by relaxing the large velocity gradients there. The 
weakening of the flow also makes the effect of the outer cylinder weaker, with the ring of rotating material not extending up to 
the outer cylinder in Fig. \ref{sfig: dissipation 400 s0.4}. Finally, one can notice in Fig.\ \ref{sfig: dissipation 400 base} 
(and in other low slip cases) that there is some material trapped in the corners between the bulge and the shaft; but in Fig.\ 
\ref{sfig: dissipation 400 s0.4} (and also in Fig.\ \ref{sfig: dissipation 200 Bn320}, where slip is again large) there is no 
such entrapment. This is remniscent of the unyielded cups which are observed at the poles of a sphere falling through a 
viscoplastic material \cite{Beris_1985}; actually, Fig.\ \ref{fig: base flow} shows that the material at the bulge corners is 
yielded, but the low rates of deformation suggested there by Figs.\ \ref{fig: dissipation per Bn} and \ref{fig: dissipation 
various} indicate that the material is close to the unyielded state. It would not be unreasonable to suspect that increasing the 
grid resolution locally might reveal small amounts of uyielded material at the corners.

\begin{figure}[t]
  \centering
  \subfigure[$\tilde{\beta} = 0.00625$] {\label{sfig: work s0.00625}
      \includegraphics[scale=0.7]{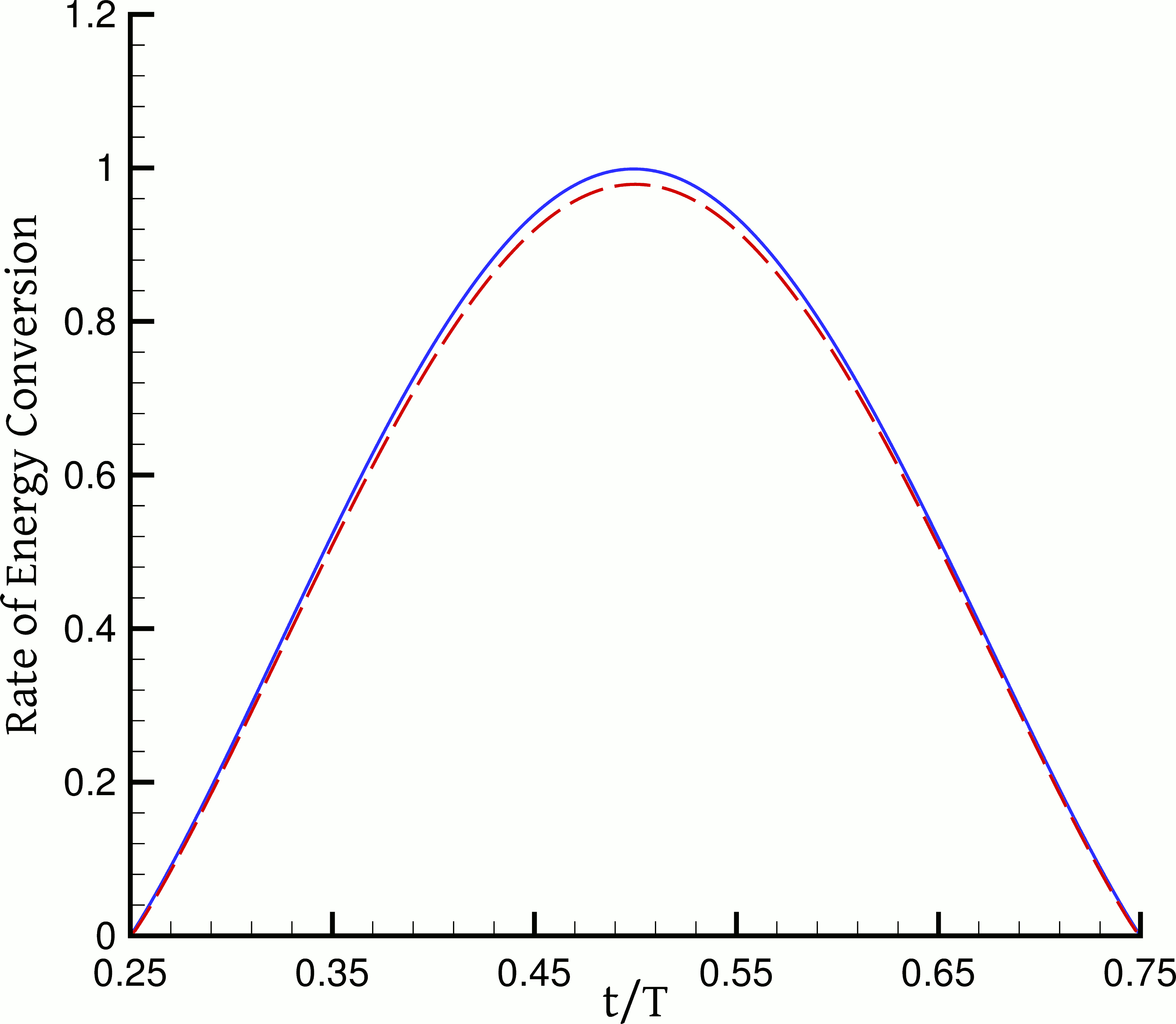}}
  \subfigure[$\tilde{\beta} = 0.025$] {\label{sfig: work Bn20}
      \includegraphics[scale=0.7]{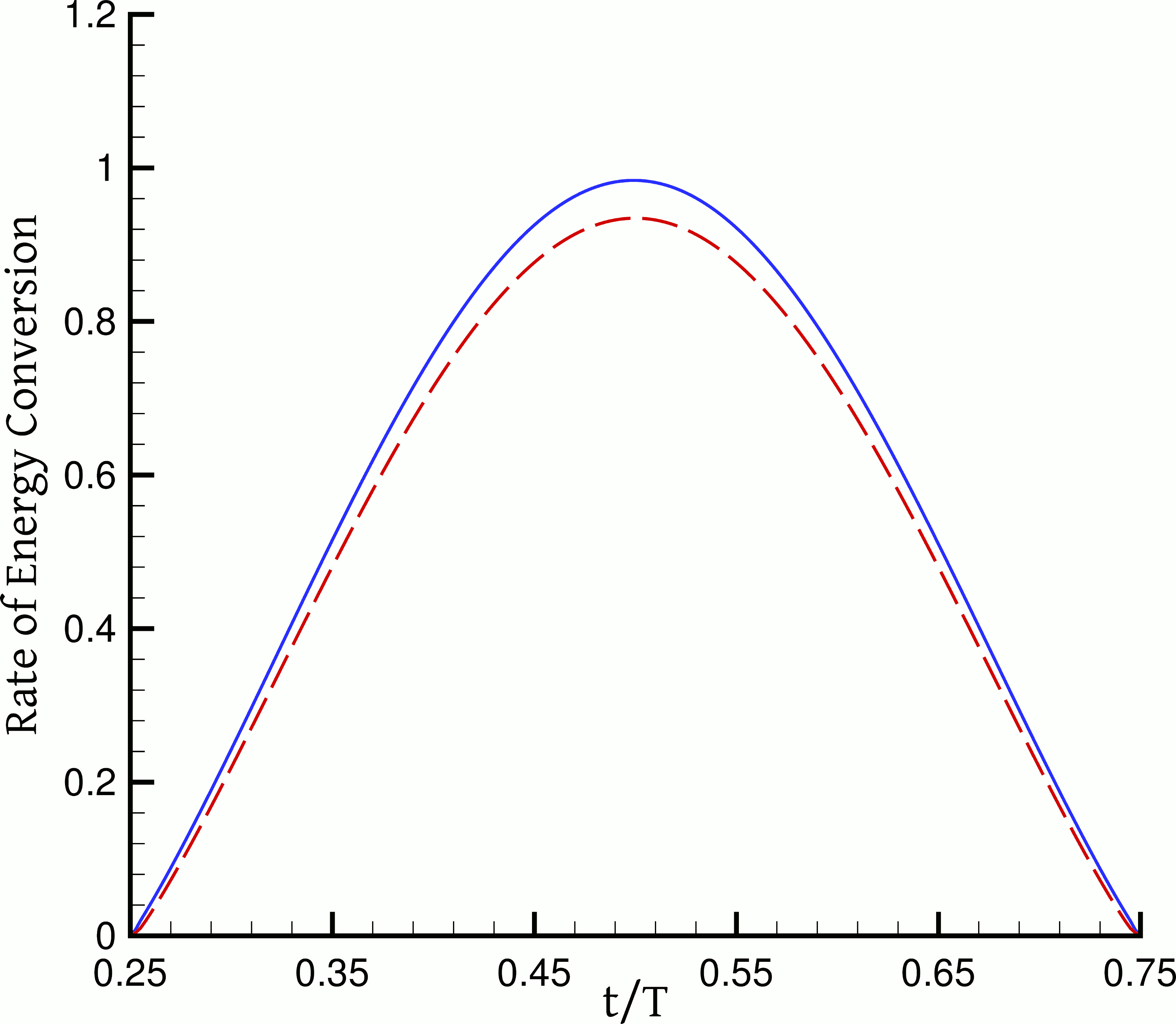}}
  \subfigure[$\tilde{\beta} = 0.4$] {\label{sfig: work s0.4}
      \includegraphics[scale=0.7]{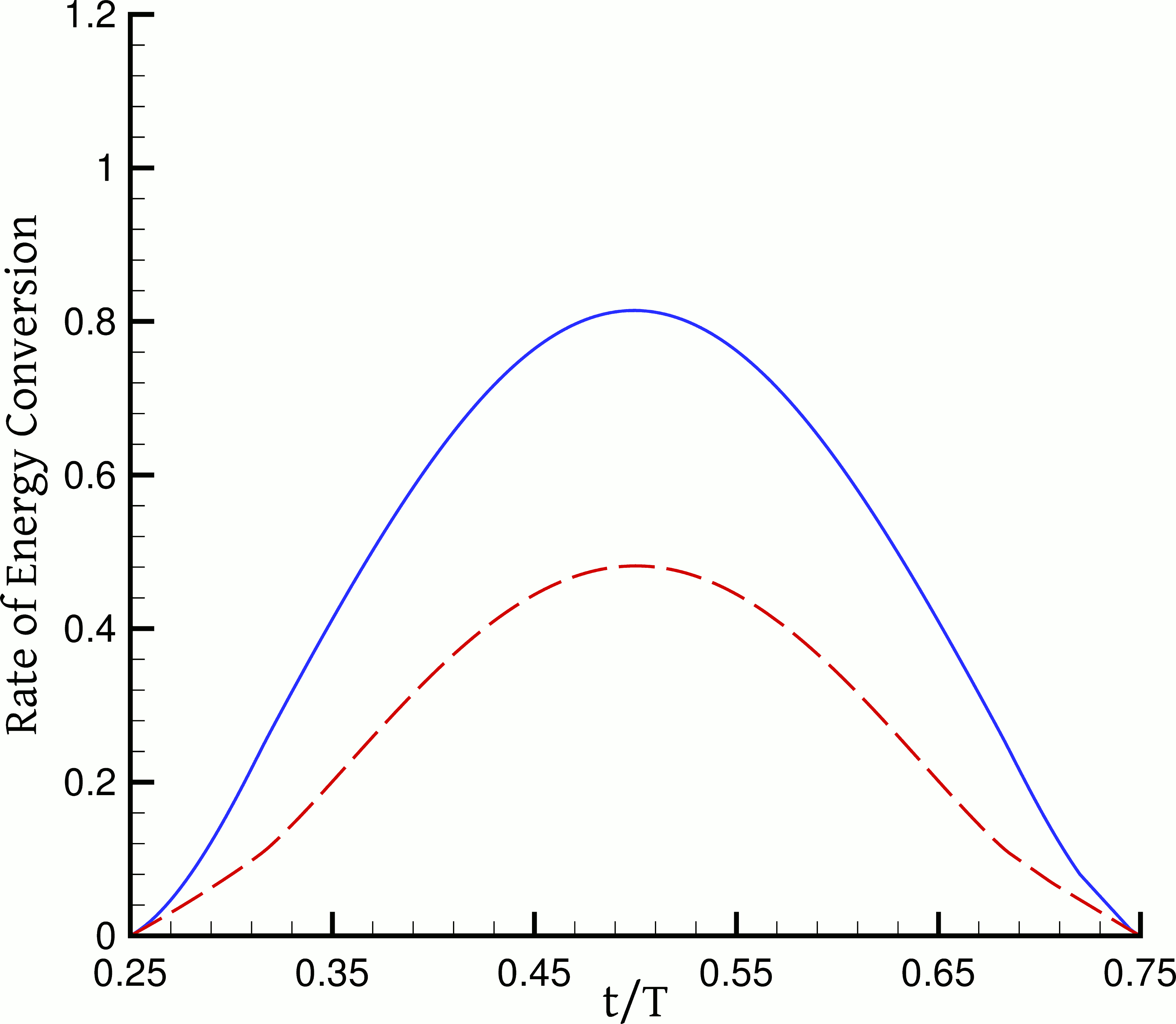}}
  \subfigure[$\tilde{\beta} = 0.00625$, no bulge] {\label{sfig: work s0.00625 nb}
      \includegraphics[scale=0.7]{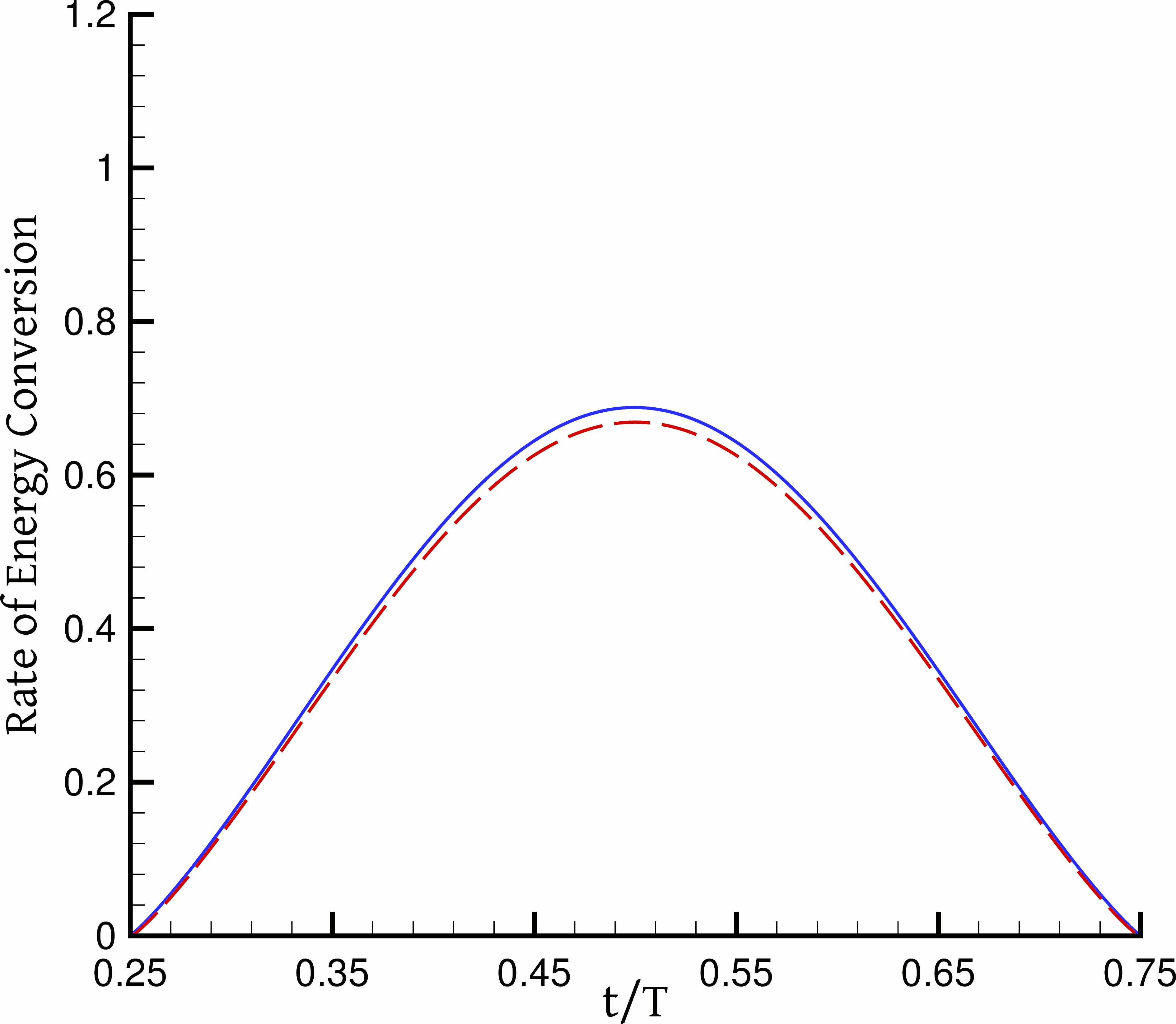}}
  \subfigure[$\tilde{\beta} = 0.025$, no bulge] {\label{sfig: work Bn20 nb}
      \includegraphics[scale=0.7]{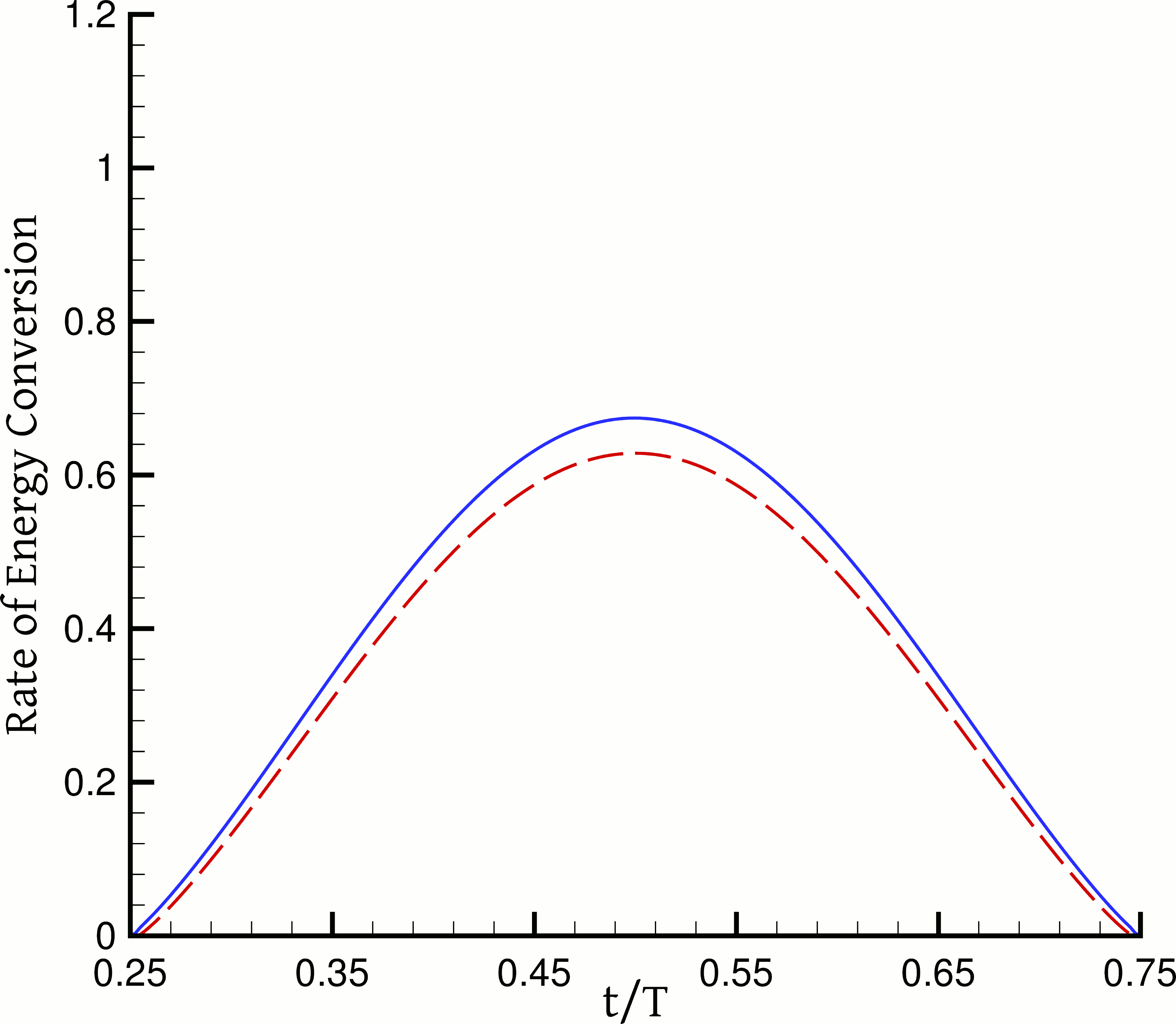}}
  \subfigure[$\tilde{\beta} = 0.4$, no bulge] {\label{sfig: work s0.4 nb}
      \includegraphics[scale=0.7]{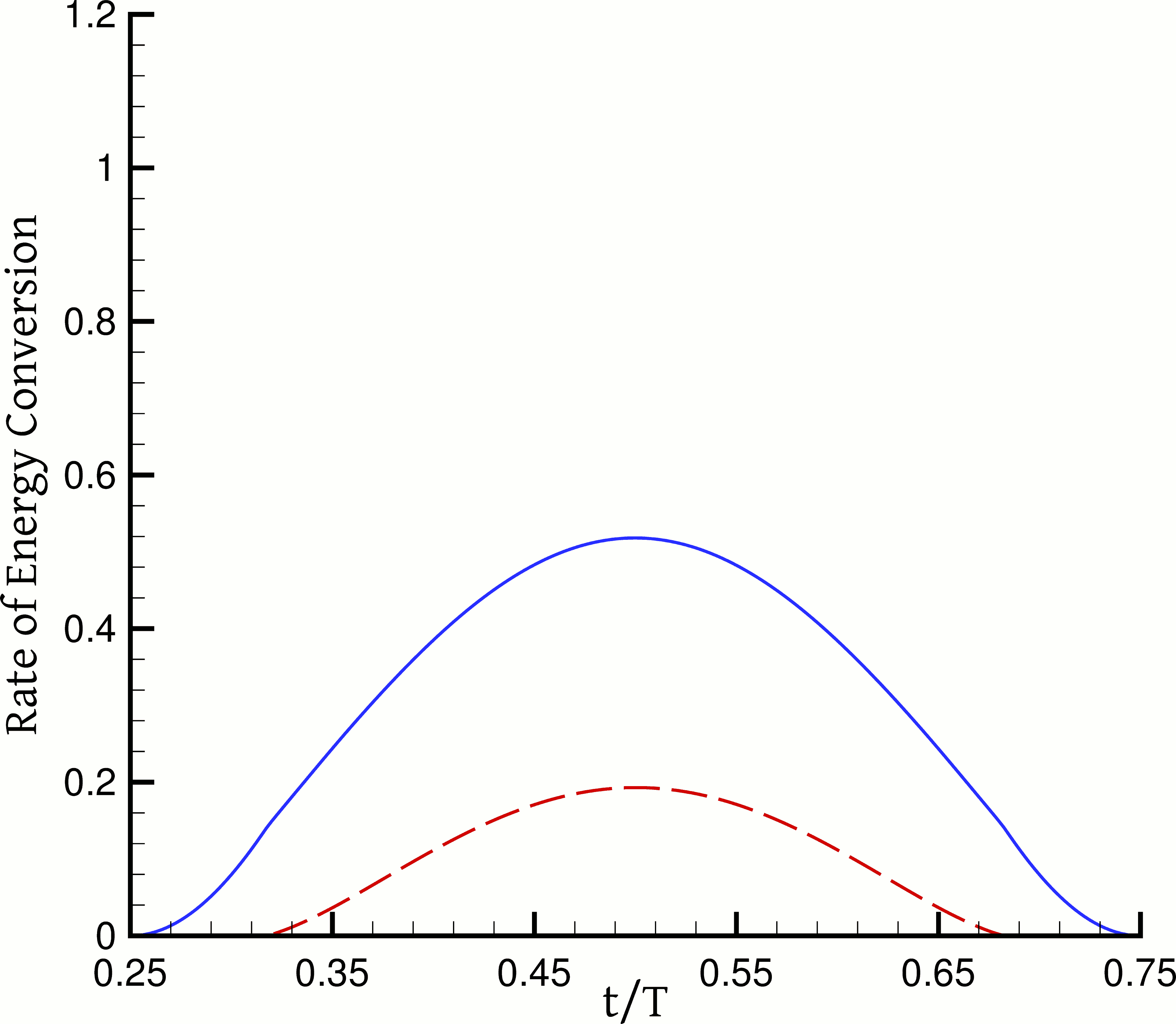}}
  \caption{Rate of energy dissipation for various values of the dimensionless slip coefficient $\tilde{\beta}$, with (top) and 
without (bottom) a bulge. The continuous blue line is the instantaneous total power consumption of the damper, calculated as 
$-F_R u_{sh}$. The dashed red line is the rate of dissipation of mechanical energy into heat due to fluid deformation, i.e.\ the 
integral of the dissipation function (\ref{eq: dissipation function}) over the computational domain. Both the power consumption 
and the dissipation rate are normalised by $\phi_{\mathrm{ref}} \Omega_{\mathrm{tot}}$ where $\Omega_{\mathrm{tot}}$ is the 
total 
volume occupied by the fluid and $\phi_{\mathrm{ref}}$ is defined by Eq.\ \eqref{eq: phi ref}. The flow parameters are as listed 
in Table \ref{table: base case}, except for the slip coefficient which is varied to obtain the values of $\tilde{\beta}$ shown.}
  \label{fig: work per slip}
\end{figure}

Figures \ref{sfig: dissi 80 Bn320} and \ref{sfig: dissipation 400 s0.4} do not show the whole picture as far as energy 
dissipation is concerned. The dissipation function only accounts for the mechanical energy that is converted into heat due to 
fluid deformation. But, whenever there is slip, mechanical energy is also converted into heat by the sliding friction between the 
shaft and the material. Figure \ref{fig: work per slip} shows two curves on each plot: the rate of work done by the shaft and the 
rate of energy dissipation within the material. The area between the two curves is the energy converted to heat due to sliding 
friction. When the slip coefficient is small, almost all of the energy is dissipated within the bulk of the material due to fluid 
deformation, and the sliding friction plays a very minor role. On the contrary, when the slip coefficient is large, sliding 
friction plays a crucial role, converting mechanical energy into heat directly on the fluid / shaft interface, whereas energy 
dissipation in the bulk of the material is weak. In Fig.\ \ref{sfig: work s0.4 nb}, which corresponds to a bulgeless shaft, the 
energy dissipation in the bulk of the material (red line) is zero over the time intervals during which the material is completely 
unyielded ($u_{sh} < u_{sh}^y$). On the contrary, in Fig.\ \ref{sfig: work s0.4} (bulged shaft) this is never zero except when 
the shaft is stopped, as otherwise the bulge always causes some yielding, as discussed previously. We note, finally, that Figs.\ 
\ref{sfig: work s0.00625} and \ref{sfig: work s0.00625 nb} provide further evidence that the flow is in quasi steady state, since 
all the instantaneous shaft work is dissipated, eventually by viscous forces. The work done in accelerating the fluid, i.e.\ 
increasing its kinetic energy, is negligible. This shows that the inertia of the system is negligible as well. Cases with 
increased significance of inertia will be examined later.

\subsection{Effect of the damper geometry}
\label{ssec: results geometry}

\begin{figure}[!p]
  \centering
  \subfigure[] {\label{sfig: Ftot vs disp per Ro}
      \includegraphics[scale=0.77]{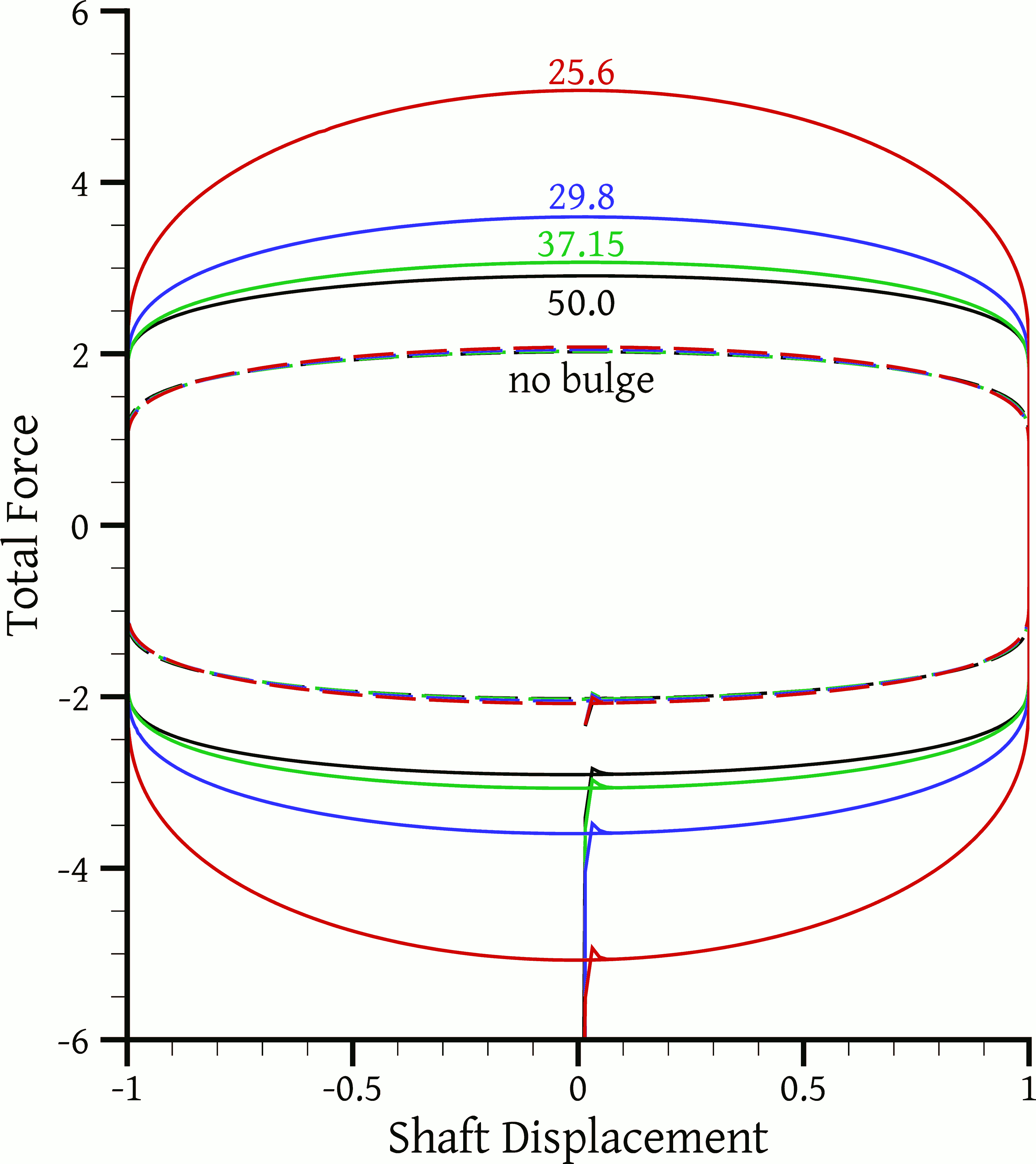}}
  \subfigure[] {\label{sfig: Fvisc vs disp per Ro}
      \includegraphics[scale=0.77]{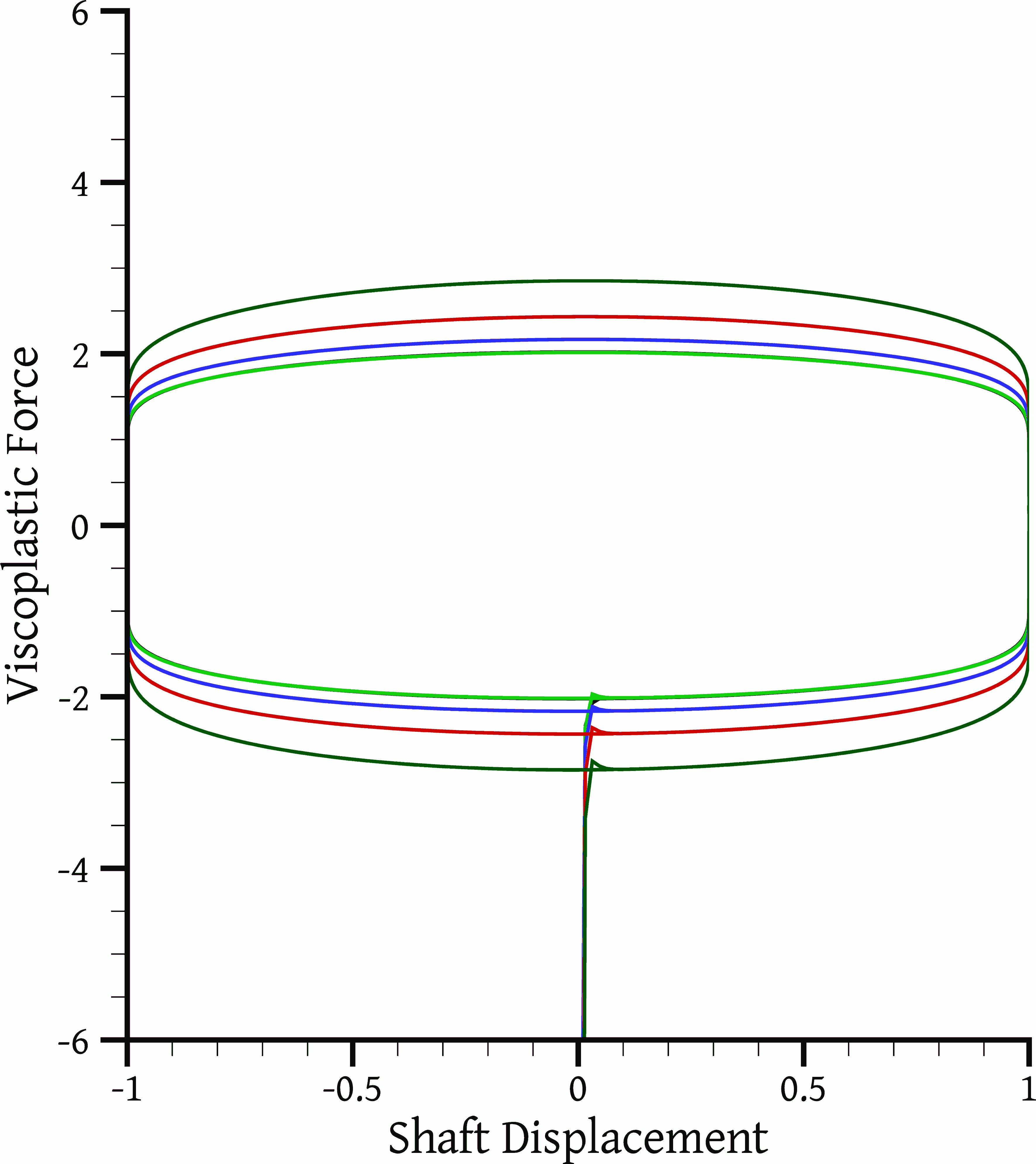}}
  \subfigure[] {\label{sfig: Fpres vs veloc per Ro}
      \includegraphics[scale=0.77]{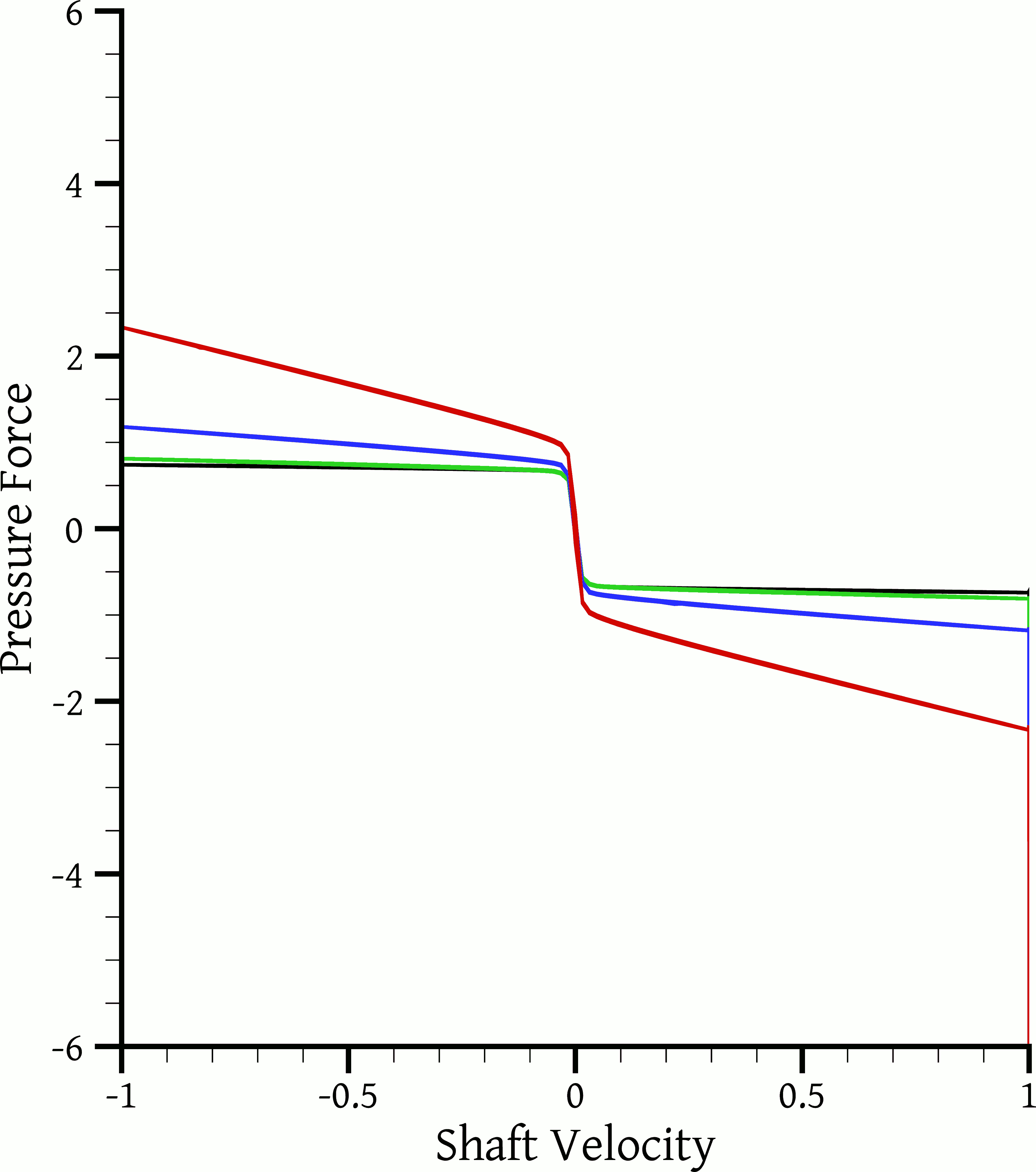}}
  \caption{The reaction force and its components as a function of shaft displacement or velocity, for various bore radii, $R_o$ 
= 50, 37.15, 29.8 and 25.6 mm. The rest of the parameters are as listed in Table \ref{table: base case}. In \subref{sfig: Ftot 
vs disp per Ro} and \subref{sfig: Fvisc vs disp per Ro} the dashed lines depict results for bulgeless shafts.}
  \label{fig: force per Ro}
\end{figure}

The effect of the bore radius $R_o$ on the reaction force is illustrated in Fig.\ \ref{fig: force per Ro}. The radii selected are 
50 (base case), 37.15, 29.8 and 25.6 mm, while the rest of the dimensional parameters have the values displayed in Table 
\ref{table: base case}. The selected values of $R_o$ are such that the length of the gap $R_o - R_b$ decreases by a constant ratio 
of 1.75. Since lengths are dedimensionalised by $H = R_o - R_i$, changing $R_o$ affects all the dimensionless parameters. In 
particular, compared to the base case of $R_o$ = 50 mm, in the $R_o$ = 25.6 mm case $Bn$ has fallen from 20 to 7.8, $Re^*$ has 
fallen from 0.12 to 0.11 (but $Re$ has fallen from 2.51 to 0.98), $Sr$ has increased from 3.14 to 8.05, and $\tilde{\beta}$ has 
increased slightly from 0.025 to 0.027. Therefore, judging from these numbers, reducing the bore radius in the present 
configuration should in general reduce the viscoplasticity of the flow and also mildly reduce its inertial character.

\begin{figure}[!p]
  \centering
  \subfigure[$R_o$ = 25.6 mm, no bulge] {\label{sfig: flow field Ro256 nb} 
            \includegraphics[scale=1.25]{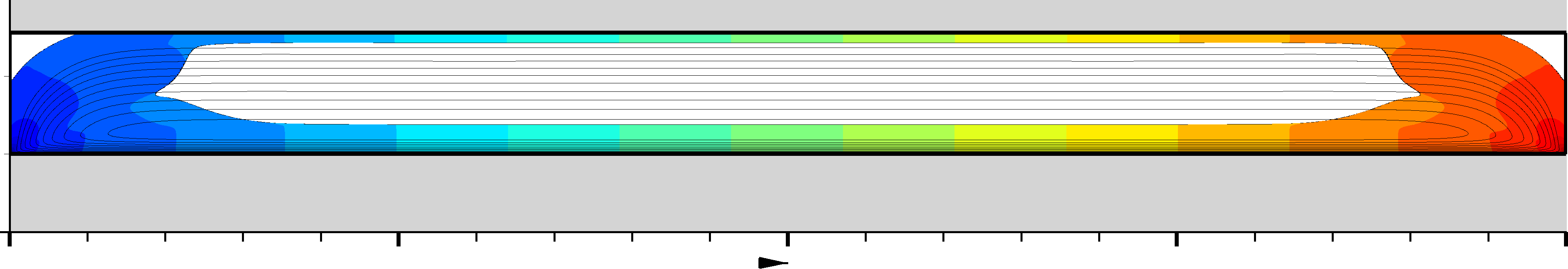}}
  \subfigure[$R_o$ = 25.6 mm] {\label{sfig: flow field Ro256}
            \includegraphics[scale=1.25]{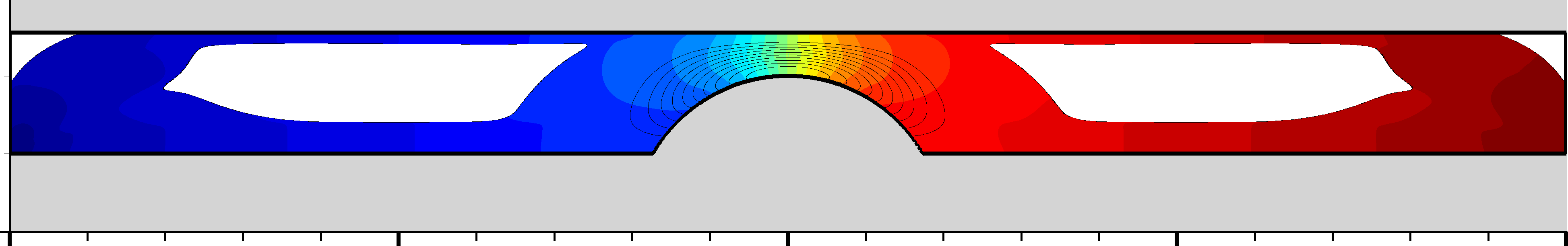}}
  \subfigure[Close-up of \subref{sfig: flow field Ro256}] {\label{sfig: flow field Ro256 zoom} 
            \includegraphics[scale=1.25]{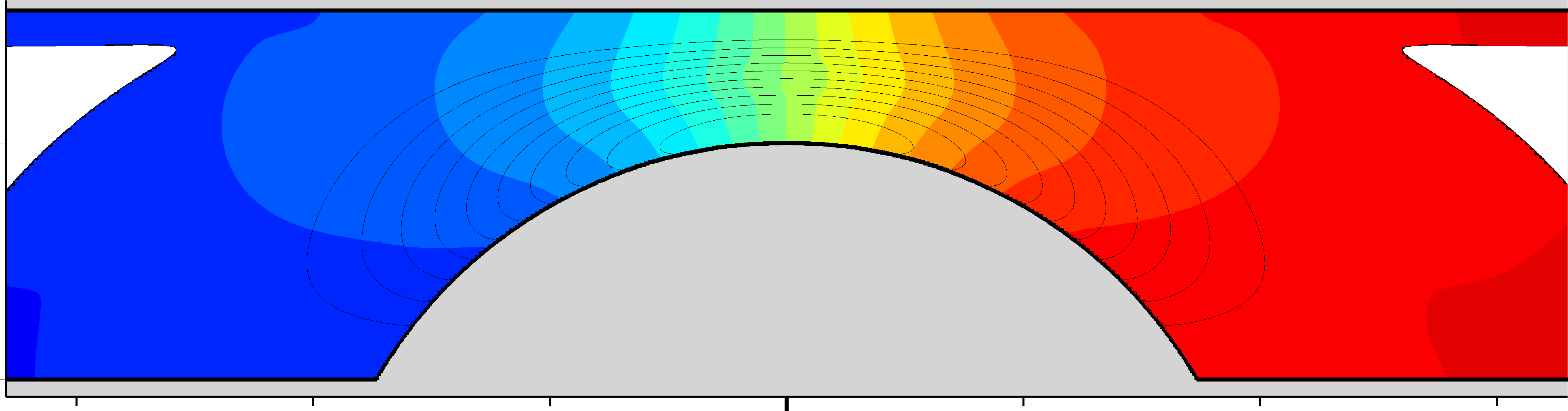}}
  \subfigure[Close-up for the base case ($R_o$ = 50 mm)] {\label{sfig: flow field Ro500 zoom} 
            \includegraphics[scale=1.25]{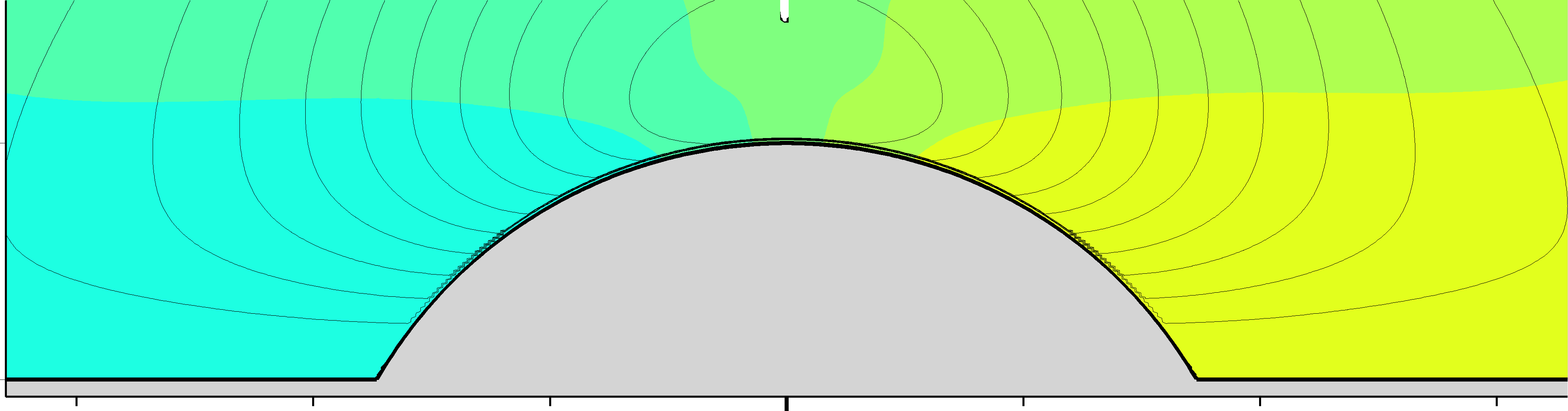}}
  \caption{Comparison between the flow fields for $R_o$ = 25.6 mm and 50 mm (the base case), at $t = T$. The rest of the 
parameters are as listed in Table \ref{table: base case}. The flow is visualised as described in the caption of Fig.\ \ref{fig: 
base flow}, only that the dimensionless pressure contours have a step of $2.4$.}
  \label{fig: flow 400 Ro256}
\end{figure}

We first examine the case where the shaft has no bulge. Figure \ref{fig: force per Ro} includes results for bulgeless shafts, 
drawn in dashed lines, for all the selected $R_o$ values. This is hard to see in the Figure though, because all the dashed curves 
nearly coincide. Therefore, for the range of values considered, the normalised force $\tilde{F}_R$ is nearly independent of $R_o$ 
in the absence of a bulge; the actual force $F_R$ increases slightly because $\tilde{F}_R$ is normalised by $F_{\mathrm{ref}}$ 
which is proportional to $\tau_{\mathrm{ref}} = \tau_y + \mu U/H$, which increases from about 33 Pa at $R_o$ = 50 mm to about 35.5 
Pa at $R_o$ = 25.6 mm. However, if $Bn$ is an appropriate indicator of the viscoplasticity of the flow, one would expect a greater 
difference between the force curve for $R_o$ = 50 mm ($Bn$ = 20) and that for $R_o$ = 25.6 mm ($Bn$ = 7.8) -- compare for example 
the curves for $Bn = 20$ and $Bn = 5$ in Fig.\ \ref{sfig: Fvisc vs dx per Bn}. As it is easily deduced from Figs.\ \ref{sfig: flow 
field Ro256 nb} and \ref{sfig: base flow 200 nb}, despite the Bingham number being lower, a larger percentage of the material is 
unyielded when $R_o$ = 25.6 mm than when $R_o$ = 50 mm. This can be attributed to the geometrical confinement of the former case, 
which forces the streamlines to be straight over a longer distance, thus reducing the deformation rates and favouring the 
unyielded state.

In order to obtain more insight, we find it useful to discuss a one-dimensional flow that shares some similarity with the present 
flow, that of annular Couette flow where the inner cylinder moves with a constant velocity and the outer one is stationary. This 
flow is described in the Appendix, where it is shown that the outer radius $R_o$ is important only if the flow is completely 
yielded, which occurs if $R_o$ does not exceed a critical value $R_y$ (given by Eq.\ \eqref{eq: ACF yield line 1} in the 
Appendix), that depends on the dimensionless number 

\begin{equation} \label{eq: B}
B = \frac{\tau_y R_i}{\mu U}
\end{equation}
(an alternative definition of the Bingham number, depending only on $R_i$ and not on $R_o$). If $R_o$ exceeds $R_y$ then the 
material from $R_i$ to $R_y$ is yielded with its velocity independent of $R_o$, and from $R_y$ to $R_o$ it is unyielded with zero 
velocity. Thus in this case it would be misleading to use the Bingham number $Bn$ as an indicator about the flow; the alternative 
Bingham number $B$ conveys all the relevant information (Eqs. \eqref{eq: ACF yield line 1}, \eqref{eq: ACF u partly yielded}).

\begin{figure}[!b]
  \centering
  \subfigure[Profiles for various $R_o$ (mm)] {\label{sfig: u profiles per Ro}
      \includegraphics[scale=1.1]{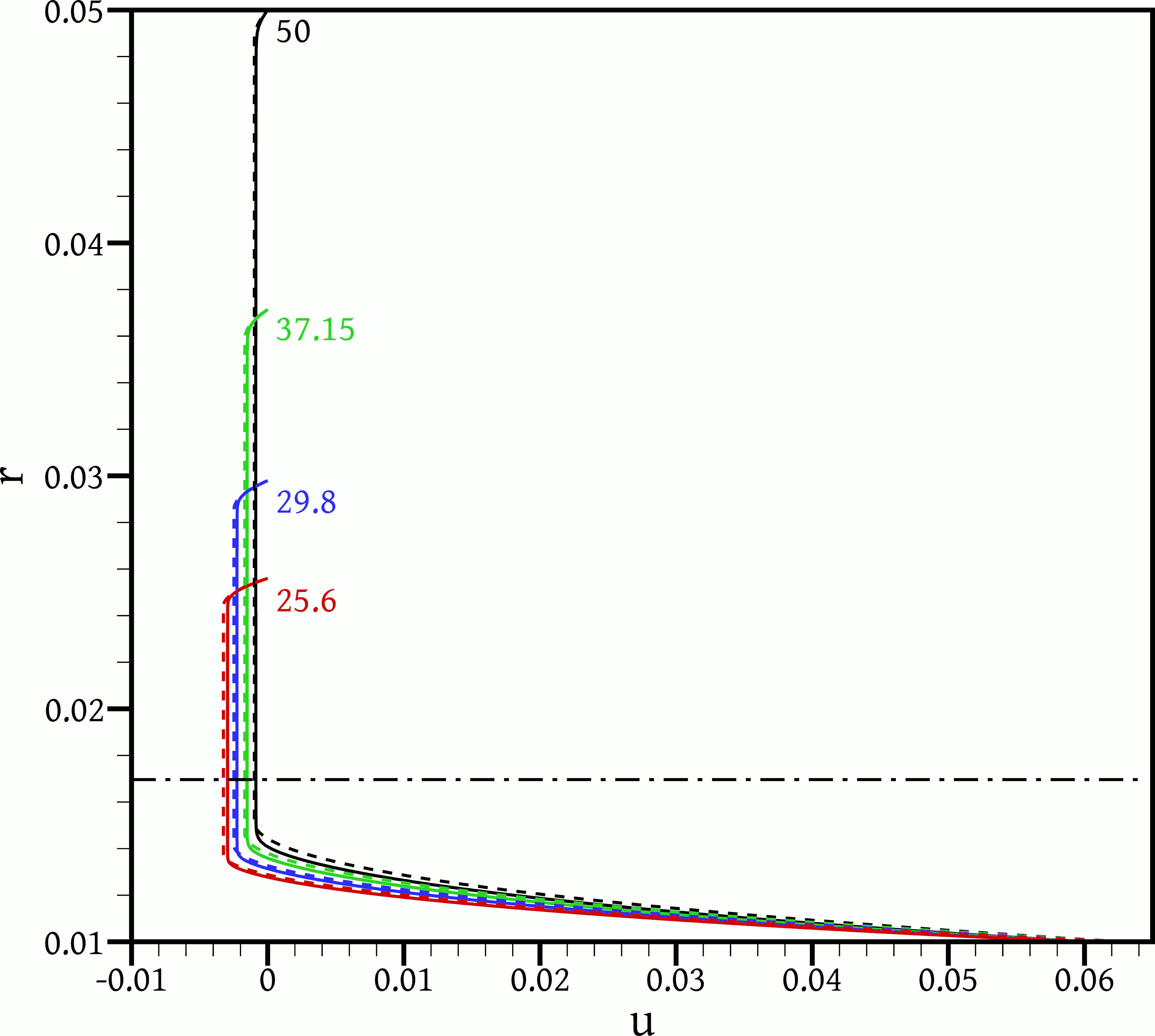}}
  \subfigure[Profiles for various $Bn$] {\label{sfig: u profiles per Bn}
      \includegraphics[scale=1.1]{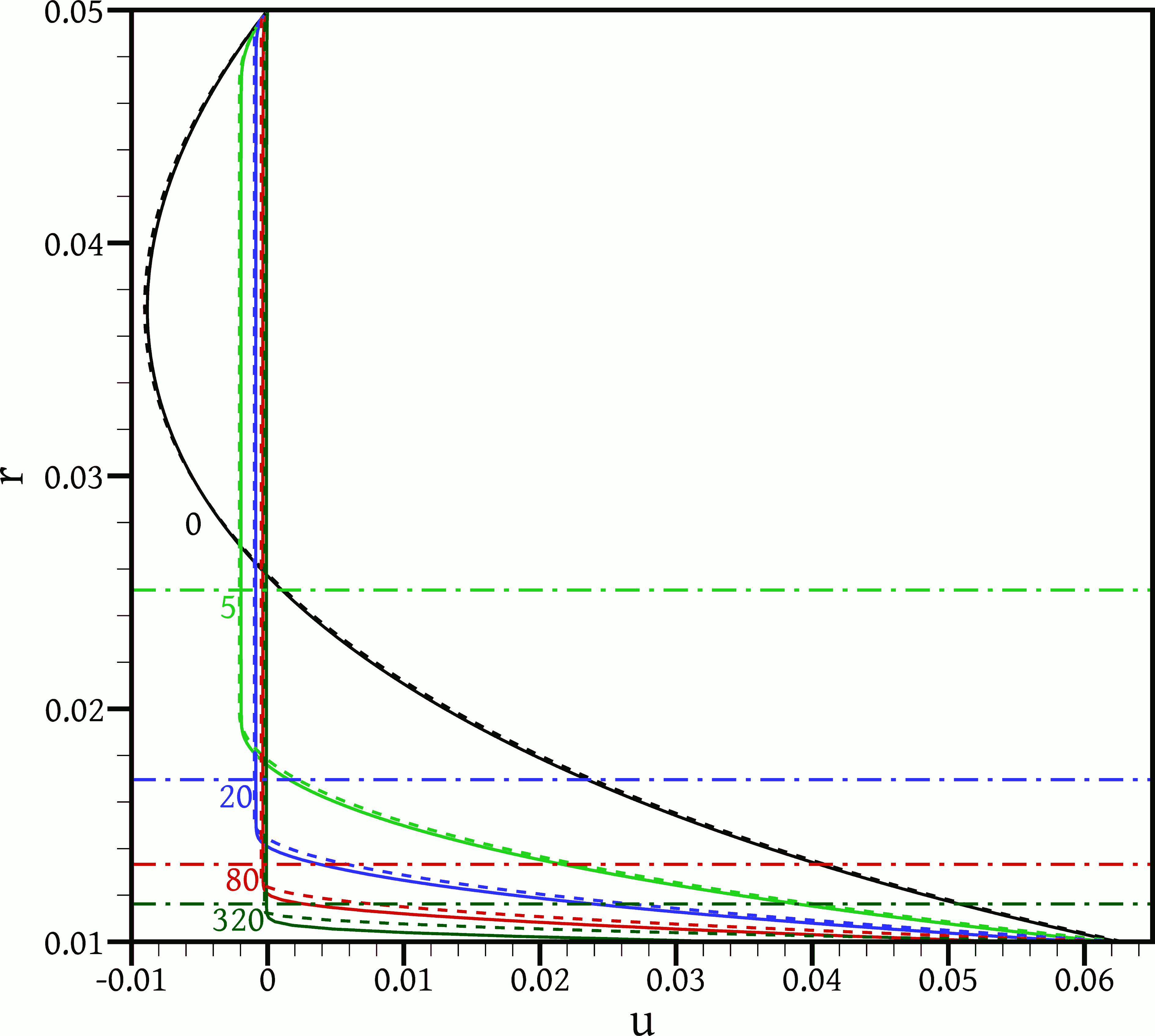}}
  \caption{Solid lines depict profiles of the axial velocity $u$ along the radial direction, at $x$ = $L/2$ and $t = T$ ($t = 
2T$ 
in the Newtonian case), for bulgeless shafts. Dashed lines depict the corresponding profiles for steady annular 
Couette-Poiseuille flow with zero net flow (see Appendix), of the same $R_i$, $R_o$, $U$ $\tau_y$ and $\mu$ as the corresponding 
damper cases. Dash-dot lines are the yield lines of corresponding steady annular Couette flow (again, see Appendix), calculated 
from Eq.\ \eqref{eq: ACF yield line 1}. In \subref{sfig: u profiles per Ro} profiles are shown for various values of $R_o$, while 
the rest of the dimensional parameters have the values shown in Table \ref{table: base case}; there is only one dash-dot line 
because it is independent of $R_o$. In \subref{sfig: u profiles per Bn} profiles are shown for various values of $Bn$, obtained 
by varying $\tau_y$ and keeping the rest of the dimensional parameters as listed in Table \ref{table: base case}.}
  \label{fig: u profiles at cline}
\end{figure}

Figure \ref{fig: u profiles at cline} shows that something similar happens in the bulgeless damper cases, in the middle of the 
bore length. Figure \ref{sfig: u profiles per Ro} shows that the velocity gradient at the inner cylinder, and therefore also the 
force $F_R$, is relatively independent of $R_o$; thus $F_R$ is relatively insensitive to changes in $Bn$ that are due to changes 
in $R_o$, as Figs.\ \ref{sfig: Ftot vs disp per Ro} -- \ref{sfig: Fvisc vs disp per Ro} also show. On the other hand, Fig.\ 
\ref{sfig: u profiles per Bn} shows that the velocity gradient at the inner cylinder, and therefore also the force, depends 
strongly on $\tau_y$; thus $F_R$ is sensitive to changes in $Bn$ that are caused by changes in $\tau_y$, as shown in Fig.\ 
\ref{sfig: Fvisc vs dx per Bn}. So, one must be careful when using $Bn$ to assess the viscoplasticity of the flow. Figure 
\ref{fig: u profiles at cline} includes the corresponding yield lines for annular Couette flow (dash-dot lines). Obviously, 
there are differences from the damper cases, but the trends are similar: $R_o$ has a minimal effect on the yield surfaces, while 
the effect of $\tau_y$ is much more important.

A one-dimensional flow that is even closer, although not as enlightening, is annular Couette-Poiseuille flow in which the 
pressure gradient is precisely that which results in zero overall flow between the two cylinders. The equations are given again 
in the Appendix, and the velocity profiles are drawn in dashed lines in Fig.\ \ref{fig: u profiles at cline}. The similarity with 
the annular cavity flow is striking; the profiles are nearly identical, and any differences can be attributed to the boundary 
conditions: for annular Couette-Poiseuille flow we used no-slip conditions. This explains why the discrepancy becomes larger with 
the Bingham number, since viscoplasticity leads to more slip as was discussed earlier. It is expected that annular 
Couette-Poiseuille flow is a good approximation for annular cavity flow away from the cavity sides, especially for long cavities, 
when inertia effects are weak. This has not been investigated further, although it could be useful for certain practical 
applications.

In the case with a bulge, $R_o$ has a significant impact, for the cases studied. It is evident from Fig.\ \ref{fig: force per Ro} 
that the narrower the cylinder, the greater the force, and the less ``viscoplastic'' (flat) the shape of its graph. An 
explanation can be sketched with the help of Fig.\ \ref{fig: flow 400 Ro256}. As the gap between the bulge and the outer cylinder 
becomes narrower, larger fluid deformations and shear stresses develop there. This causes a moderate increase in the total 
viscoplastic component of $F_R$, as seen in Fig.\ \ref{sfig: Fvisc vs disp per Ro}, because the extent of this high-shear area is 
rather small. However, these high localised stresses make it more difficult for the material to flow through the constriction, 
and this requires higher pressure gradients to push it through. This is evident by comparing Figs.\ \ref{sfig: flow field Ro256 
zoom} and \ref{sfig: flow field Ro500 zoom}. The increased pressure gradient does not just have a localised effect, but it 
increases the pressure differences across the whole bulge resulting in a significant increase of the pressure force (Fig.\ 
\ref{sfig: Fpres vs veloc per Ro}). Also, since the pressure gradient has to counteract the viscous stresses that oppose the 
fluid flow through the constriction, and the latter have a large $\mu \dot{\gamma}$ component (compared to their $\tau_y$ 
component) due to the narrowness of the constriction, the resulting pressure force is more proportional to the shaft velocity 
(more ``Newtonian-like'') the narrower the constriction is (again, see Fig.\ \ref{sfig: Fpres vs veloc per Ro}).

\begin{figure}[!b]
  \centering
  \subfigure[] {\label{sfig: Ftot vs veloc per Rb}
      \includegraphics[scale=0.77]{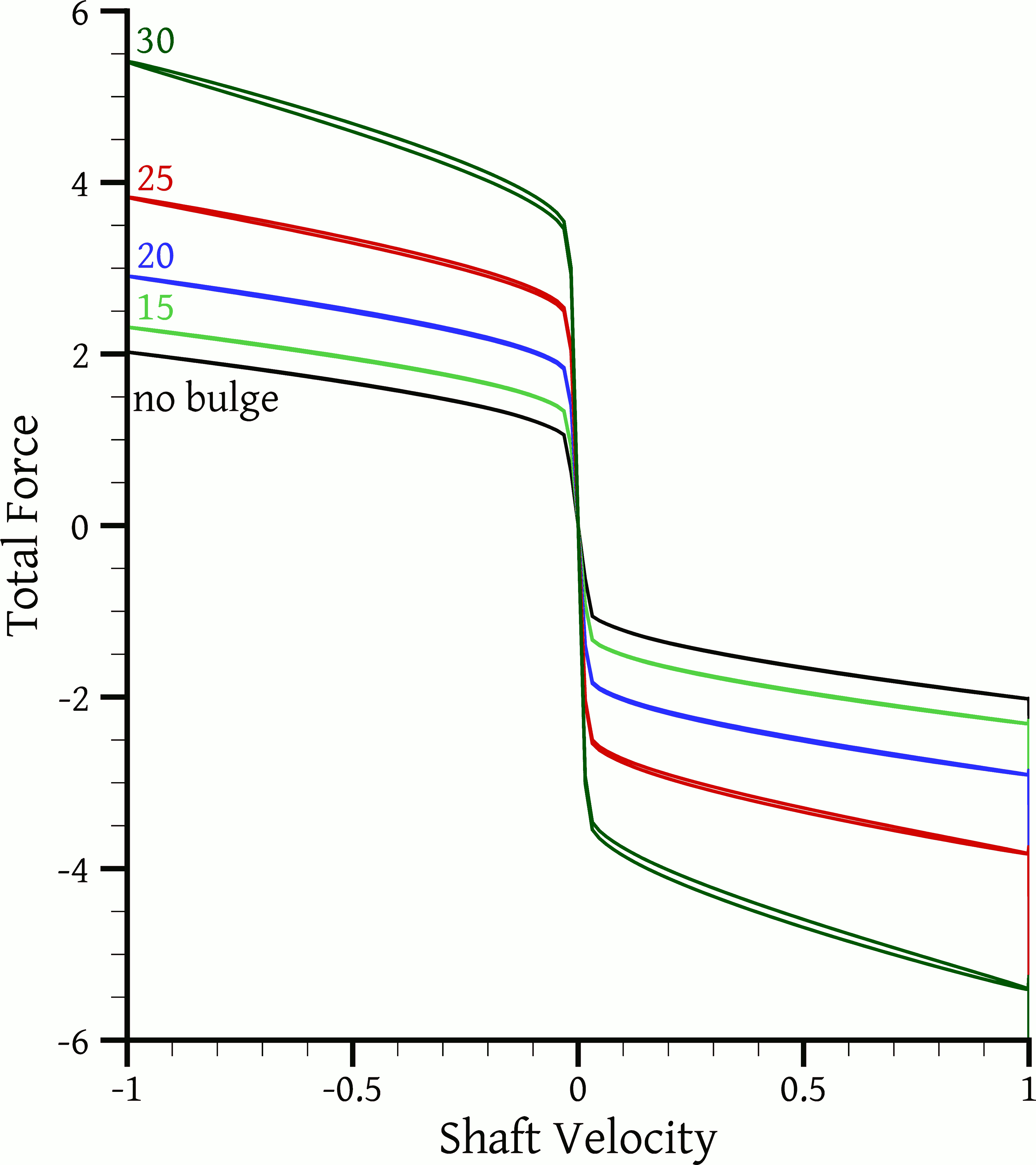}}
  \subfigure[] {\label{sfig: Fvisc vs veloc per Rb}
      \includegraphics[scale=0.77]{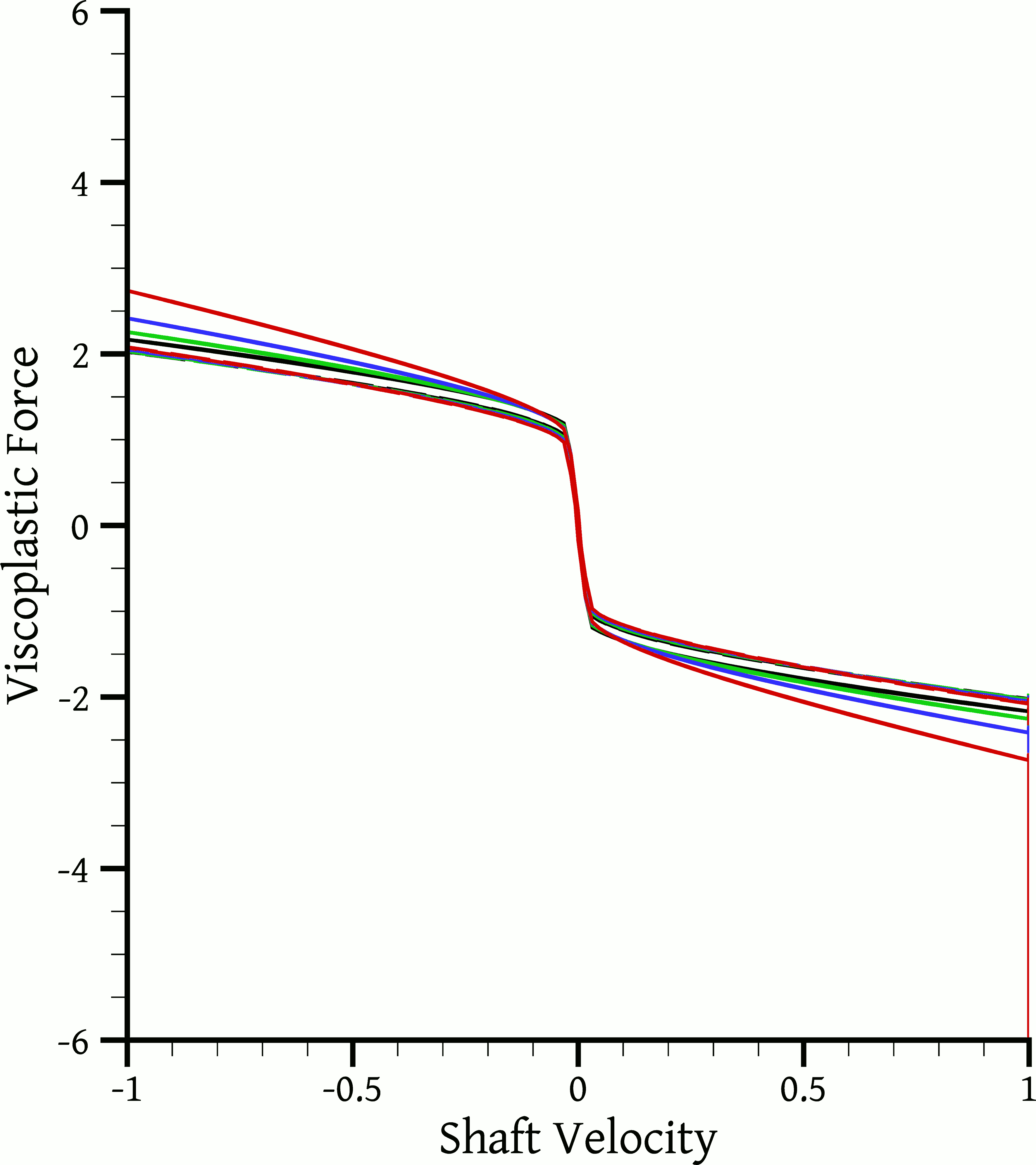}}
  \subfigure[] {\label{sfig: Fpres vs veloc per Rb}
      \includegraphics[scale=0.77]{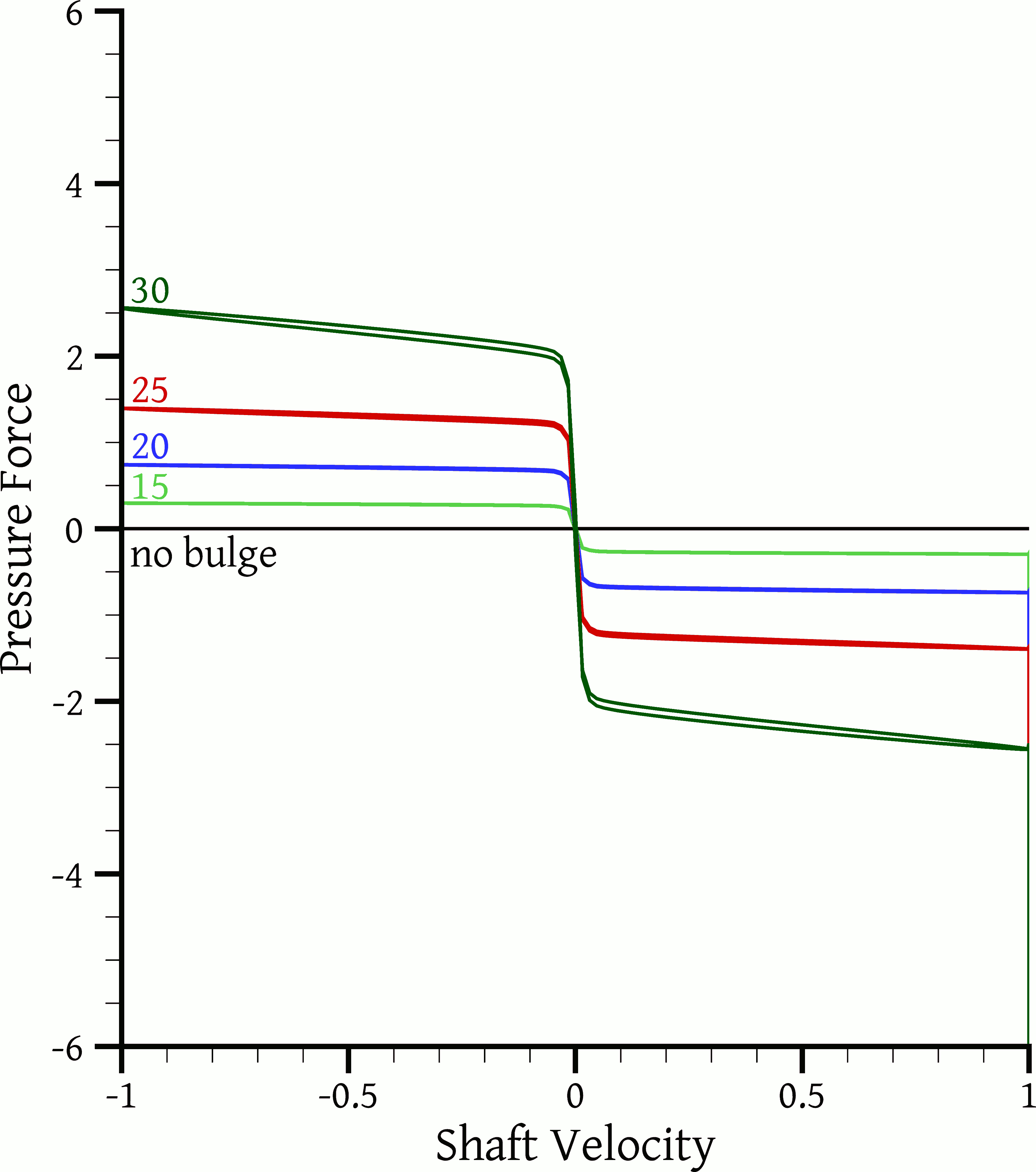}}
  \caption{The reaction force and its components as a function of shaft displacement or velocity, for various bulge radii, $R_b$ 
= 0 (no bulge), 15, 20, 25 and 30 mm. The rest of the parameters are as listed in Table \ref{table: base case}.}
  \label{fig: force per Rb}
\end{figure}

\begin{figure}[t]
  \centering
  \subfigure {\includegraphics[scale=0.65]{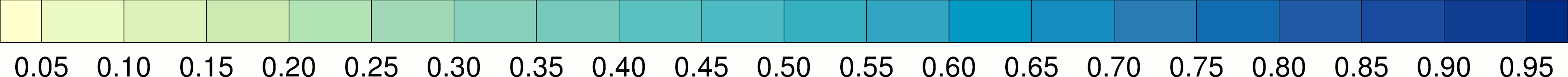}}
  \subfigure[$R_b$ = 20 mm (base case)] {\label{sfig: Vmag Rb20}
            \includegraphics[scale=0.65]{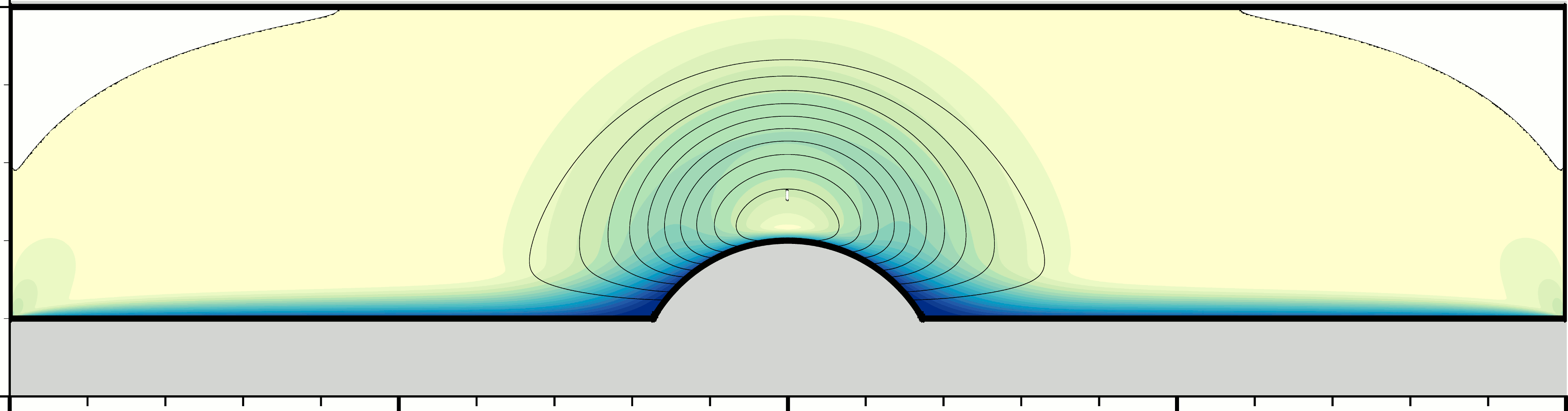}}
  \subfigure[$R_b$ = 30 mm] {\label{sfig: Vmag Rb30}
            \includegraphics[scale=0.65]{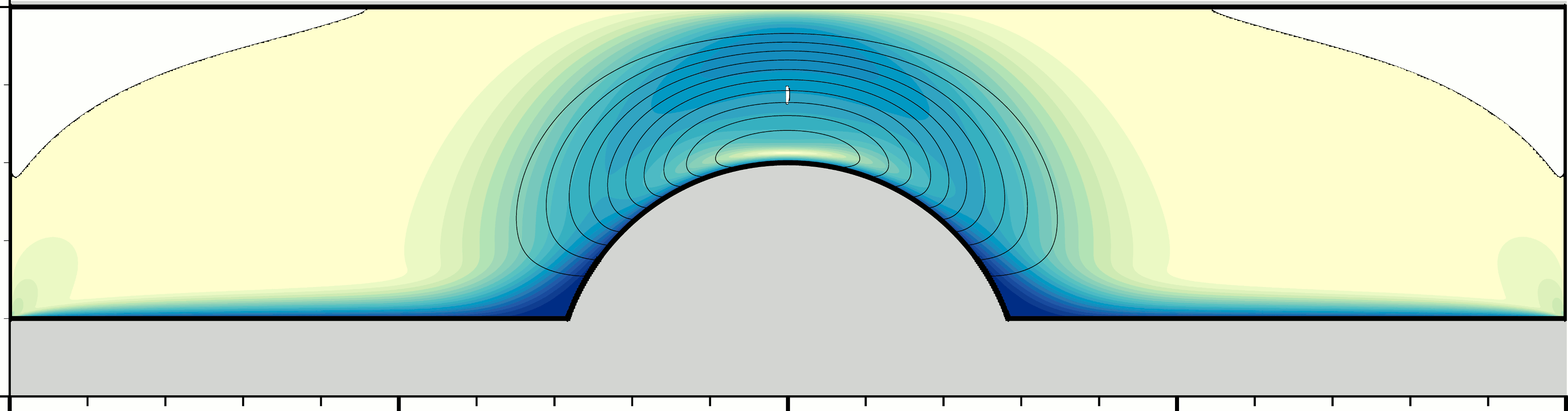}}
  \caption{Non-dimensional velocity magnitude $\|\vf{u}\|/U$ at $t = T$ for two different bulge radii $R_b$. The rest of the 
flow parameters are as listed in Table \ref{table: base case}. Unyielded material is shown in white. The streamlines drawn 
correspond to equispaced values of the streamfunction from zero to the maximum value of each case.}
  \label{fig: u/U magnitude}
\end{figure}

In another set of simulations, the bore radius $R_o$ is held constant while the bulge radius $R_b$ is varied. This also has the 
effect of varying the narrowness of the constriction, but without changing any of the dimensionless numbers characterising the 
flow, except the geometric ratio $R_b/R_o$ (Table \ref{table: base case}). Figure \ref{fig: force per Rb} shows that again, like 
when $R_o$ was varied, constricting the stenosis increases $F_R$ and it does so mostly through the pressure component. The 
explanation is the same as for the variation of $R_o$. It is interesting to note in Figs.\ \ref{sfig: Ftot vs veloc per Rb} and 
\ref{sfig: Fpres vs veloc per Rb} that some hysteresis is exhibited for $R_b$ = 30 mm, meaning that the relative magnitude of 
inertia forces increases with $R_b$, despite the Reynolds number being constant. Figure \ref{fig: u/U magnitude} helps to explain 
why: increasing $R_b$ results in increased velocities in a larger part of the domain as the constriction becomes narrower but 
also the bulge occupies a larger part of the axial extent of the shaft. The increased velocities imply increased velocity 
variations in time and space, and therefore increased inertia forces, as the flow is transient and the streamlines are curved.

Figures \ref{sfig: dissipation 400 Ro256} and \ref{sfig: dissipation 400 Rb30} show plots of the dissipation function when the 
shaft velocity is maximum, for the cases of minimum bore radius and maximum bulge radius tested. In \ref{sfig: dissipation 400 
Ro256} one can discern very high dissipation rates also at the cylinder bore, opposite to the bulge. In \ref{sfig: dissipation 400 
Rb30} the rate of dissipation does not reach so high values near the bore, because the gap between the bulge and the bore is 
wider than in \ref{sfig: dissipation 400 Ro256}, but there is extensive energy dissipation in a wide area of the domain.

The results of this paragraph show that when changing the damper geometry it is important not to rely too much on what happens to 
the Bingham and Reynolds numbers in order to make conjectures about the effects of the geometry change on the viscoplastic and 
inertial character of the flow.

\subsection{Effect of the frequency}
\label{ssec: results inertia}

\begin{figure}[t]
  \centering
  \subfigure[] {\label{sfig: Ftot vs displ per f}
      \includegraphics[scale=1.00]{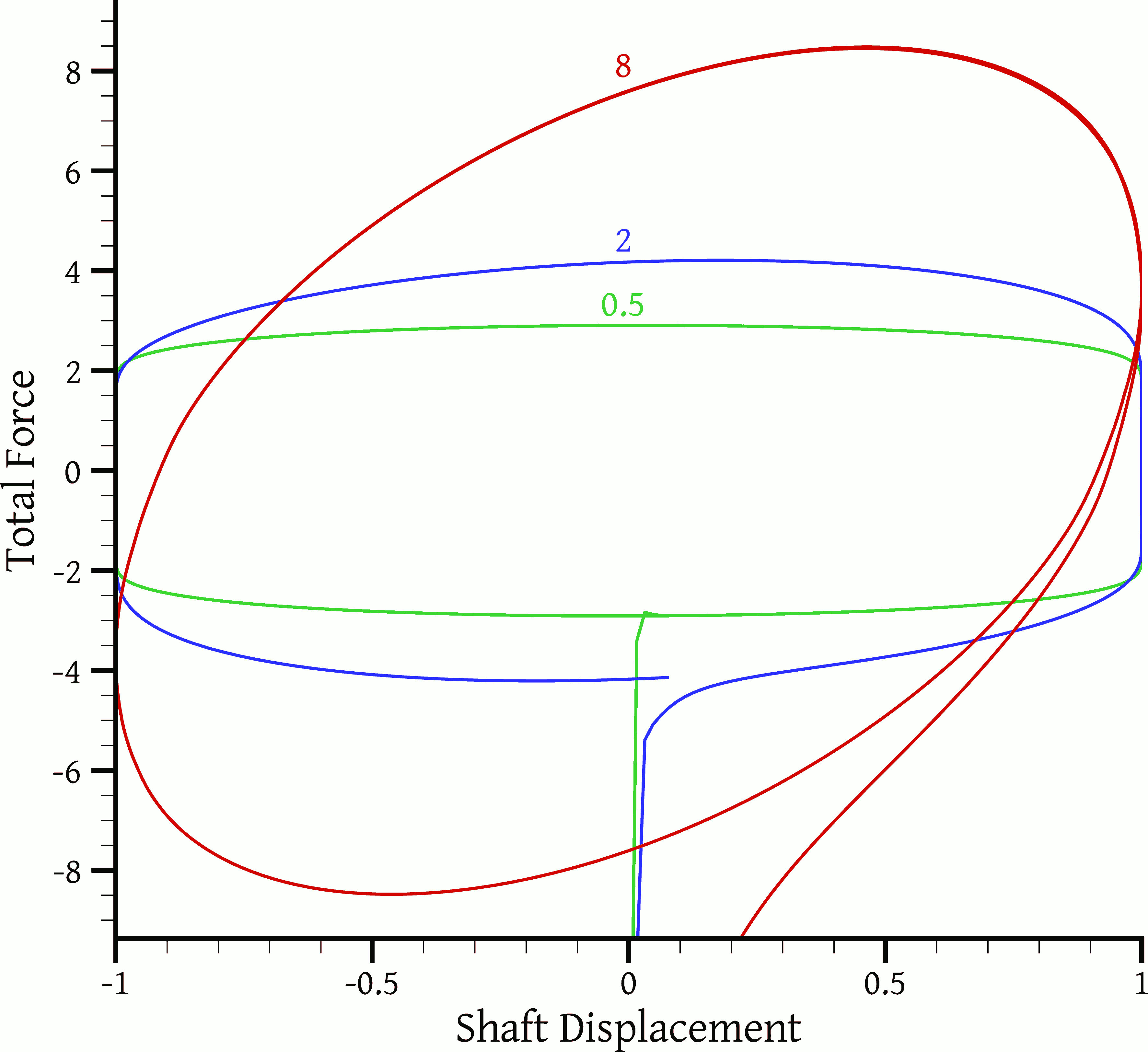}}
  \subfigure[] {\label{sfig: Ftot vs veloc per f}
      \includegraphics[scale=1.00]{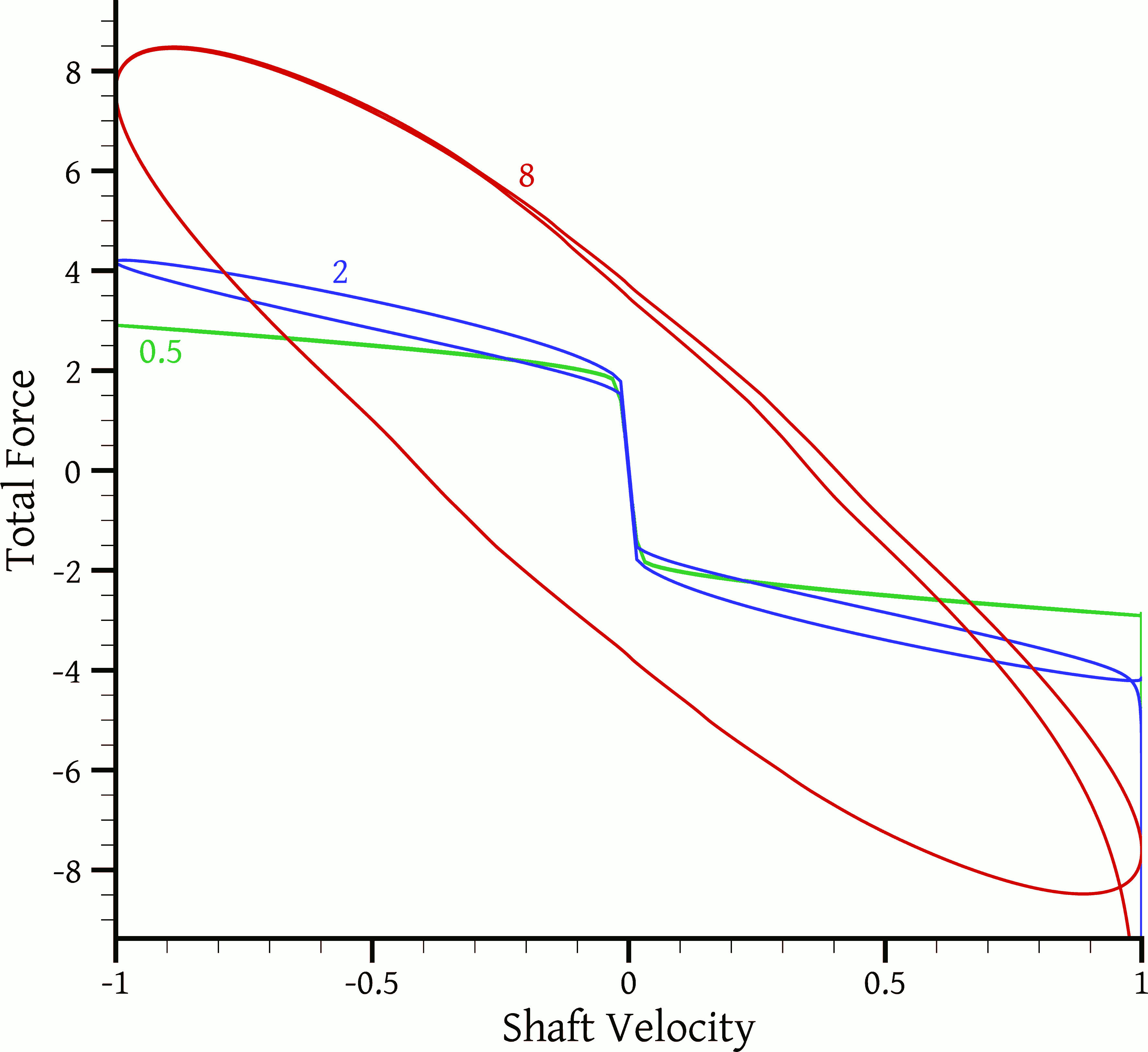}}
  \caption{The reaction force as a function of shaft displacement and velocity, for various oscillation frequencies, $f$ = 0.5 
(base case), 2 and 8 Hz. The rest of the parameters are as listed in Table \ref{table: base case}.}
  \label{fig: force per f}
\end{figure}

Finally, we study the effect of the oscillation frequency on the damper response. In particular, in addition to the $f$ = 0.5 Hz 
base case, we performed simulations for $f$ = 2 and 8 Hz, while keeping the rest of the dimensional parameters of Table 
\ref{table: base case} unaltered. The variation of the reaction force $F_R$ with respect to shaft displacement and velocity is 
plotted in Fig.\ \ref{fig: force per f}. For $f$ = 2 Hz it was observed that all the material became unyielded when the shaft 
stopped, and so the simulation duration was set to $t \in [0,T]$, like for most other simulations; but for $t$ = 8 Hz the 
material continues to flow even at the instances when the shaft is still, and so the simulation duration was extended to $t \in 
[0, 2T]$, which, as the results show (Fig.\ \ref{sfig: Ftot vs displ per f}), is more than enough to attain the periodic state.

Increasing the frequency while holding the amplitude of oscillation constant means that the maximum velocity $U$ is increased 
proportionally. This results in a reduction of the Bingham number $Bn$ and in an increase of the Reynolds number $Re^*$. Unlike 
the situation presented in Section \ref{ssec: results geometry}, now the geometrical parameters of the problem do not change 
between the different frequency cases studied, and therefore $Bn$ and $Re^*$ are appropriate indicators of the viscoplastic and 
inertial character of the flow, respectively. As far as the rest of the dimensionless numbers are concerned, the Strouhal number, 
being proportional to the dimensionless amplitude of oscillation, is not affected, while the slip coefficient $\tilde{\beta}$ 
drops, approaching its Newtonian value $\beta \mu / H$.

Evidently, as $f$ increases, the invariability of the reaction force, which is characteristic of viscoplasticity, is lost. Figure 
\ref{sfig: Ftot vs veloc per f} shows that the relationship between $F_R$ and the shaft velocity becomes more linear as $f$ 
increases, a sign that the $\mu \dot{\gamma}$ component of stress becomes dominant over the $\tau_y$ component. This is reflected 
in the reduction of the Bingham number, which falls from 20 at $f$ = 0.5 Hz, to 5 at $f$ = 2 Hz, and to 1.25 at $f$ = 8 Hz.

Similarly, the skewness of the force curve for $f$ = 8 Hz in Fig.\ \ref{sfig: Ftot vs displ per f} and the hysteresis of the 
corresponding curve in Fig.\ \ref{sfig: Ftot vs veloc per f} reveal that when $f$ is increased inertia becomes more important. 
This is reflected in the increase of the Reynolds number $Re^*$, which increases from 0.12 at $f$ = 0.5 Hz, to 1.68 at $f$ = 2 Hz, 
and to 17.9 at $f$ = 8 Hz. This is in agreement with previous studies \cite{Nguyen_2009, Yu_2013}. We note that with the present 
modelling assumptions hysteresis is only associated with inertia effects. In the literature it is often reported that in ER/MR 
dampers hysteresis is exhibited also under low inertia conditions, when the displacement approaches its extreme values (e.g.\ 
\cite{Symans_1999, Li_2000, Wang_2007, Parlak_2012}). It has been proposed that this is due to the fluid exhibiting pre-yield 
elastic behaviour \cite{Li_2000} or to compressibility effects \cite{Wang_2007}, neither of which are accounted for by the 
present Bingham model.

\begin{figure}[t]
  \centering
  \subfigure[$f$ = 2 Hz] {\label{sfig: work f2}
      \includegraphics[scale=1.00]{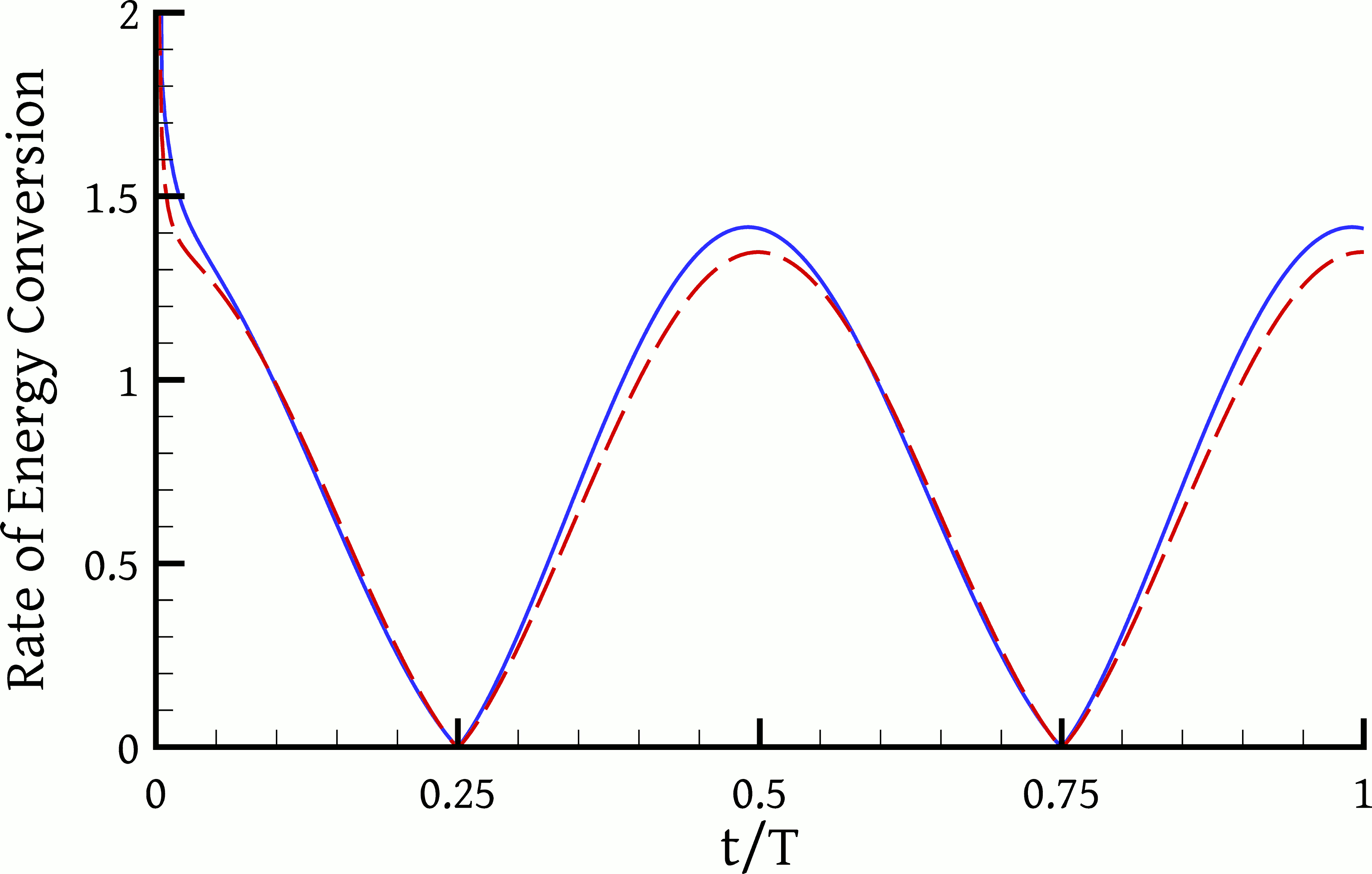}} \\
  \subfigure[$f$ = 8 Hz] {\label{sfig: work f8}
      \includegraphics[scale=1.00]{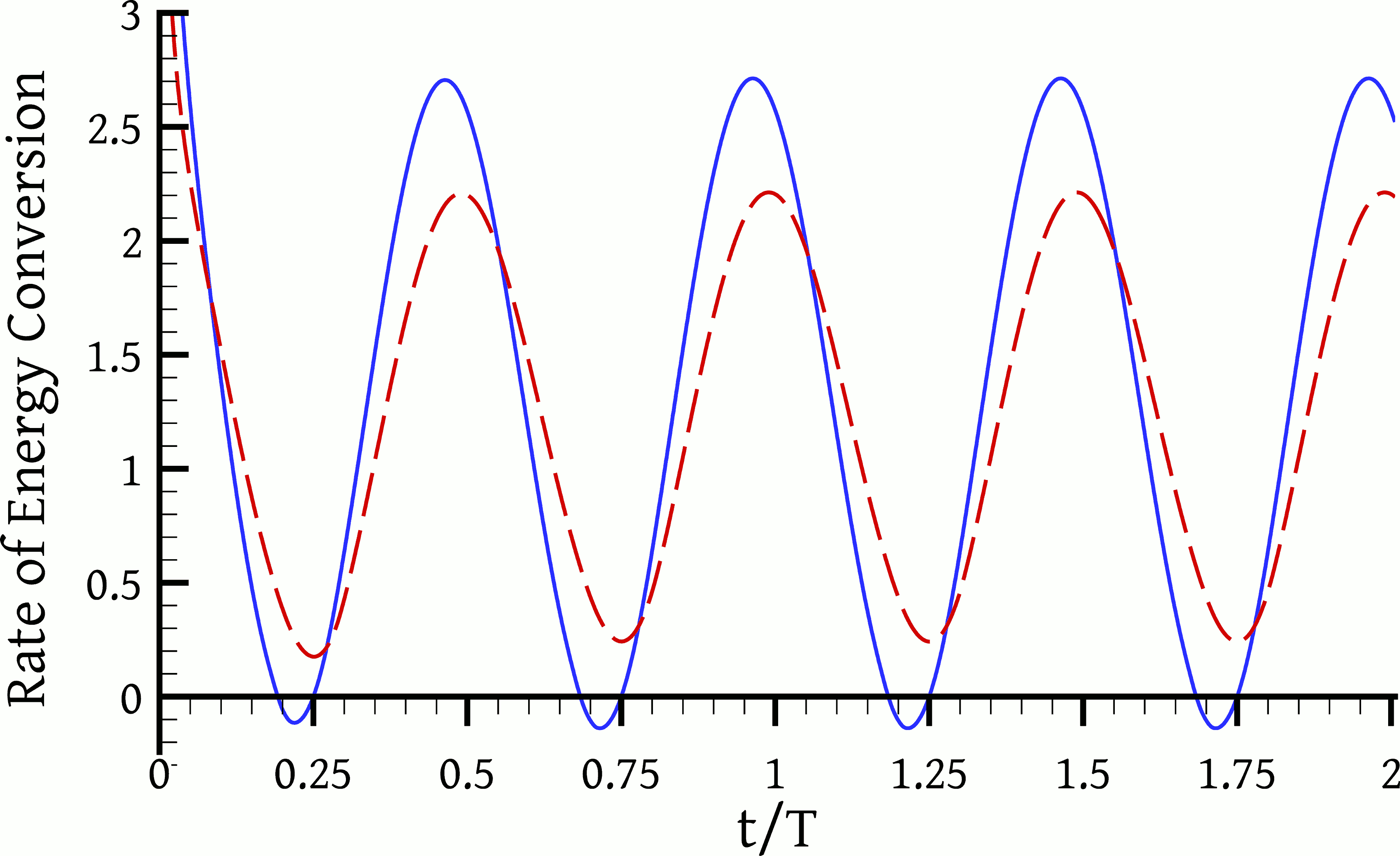}}
  \caption{Instantaneous power consumption of the damper (blue, solid line) and rate of energy dissipation due to fluid 
deformation (red, dashed line) for different oscillation frequencies. For details see the caption of Fig.\ \ref{fig: work per 
slip}.}
  \label{fig: work per f}
\end{figure}

Figure \ref{fig: work per f} shows the time history of the rate of energy absorption by the damper, $-F_R u_{sh}$, together with 
the rate of energy dissipation in the bulk of the material due to its deformation, calculated as the integral of the dissipation 
function, for $f$ = 2 and 8 Hz. The corresponding plot for $f$ = 0.5 Hz is shown in Fig.\ \ref{sfig: work Bn20}. Obviously, as 
the frequency increases, there develops a phase difference between the total rate of energy absorption and the rate of energy 
dissipation due to fluid deformation. In Fig.\ \ref{sfig: work Bn20} they are in phase with each other and with the shaft 
velocity. However, in Fig.\ \ref{sfig: work f2}, and even more so in Fig.\ \ref{sfig: work f8}, the variation of the total rate 
of energy absorption is shifted towards earlier times, while the dissipation due to fluid deformation remains in phase with the 
shaft velocity. This can be attributed to the role of inertia, which becomes more important when the frequency is increased. In 
what follows, a simple explanation for this will be presented which results in an algebraic formula that describes well the main 
characteristics of the damper response.

Taking the dot product of the velocity vector with the momentum equation and integrating over the whole volume $\Omega$ of the 
viscoplastic material, one obtains, after some manipulation, an energy balance for the whole of that material \cite{Winter_1987}:

\begin{equation} \label{eq: energy balance}
 \underbrace{
 \int_{\partial \Omega} \left( -p \vf{n} \cdot \vf{u} \;+\; \vf{n} \cdot \tf{\tau} \cdot \vf{u} \right) \mathrm{d}\!A
 }_{\substack{\text{Rate of work done by external forces} \\ = -F_R u_{sh}}}
 \;=\;
 \underbrace{
 \int_{\Omega} \tf{\tau} : \nabla \vf{u} \, \mathrm{d}\Omega
 }_{\substack{\text{Rate of energy dissipation} \\ \text{due to fluid deformation}}}
 \;+\;
 \underbrace{
 \int_{\Omega} \rho \vf{u} \cdot \frac{\mathrm{D}\vf{u}}{\mathrm{D}t} \, \mathrm{d}\Omega
 }_{\substack{\text{Rate of increase of the} \\ \text{kinetic energy of the fluid}}}
\end{equation}
where $\partial \Omega$ is the boundary of the material, consisting of its interface with the shaft and the containing cylinder. 
The vector $\vf{n}$ is the outward unit vector normal to this surface, $\mathrm{d}\!A$ is an infinitesimal area of the surface, 
and $\mathrm{D}/\mathrm{D}t = \partial/\partial t + \vf{u} \cdot \nabla$ is the substantial time derivative. The above equation 
says that all the work done on the fluid by the motion of the shaft is either dissipated or stored as kinetic energy of the 
fluid. Actually, the left hand side is only equal to $-F_R u_{sh}$ in the absence of slip; otherwise it is smaller. But for 
this simplified analysis we will neglect slip. The goal is to make conjectures about the temporal variation of the terms of the 
right-hand side of Eq.\ \eqref{eq: energy balance} and combine them to estimate the overall temporal variation of the damper work.
To proceed, we will assume that the terms of the right-hand side can be expressed as products of a characteristic force, 
viscoplastic or inertial, respectively, times a characteristic fluid velocity.

Velocities and velocity gradients in the fluid can be assumed to be roughly proportional to the shaft velocity $u_{sh} = U 
\cos(\omega t)$. Then, if the flow is viscoplastic, the viscoplastic forces would be expected to be of the form $F_V = F_V^0 
(b\, \mathrm{sign}(\cos(\omega t)) + \cos(\omega t)) / (b+1)$, for some constant $b$ proportional to the Bingham number. But the 
function $b \, \mathrm{sign}(\cos(\omega t))$, which is a square wave of the same frequency as $\cos(\omega t)$, bears some 
resemblance to $\cos(\omega t)$ and their sum can be replaced by just $(b+1)\cos(\omega t)$ for the purposes of this simplified 
analysis. A typical viscoplastic force would then have the form $F_V = F_V^0 \cos(\omega t)$. The maximum value $F_V^0$ would 
increase with the maximum shaft velocity $U = \alpha \omega$ but not proportionally, due to the constant plastic component of 
the force; but it should tend to become proportional to $U$ (and $\omega$) at high frequencies.

Similarly, we assume that accelerations in the fluid are proportional to the shaft acceleration, $\dot{u}_{sh} = -\omega U 
\sin(\omega t)$, so that a typical inertia force such as that appearing in the last term of Eq.\ \eqref{eq: energy balance} has 
the form $F_I = \rho \mathrm{D}\vf{u} / \mathrm{D}t = -F_I^0 \sin(\omega t)$. The maximum value $F_I^0$ would be proportional to 
the maximum shaft acceleration $\omega U = \omega^2 \alpha$. Thus, increasing the frequency favours inertial forces over viscous 
forces: the ratio $F_I^0 / F_V^0$ tends to become proportional to $\omega$.

Under these assumptions Eq.\ \eqref{eq: energy balance} can be approximated by

\begin{align}
 \nonumber
 - F_R u_{sh} \;&=\; c \left( F_V^0 \cos(\omega t) \;-\; F_I^0 \sin(\omega t) \right) U \cos(\omega t)
\\
 \nonumber
             &=\; \underbrace{c U \sqrt{(F_V^0)^2 + (F_I^0)^2}}_{c'}
                  \left( \frac{F_V^0}{\sqrt{(F_V^0)^2 + (F_I^0)^2}} \cos(\omega t) \;-\;
                         \frac{F_I^0}{\sqrt{(F_V^0)^2 + (F_I^0)^2}} \sin(\omega t) \right) \cos(\omega t)
\\
 \nonumber
             &=\; c' \left( \cos\delta \cos(\omega t) \;-\; \sin\delta \sin(\omega t) \right) \cos(\omega t)
\\
 \nonumber
             &=\; c' \cos \left( \omega t \,+\, \delta \right) \cos(\omega t)
\\
 \label{eq: Fu trigonometric}
             &=\; (c'/2) \, \left[ \cos\delta \;+\; \cos(2 \omega t \,+\, \delta) \right]
\end{align}
where we have used some simple trigonometric identities, $c$ and $c'$ are constants (actually, $c'$ depends on $\omega$ but 
not on $t$), and $\delta$ is the angle adjacent to the side of length $F_V^0$ of a right triangle whose perpendicular sides have 
lengths $F_V^0$ and $F_I^0$. Thus $\delta = \arctan(F_I^0 / F_V^0)$, and when the viscous forces dominate ($F_I^0 / F_V^0 
\rightarrow 0$), i.e.\ when the Reynolds number is very small, then $\delta$ is close to zero; and when the inertia forces are 
much larger than the viscous forces ($F_I^0 / F_V^0 \rightarrow \infty$), i.e.\ when the Reynolds number is very large, then 
$\delta$ tends to $\pi/2$.

According to Eq.\ \eqref{eq: Fu trigonometric}, the rate of energy absorption by the damper is proportional to $\cos\delta + 
\cos(2 \omega t + \delta)$; this has two parts: a constant part, $\cos\delta$, and a time-varying part, $\cos(2 \omega + \delta)$. 
It follows that the rate of energy absorption by the damper varies with a frequency of $2\omega$, twice that of the shaft 
oscillation. This is confirmed by Figs.\ \ref{fig: work per slip} and \ref{fig: work per f}, and is easily explained by the fact 
that each shaft oscillation can be split into two half-periods, one when the shaft is moving from left to right and one when the 
shaft is moving from right to left. The variation of the rate of energy absorption is exactly the same in both half-periods, due 
to flow symmetry: $F_R(t+T/2) u_{sh}(t+T/2) = (-F_R(t)) (-u_{sh}(t)) = F_R(t) u_{sh}(t)$. Therefore each of these two 
half-periods is a full period of the variation of the rate of energy absorption.

When $Re^*$ is small and inertia is negligible then $\delta \approx 0$ and $\cos(\delta) \approx 1$ and Eq.\ \eqref{eq: Fu 
trigonometric} predicts that $-F_R u_{sh}$ is proportional to $1 + \cos(2\omega)$, which is always positive or zero. This is 
confirmed by Figs.\ \ref{sfig: work Bn20} and \ref{sfig: work f2}. It is also in phase with the shaft oscillation, albeit at twice 
the frequency; when the shaft moves with maximum velocity (either positive or negative) $-F_R u_{sh}$ is maximum, and when the 
shaft momentarily becomes still $-F_R u_{sh}$ drops to zero. This is because the only forces of importance are the viscous 
forces, and they are proportional to the shaft velocity, according to our assumptions. Figure \ref{fig: work per slip}, 
corresponding to a low Reynolds number, confirms this.

On the other hand, if the relative magnitude of the inertia forces is increased, i.e.\ at higher $Re^*$, Eq.\ \eqref{eq: Fu 
trigonometric} predicts that the variation of the rate of energy absorption $-F_R u_{sh}$ will precede the variation of shaft 
velocity by an increasing phase difference $\delta$ (which however will never exceed the value $\pi/2$). This is confirmed by 
Figs.\ \ref{sfig: work Bn20}, \ref{sfig: work f2} and \ref{sfig: work f8}, where higher frequencies are seen to correspond to 
larger $\delta$. A consequence is that the maximum energy absorption occurs not when the shaft velocity is maximum, like in the 
low $Re^*$ cases, but earlier. It is a matter of balance between viscous and inertia forces: as the shaft accelerates from a still 
position (maximum displacement) to its maximum velocity position (zero displacement) the velocity rises but the acceleration 
drops. Accordingly, viscous forces rise from zero to their maximum, while inertial forces drop from their maximum to zero; the 
maximum rate of energy absorption occurs somewhere in between. This situation is similar to that described by Iwatsu et al.\
\cite{Iwatsu_1992} for the oscillating lid driven cavity problem, who report that the time lag between the lid force and the lid 
velocity increases with frequency.

Another consequence of the phase difference $\delta$, which can be seen clearly only in Fig.\ \ref{sfig: work f8}, is that, 
roughly during the shaft acceleration phase, the rate of energy absorption (blue curve) is larger than the rate of viscous 
dissipation (red curve) because some of the absorbed energy becomes kinetic energy of the fluid rather than being dissipated.
Conversely, during the shaft deceleration phase, the rate of viscous dissipation is larger than the rate of energy absorption by 
the damper, as it is not only this absorbed energy but also the kinetic energy of the contained fluid that are dissipated. But 
the integrals of both lines in Fig.\ \ref{sfig: work f8} over an integer number of cycles must be equal, because the kinetic 
energy at $t=t_0$ is equal to that at $t=t_0 + kT$ for $k$ integer and therefore all the absorbed energy has been 
converted to heat.

The fact that $\delta > 0$ also means that $\cos\delta + \cos(2\omega t + \delta)$ will necessarily become negative during 
certain time intervals, because $\cos\delta < 1$. Indeed, this can be seen in Fig.\ \ref{sfig: work f8}, where $-F_R u_{sh}$ 
becomes negative during short time intervals just before the shaft becomes still. During these time intervals the flow of energy 
is reversed, i.e.\ instead of going from the shaft to the fluid it returns from the fluid (kinetic energy) to the shaft 
(mechanical energy). The fact that $-F_R u_{sh} < 0$ means that $F_R$ and $u_{sh}$ have the same sign, so that during such a time 
interval as the shaft is decelerating, instead of having to push away the fluid in front of it, it is pushed forward by the fluid 
behind it. This is because the fluid has acquired momentum in the direction of the shaft motion, and the inertia of the fluid is 
significant.

The maximum rate of work is roughly proportional to the constant $c'$ in \eqref{eq: Fu trigonometric}, which increases with 
$\omega$, so that increasing the frequency results in higher rates of energy absorption. This can be seen in Figs.\ \ref{sfig: 
work Bn20}, \ref{sfig: work f2} and \ref{sfig: work f8}, but the exact relationship between the magnitude of energy absorption 
and $\omega$ is a bit complicated and things are made even more complicated by the fact that in the figures the rate of work is 
normalised by $\phi_{\mathrm{ref}} \Omega_{\mathrm{tot}}$ (see caption of Fig.\ \ref{fig: work per slip}) which also depends on 
$\omega$, through the velocity $U$ (Eq.\ \eqref{eq: phi ref}).

This simplified analysis is useful, but it has its limitations. In Fig.\ \ref{sfig: work f8} it may be seen that for $f$ = 8 Hz 
the integral of the dissipation function is not exactly proportional to the shaft velocity. In particular, its value is minimum 
but non-zero when the shaft velocity is zero. In fact the dissipation function is never zero because the fluid never ceases to 
flow, due to inertia, even when the shaft is still. This is demonstrated in Figure \ref{sfig: dissipation 1400 f8}, which shows a 
plot of the dissipation function for $f$ = 8 Hz at a time instance when the shaft is still. On the other hand, maximum energy 
dissipation occurs when the shaft velocity is maximum, as in Fig.\ \ref{sfig: dissipation 1600 f8}, where one can notice the 
asymmetry that is due to the substantial inertia of the fluid. One can also notice in the same Figure the increased importance of 
the regions near the shaft endpoints in terms of energy dissipation, where the increased velocities at high frequencies produce 
high velocity gradients, and reduce the role of viscoplasticity.

\section{Conclusions}
\label{sec: conclusions}

In this paper we studied numerically the viscoplastic flow in an extrusion damper where a sinusoidal displacement is forced on 
the damper shaft. The flow is assumed axisymmetric, and, except when the shaft is bulgeless, the shape of the domain changes 
with time. To cope with this, a finite volume method applicable to moving grids was employed. As the calculation of the force on 
the shaft is crucial, the usual no-slip boundary condition is inappropriate due to the velocity discontinuities at the shaft 
endpoints, and the Navier slip boundary condition was employed instead. A series of simulations was performed, where several 
parameters were varied, in order to study the effects of viscoplasticity, slip, damper geometry, and oscillation frequency on the 
damper response.

The reciprocating motion of the bulged shaft creates a ring-shaped flow around the bulge, as the latter pushes away the fluid in 
front of it; away from the bulge the fluid motion is very weak. The bulge creates a stenosis through which the fluid is pushed 
(``extruded''). In order to overcome the resisting viscous stresses and push the fluid across, high pressure gradients develop; in 
turn these result in pressure differences between the two sides of the bulge that give rise to significant pressure forces. These 
pressure forces are the major contribution of the bulge to the total reaction force, and they are larger when the constriction is 
narrower, i.e.\ when the bulge diameter is larger or when the outer cylinder diameter is smaller.

The bulgeless case, an annular analogue of the lid-driven cavity problem, was studied as well. It was shown that at low Reynolds 
numbers, away from the damper end walls, the flow can be approximated very accurately by annular Couette-Poiseuille flow where the 
pressure gradient is precisely that required for zero net flow across the annulus. The flow pattern is different from the bulged 
case, with most of the flow occurring in a thin layer surrounding the shaft; the radius of the outer cylinder plays a minor role 
in 
this case.

The more viscoplastic the flow is the less the force varies during the damper operation. This is often desirable, because it 
maximises the energy absorbed for a given force capacity. In combination with slip, viscoplasticity can result in a situation 
where the shaft moves while the fluid is unyielded and stationary. In this case, but also in every case that there is slip, even 
if the fluid is yielded, mechanical energy is dissipated not only in the bulk of the fluid due to deformation, but also 
directly at the fluid-shaft interface due to friction. The percentage of energy lost in this way can be significant if the slip 
coefficient is large, but also if the yield stress is high (these two are combined in the dimensionless slip coefficient 
$\tilde{\beta}$).

Increasing the frequency of oscillation makes the inertia of the system more significant and introduces hysteresis into the 
damper response to the sinusoidal forcing. It also brings the kinetic energy of the fluid into the energy balance, introducing a 
phase shift of the energy absorption rate relative to the sinusoidal forcing. Furthermore, increasing the frequency weakens the 
viscoplastic character of the damper response, i.e.\ it results in greater dependency of the reaction force on the shaft velocity.

Plots of the dissipation function reveal that most of the energy dissipation occurs near the outer part of the bulge and near the 
shaft ends. When the constriction between the bulge and the outer cylinder is narrow, high dissipation occurs also at the outer 
cylinder, at the region opposite to the bulge. At the shaft ends there develop high velocity gradients, even when slip occurs. 
However, these high velocity gradients contribute less to the overall force when the flow is more viscoplastic.

Overall, a two-dimensional simulation can reveal more details about the operation of a damper than a simplified one-dimensional 
analysis. The present work investigated only dampers operating with Bingham fluids, but in practice the fluids used may exhibit 
more rheologically complex behaviour, possibly with temperature effects, shear-thinning, viscoelasticity, and thixotropy. The 
present methodology could be extended to cover these cases as well.

\section*{Acknowledgements}
This research was funded by the Thales Project ``COVISCO'' (project number 648) and the bilateral Greece - Israel Project named 
PHARMAMUDS (project number 3163), which are co-funded by Greece and the European Union.



\begin{appendices}
\renewcommand\theequation{\thesection.\arabic{equation}}
\setcounter{equation}{0}

\section{Annular Couette and Couette-Poiseuille flow of a Bingham fluid}
\label{appendix: couette}

Consider first the steady, annular Couette flow of a Bingham fluid between two concentric cylinders of infinite length, of which 
the inner one, of radius $R_i$, moves with a constant velocity $U$ in the axial direction while the outer one, of radius $R_o$, 
is stationary. The pressure gradient is zero, the only non-zero velocity component is the axial component $u$, and the only 
non-zero stress component is $\tau_{rx}$. The flow is one-dimensional and steady so that $u = u(r)$ and $\tau_{rx} = 
\tau_{rx}(r)$ are functions only of the radial coordinate $r$. No-slip boundary conditions are assumed. This flow has an 
analytical solution which is presented in \cite{Bird_1982}, but the location of the yield line is not given there explicitly in 
closed form. Here we will do so with the help of the Lambert $W$ function \cite{Corless_1996}.
For this flow, the momentum equation simplifies to

\begin{equation} \label{eq: ACF tau variation}
 \tau_{rx} \;=\; \frac{c}{r}
\end{equation}
for some constant $c$. Therefore, the stress decreases monotonically from $r = R_i$ to $r = R_o$ and thus it is maximum at 
$R_i$. Since the relative motion between the cylinders implies that yielding is always present, the inner cylinder is always in 
contact with yielded material and the stress there exceeds the yield stress. Substituting the one-dimensional version of the 
constitutive equation \eqref{eq: Bingham_constitutive} into Eq.\ \eqref{eq: ACF tau variation}, integrating, and using the 
boundary condition that $u(R_i) = U$, we arrive at the following equation which is valid from $r = R_i$ up to any radius where 
the material is yielded:

\begin{equation} \label{eq: ACF ODE integrated}
 \frac{u}{U} \;=\; 1 \;-\; \frac{c}{\mu U} \ln \left( \frac{r}{R_i} \right) \;+\; \frac{\tau_y}{\mu U} (r - R_i)
\end{equation}

Let us assume at first that the yielded region extends up to the outer cylinder, i.e.\ that $R_o$ is not large enough for 
$\tau_{rx}$ to fall below $\tau_y$. Using the boundary condition $u(R_o) = 0$ we can determine the constant $c$:

\begin{equation} \label{eq: ACF c}
 c \;=\; \frac{\tau_y (R_o - R_i) \,+\, \mu U}{\ln (R_o/R_i)}
\end{equation}
This can then be substituted in Eq. \eqref{eq: ACF ODE integrated} to obtain the velocity:

\begin{equation} \label{eq: ACF u fully yielded}
 \frac{u}{U} \;=\; 1 \;-\; \frac{\ln(\tilde{r})}{\ln(\tilde{R}_o)} 
                     \;-\; Bn \left[ \frac{\ln(\tilde{r})}{\ln(\tilde{R}_o)} \,-\, \frac{\tilde{r}-1}{\tilde{R}_o-1} \right]
\end{equation}
where $\tilde{r} \equiv r/R_i$, $\tilde{R}_o \equiv R_o/R_i$ and $Bn$ is the familiar Bingham number, Eq.\ \eqref{eq: Bn} (with 
$H = R_o - R_i$, as usual). The term in square brackets in \eqref{eq: ACF u fully yielded} is always positive, or zero for $r = 
R_i$ ($\tilde{r} = 1$) and $r = R_o$ ($\tilde{r} = \tilde{R}_o$), so that increasing the Bingham number reduces the velocity. 
Equation \eqref{eq: ACF u fully yielded} is valid as long as the stress at $R_o$ has not fallen below $\tau_y$. The larger $R_o$ 
the lower $\tau_{rx}(R_o)$ will be; for all the material to be yielded $R_o$ must not exceed a value, say $R_y$, such that 
$\tau_{rx}(R_y) = \tau_y$, or, using Eqs.\ \eqref{eq: ACF tau variation} and \eqref{eq: ACF c}:

\begin{equation*}
 \frac{\tau_y (R_y - R_i) \,+\, \mu U}{\ln(R_y/R_i)} \cdot \frac{1}{R_y} \;=\; \tau_y
\end{equation*}
Employing the alternative Bingham number

\begin{equation} \label{eq: ACF B}
 B \equiv \tau_y R_i / (\mu U)
\end{equation}
which is based on the radius $R_i$ instead of the gap $H = R_o - R_i$, as well as the ratio

\begin{equation}
  \tilde{R}_y \equiv R_y/R_i
\end{equation}
after some rearrangement, one obtains:

\begin{equation*}
 \tilde{R}_y \left( \ln \tilde{R}_y - 1 \right) \;=\; B^{-1} - 1
\end{equation*}
This equation can be solved using the Lambert $W$ function, which is the inverse function of $f(x) = x\mathrm{e}^x$: 
$x\mathrm{e}^x = y \Leftrightarrow x = W(y)$. Noting that $1 = \ln \mathrm{e}$, we can manipulate the above equation
to get:

\begin{equation} \label{eq: ACF yield line 1}
  \tilde{R}_y \;=\; \mathrm{e}^{W \left( \frac{B^{-1} - 1}{\mathrm{e}} \right) + 1}
\end{equation}
The Lambert $W$ function is double-valued on the interval $(-1/e,0)$, where we follow its upper branch because the lower branch 
results in $\tilde{R}_y < 1$, an unrealistic result.

So, for $R_i < R_o \leq R_y$ the velocity is given by Eq.\ \eqref{eq: ACF u fully yielded}. What happens when $R_o > R_y$? In 
that case, at the outer cylinder $\tau_{rx} < \tau_y$ and therefore that cylinder is in contact with a layer of unyielded 
material, where the velocity is zero due to the no-slip boundary condition and the fact that the outer cylinder is stationary. So, 
in this case there are two layers of fluid: a yielded one in contact with the inner cylinder, and an unyielded one in contact with 
the outer cylinder. The velocity variation in the yielded layer together with the location of the interface between the two layers 
can be found from Eq.\ \eqref{eq: ACF ODE integrated} by using the boundary conditions $u = 0$ and $\tau_{rx} = \tau_y$ at the 
interface. But we have already done that; the location of the interface is $R_y$, Eq.\ \eqref{eq: ACF yield line 1}, and the 
velocity is given by Eq.\ \eqref{eq: ACF u fully yielded} with $R_o$ replaced by $R_y$. With a little manipulation the result for 
this partially yielded case is

\begin{equation}
\label{eq: ACF u partly yielded}
\begin{cases}
 \displaystyle
 \frac{u}{U} \;=\; 1 \;-\; \frac{\ln(\tilde{r})}{\ln(\tilde{R}_y)} 
                     \;-\; B (\tilde{R}_y - 1) \left[ \frac{\ln(\tilde{r})}{\ln(\tilde{R}_y)} 
                                               \,-\, \frac{\tilde{r}-1}{\tilde{R}_y-1} \right] , & R_i \leq r \leq R_y
\\[0.3 cm]
 u \;=\; 0 , & R_y \leq r \leq R_o
\end{cases}
\end{equation}
The thickness of the yielded layer $\tilde{R}_y - 1$ is a strictly decreasing function of $B$. We note that according to Eqs.\ 
\eqref{eq: ACF yield line 1} and \eqref{eq: ACF u partly yielded} the thickness and the velocity of the yielded layer are 
independent of the outer cylinder diameter, contrary to the fully yielded case (Eq.\ \eqref{eq: ACF u fully yielded}). This is 
reflected in the use of $B$ instead of $Bn$ in the partially yielded case.

Next, consider the case of annular Couette-Poiseuille flow, i.e.\ let there also be an axial pressure gradient $\mathrm{d}p / 
\mathrm{d}x \neq 0$. This case can be solved in a similar manner, but it is more complex and there are many possible flow types 
depending on the importance of the pressure gradient relative to the inner cylinder velocity. All the possibilities are reported 
by Liu and Zhu \cite{Liu_2010}, but here we are only interested in the case where the pressure gradient opposes the cylinder 
motion and causes a zero net flow through any annular section. This may be a good approximation to the flow in an annular cavity, 
where the sides of the cavity restrict the flow in the axial direction. This case falls under ``Case I'' of Liu and Zhu 
\cite{Liu_2010}, and the flow pattern consists of two yielded layers adjacent to the two cylinders, with an unyielded layer in 
between. The inner yielded layer moves mostly along with the inner cylinder, but its outer part moves in the opposite direction; 
the unyielded layer and the outer yielded layer move opposite to the inner cylinder. Suppose the yield lines are at $r = y_1$ and 
$y_2$ ($\tilde{r} = {y}_1$ and $\tilde{y}_2$). Then the velocity is given by

\begin{equation}
 \label{eq: ACPF u}
 \begin{cases}
    \dfrac{u}{U} \;=\; 1 \;+\; B(\tilde{r} - 1) \;+\; \frac{1}{4} P (\tilde{r}^2 - 1) 
                         \;-\; \left( B\tilde{y}_1 \,+\, \frac{1}{2} P \tilde{y}_1^2 \right) \ln(\tilde{r}) \;,
    & R_i \leq r \leq y_1
\\[0.3 cm]
    u \;=\; u(y_1) \;=\; u(y_2) \;,
    & y_1 < r < y_2
\\[0.3 cm]
    \dfrac{u}{U} \;=\; B(\tilde{R}_o - \tilde{r}) \;-\; \frac{1}{4} P (\tilde{R}_o^2 - \tilde{r}^2)
                 \;+\; \left( B\tilde{y}_2 \,-\, \frac{1}{2} P \tilde{y}_2^2 \right)
                       \ln\left( \dfrac{\tilde{r}}{\tilde{R}_o} \right) \;,
    & y_2 \leq r \leq R_o
 \end{cases}
\end{equation}
where $P$ is a dimensionless pressure gradient

\begin{equation}
P \;\equiv\; \frac{\mathrm{d}p}{\mathrm{d}x} \frac{R_i^2}{\mu U}
\end{equation}
The yield lines can be found from the fact that the velocities are equal there, $u(y_1) = u(y_2)$, using also the relation 

\begin{equation} \label{eq: ACPF y2}
\tilde{y}_2 = \tilde{y}_1 + 2B/P
\end{equation}
which derives from the momentum balance on the unyielded layer. The result is

\begin{equation} \label{eq: ACPF y1}
 1 \;+\; B \left( \frac{B}{P} -\tilde{R}_o - 1 \right) \;+\; \frac{1}{4} P (\tilde{R}_o^2 - 1) \;+\; B\tilde{y}_1
   \;+\; \left( B\tilde{y}_1 \,+\, \frac{1}{2} P \tilde{y}_1^2 \right) 
         \ln\left( \frac{\tilde{y}_1 + 2B/P}{\tilde{R}_o\tilde{y}_1} \right)
   \;=\; 0
\end{equation}
Equation \eqref{eq: ACPF y1} can be solved numerically to obtain $\tilde{y}_1$, and then $\tilde{y}_2$ is obtained from \eqref{eq: 
ACPF y2}. Thus Eq.\ \eqref{eq: ACPF u} contains no unknown terms and can be integrated to obtain the flow rate $Q = 2\pi 
\int_{R_i}^{R_o} u r \mathrm{d}r$. For a given geometry, fluid, and inner cylinder velocity, the flow rate depends on the pressure 
gradient, $Q = Q(P)$. We seek the pressure gradient that results in $Q(P) = 0$. This is solved numerically in the present work, 
using the Newton-Raphson method, with $\mathrm{d} Q/\mathrm{d} P$ calculated numerically by perturbing $Q$. The results shown in 
Fig.\ \ref{fig: u profiles at cline} were obtained in this manner.

For completeness, we also give the velocity when the flow is Newtonian (also shown in Fig.\ \ref{sfig: u profiles per Bn}):

\begin{equation}
 \frac{u}{U} \;=\; 1 \;+\; \frac{1}{4} P (\tilde{r}^2 - 1) \;-\; \left[ 1 \,+\, \frac{1}{4} P (\tilde{R}_o^2 - 1) \right]
                   \frac{\ln(\tilde{r})}{\ln(\tilde{R}_o)}
\end{equation}

\end{appendices}




\section*{References}
\bibliographystyle{ieeetr}
\bibliography{viscoplastic_damper}









\end{document}